# The Roadmap of New Capabilities of High-Intensity Lasers in Material Design and Manipulation

Alexander V. Bulgakov[1], Yury V. Ryabchikov[1], Yoann Levy[1,2], Nathan T. Goodfriend[3], Inam Mirza[1], Petr Hauschwitz[2], Vladimir A. Volodin[4,5], Martin Divoky[2], Carlos Doñate-Buendía[6], Bilal Gökce[7], Nadezhda M. Bulgakova[1,*]

[1] FZU - Institute of Physics of the Czech Academy of Sciences, Na Slovance 1999/2, 182 00 Prague 8, Czech Republic

[2] HiLASE Centre, FZU - Institute of Physics of the Czech Academy of Sciences, Za Radnicí 828, 25241 Dolní Břežany, Czech Republic

[3] Energy Saving Trust, Prospect House, 5 Thistle Street, Edinburgh, EH2 1DF, UK

[4] Rzhanov Institute of Semiconductor Physics SB RAS, Ac. Lavrentieva ave. 13, 630090, Novosibirsk, Russia

[5] Novosibirsk State University, Pirogova Str. 1, 630090, Novosibirsk, Russia

[6] GROC-UJI, Instituto de Nuevas Tecnologías de la Imagen, Universitat Jaume I, Castellón 12071, Spain

[7] Materials Science and Additive Manufacturing, School of Mechanical Engineering and Safety Engineering, University of Wuppertal, 42119 Wuppertal, Germany

[*]Corresponding author, e-mail: bulgakova@fzu.cz

## Abstract

One of the current trends of laser applications in material science is using high-intensity lasers to provide fast and efficient surface or volume modifications for achieving controllable material properties, synthesis of novel materials with desired functionalities, and upscaling laser technologies with industry-demanded throughputs. Depending on the parameters (wavelength, pulse energy and duration, repetition rate), lasers can offer versatile solutions for scientific and industrial applications, starting from exploring the fundamental physics of warm dense matter and molecular chemistry at ultrashort timescales to large-scale fabrication of surfaces with anti-bacterial, tribological, hydrophobic, or hydrophilic properties.

The objectives of this Chapter are to provide a review of recent advancements in several laser application fields, which involve high-intensity lasers, both ultrashort (femto- and picosecond) and short (nanosecond). After summarizing general trends in high-intensity laser processing of materials, we will first focus on the new opportunities offered by high-intensity lasers for the controlled synthesis of multielement nanoparticles for catalytic and theranostic applications. Then, the blister-based laser-induced forward transfer (BB-LIFT) technique will be presented, allowing a one-step, high-precision printing of nanomaterials, including 2D material flakes, on any substrates. The next section will discuss the selective crystallization of amorphous (as prepared) semiconductor nanoscale materials. The processes enabling high selectivity of crystallization into the desired phase using ultrashort powerful lasers will be analyzed. After that, opportunities for using high-power lasers will be discussed for upscaling surface nanostructuring with high throughput for bio-medical and industrial applications. Finally, an introduction to the Open Access program of the HiLASE Centre, which is targeted at offering users high-intensity beam time, will be given.

# 6. Open Access Program of the HiLASE Centre

    6.1 Importance of Open Access Programs of Large-Scale Laser Infrastructures

    6.2 Overview of Laser Lines for Open Access Experiments at the HiLASE Centre

    6.3 Examples of Open Access Campaigns on HiLASE Beamlines

# 7. Conclusions

# 1. Introduction

From the publication by Theodore Maiman on the first operating laser [Maiman, 1960] followed in 1961-1962 years by demonstrations of helium-neon, neodymium-glass, and gallium-arsenide lasers, a real race started on the development of lasers based on different gain media and in different spectral ranges [Hecht, 2005], which was accompanied by the race in their applications [Allen, 1966; Ready, 1997]. A significant milestone that gave the further enormous push for ultrafast laser technologies was achieved by Gerard Mourou and Donna Strickland, who developed chirped pulse amplification of femtosecond laser intensity [Strickland and Mourou, 1985]. Nowadays, most technologies involve laser sources in manufacturing processes and/or as a part of the final product. One of the examples is shown in Figure 26.1, where fabrication processes of micro-LED full-color displays involving lasers are specified as nano-processing, detection of defects, repairing and mass transfer, patterning, and morphology modification [Lai, 2023]. The race of new laser developments has not yet been finished, which in particular is directed to shorter laser pulses and higher laser powers [Reid, 2016] for specific technological applications to achieve high-throughput laser processing of materials at a high quality.

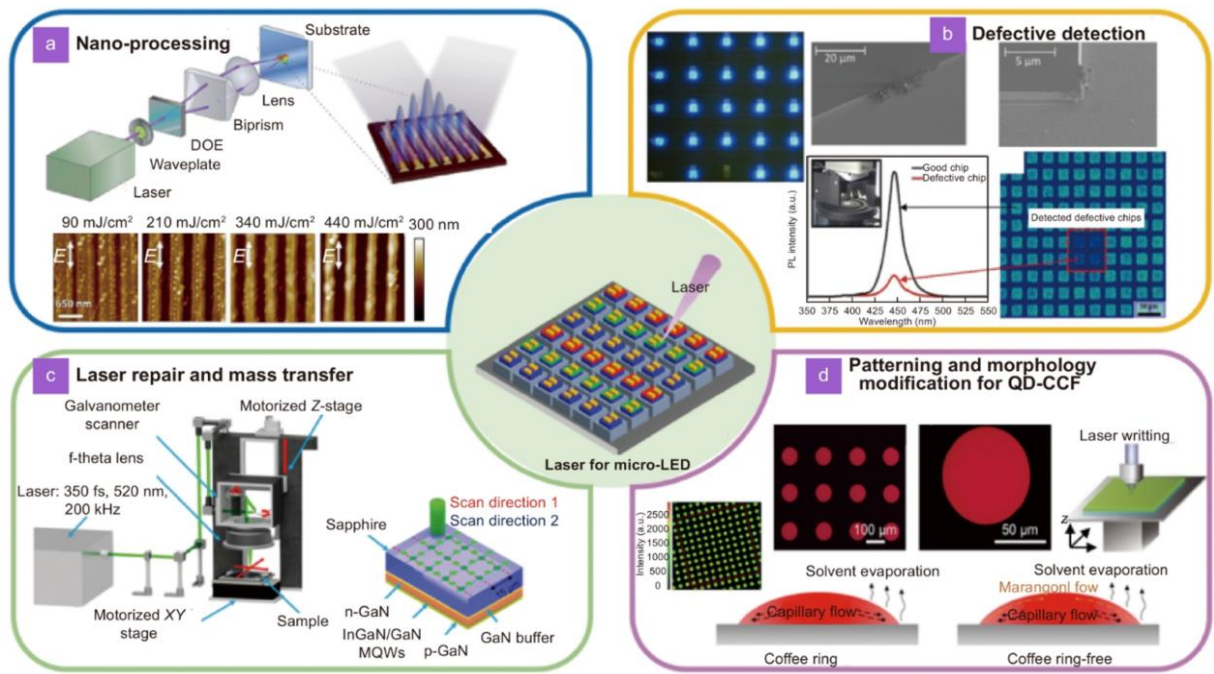

**Fig. 26.1:** The developments of laser applications in the fabrication processes of micro-LED full-color displays [Lai, 2023]: (a) Nano-processing, (b) Defective detection, (c) Laser repair and mass transfer, (d) Patterning and morphology modification for quantum-dot color conversion films. Reprinted from [Lai, 2023] with permission of Lai et al. (2023), S.Q. Lai et al. Applications of lasers: A promising route toward low-cost fabrication of high-efficiency full-color micro-LED displays. Opto-Electron. Sci. **2**:230028. Copyright 2023, Authors.

In this chapter, we give an overview of several applications of high-intensity lasers in material science and technology that already provide fast large-scale results in laser fabrication or are promising for high-throughput high-quality laser processing and material modification. This includes the synthesis of hybrid and high-entropy alloy nanoparticles in a liquid environment, selective laser annealing with nanoscale precision, blister-based laser-induced forward transfer (BB-LIFT), and large-area surface micro/nanostructuring (Figure 26.2). Most of these applications require high-power or high-intensity laser pulses, which are often unavailable in scientific/industrial laboratories and can be accessed through Open Access services in large laser facilities. Some examples of Open Access experiments at the HiLASE Centre will be provided that include the generation of alpha particles and studies of laser-induced momentum transfer in the context of space propulsion.

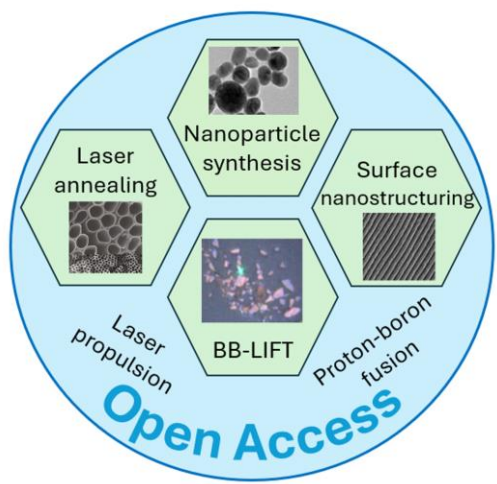

**Fig. 26.2:** Scheme of the processes overviewed in this Chapter with some examples of the Open Access experiments performed at the HiLASE Centre. Note that the activities in the green hexagon areas can also be supported by the Open Access program.

# 2. Laser Synthesis of Nanoparticles in Liquid Surroundings

Pulsed laser ablation in liquids (PLAL) is one of the widely used techniques for the synthesis of colloidal solutions of semiconductor, metallic, and bimetallic nanoparticles. It not only allows varying physical and chemical properties of nanostructures but also uses chemically clean experimental conditions requiring no chemical surfactants or stabilizing agents. However, a combination of semiconductor and metallic materials using laser synthesis is still a challenging task requiring deeper studies of physicochemical processes taking place between them. Subsection 2.1 summarizes some recent achievements in the field of synthesis of semiconductor-metallic hybrid nanostructures using the laser ablation technique. Another challenging task in PLAL is the generation of multielement nanomaterials, especially given the recent development of a new type of alloy, called high entropy alloys (HEAS), promising breakthroughs in many applications. Careful control of the composition and internal structure of nanoparticles, in addition to size and shape, is required in this case. Subsections 2.2 and 2.3 consider the main principles and recent achievements of the PLAL synthesis of multielement nanosystems with a particular focus on HEA nanoparticles.

## 2.1  Hybrid Nanoparticles for Theranostics

Formation of nanohybrids allows considerable improvement of properties of both semiconductor and metallic nanostructures due to the combination of specific properties inherent to these elements. Pulsed laser ablation in liquids (PLAL) is one of the most promising and facile routes of formation of such kinds of nanostructures using highly pure solid targets (chemical purity >99.9 %) immersed in any clean liquids (e.g., deionized water). Accompanied with controlled variation of their properties, this method is one of the most promising techniques for different healthcare applications due to easy adjustment of features of chemically clean nanoparticles for a required application task. By combining with plasmonic or magnetic nanostructures, this approach can easily extend performance of different semiconductor nanostructures such as silicon, silicon carbide, carbon and germanium showing promising applications for optical bioimaging [Yang, 2009], drug delivery (Figure 26.3) [Chen, 2024a], enhancing photodynamic therapy efficiency [Makhadmeh, 2024], controlling amino acid uptake [Sánchez-Cano, 2024], biosensing and cell imaging [Azar, 2024], cancer cell labeling [Boksebeld, 2017], antimicrobial activity [Lazarević, 2024] and near-infrared dual-modality imaging [Chen, 2024b].

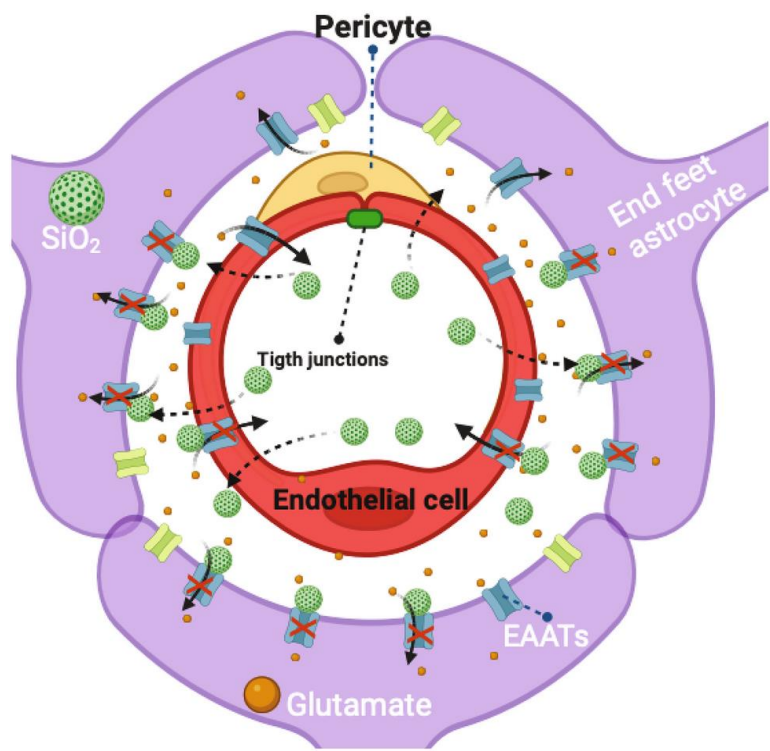

**Fig. 26.3:** Silica nanoparticles decrease glutamate uptake in blood–brain barrier components [Sánchez-Cano, 2024]. Reprinted from [Sánchez-Cano, 2024], Silica nanoparticles decrease glutamate uptake in blood–brain barrier components, F. Sánchez-Cano et al., Neurotox Res. **42**:20, Copyright 2024), with permission of Springer Nature.

Indeed, such a possibility of merging various IV group semiconductor nanoparticles with plasmonic gold elements has been shown recently using picosecond PLAL [Ryabchikov, 2024]. It allows successful formation of plasmonic nanohybrids based on these elements having quite narrow size distribution (~ 4-5 nm) with small mean size of nanoparticles (<10 nm) depending on the used PLAL approach (Figure 26.4) that can be associated with peculiarities of laser-assisted interaction between semiconductor and gold nanoclusters in surrounding water [Ryabchikov, 2024]. It can point out a considerable difference with the growth of single-element nanoparticles and the importance of interaction between nanoclusters of different nature elements. The successful merging of semiconductor and metallic species in one nanoparticle is confirmed by energy-dispersive X-ray spectroscopy (EDX) studies performed on single nanohybrids. They reveal strong size-dependent chemical composition with a lower contribution from gold for smaller nanoparticles (Figure 26.4b). Moreover, different ratios between semiconductor and metallic elements in the nanohybrids can point out different mechanisms of the nanohybrid formation in the cases of laser ablation (LA) and laser co-fragmentation (LcF) approaches.

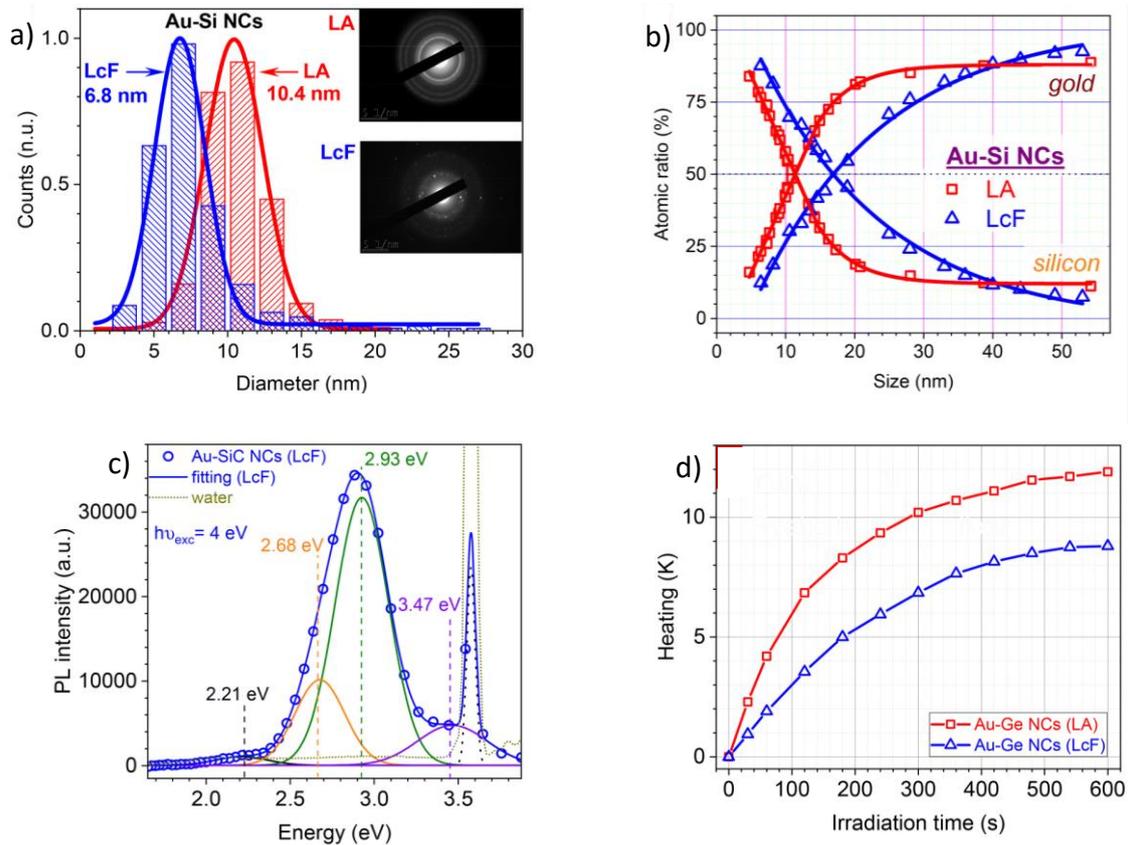

**Fig. 26.4:** (a) Size distribution and diffraction patterns of plasmonic Au-Si nanohybrids; (b) their size-dependent chemical content; (c) photoluminescence (PL) of Au-SiC nanohybrids formed by the LcF approach; (d) fs laser-induced heating of colloidal solutions of Au-Ge nanohybrids [Ryabchikov, 2024]. Reprinted and adapted from [Ryabchikov, 2024], Design of "green" plasmonic nanocomposites with multi-band blue emission for ultrafast laser hyperthermia Yu.V. Ryabchikov, Nanoscale, **16**:19453–19468, Copyright 2024. Retrieved from https://doi.org/10.1039/D4NR03120B.

The different chemical compositions of nanohybrids are also reflected in their optical properties. Thus, absorbance efficiency is affected by the PLAL approach used: either direct LA or LcF. Indeed, stronger plasmonic properties are found for all nanohybrids synthesized via the direct laser ablation approach that can be related to a larger amount of metals in them [Ryabchikov, 2024]. This point is very important for further applications of nanohybrids to molecule identification using Surface-Enhanced Raman Scattering (SERS). Additionally, the efficiency of the plasmonic properties of Au-Si and Ag-Si nanohybrids can be improved by using higher laser fluence [Saraeva, 2018]. Indeed, the intensity of localized surface plasmon resonance peaks increases with laser energy in both cases, indicating larger metallic content in them while the spectral position remains the same. Dependence of the absorption coefficient on the applied laser intensity $I_0$ obey the following law: $\alpha \propto I_0^x \ln(I_0/I_{th})$ with $x \approx 1.6$ for Ag-based and $x \approx 1.6$ for Au-based nanoparticles. Here, the factor $\ln(I_0/I_{thresh})$ gives the area of the laser ablation spot and the intersection of the fitting curve with the $x$-axis defines ablation threshold values, $I_{th}$, of 0.25 GW/cm$^2$ and 0.25 GW/cm$^2$, respectively [Saraeva, 2018].

Moreover, all nanohybrids based on various IV group semiconductors also reveal effective multi-band photoluminescence upon excitation of a Xe arc lamp at ~300 nm that can be further used for labeling purposes (Figure 26.4c). The presence of several emission bands indicates different recombination mechanisms that can be assigned to the following transitions:

$$\text{Si}^0 \rightarrow E_v \ (2.9 \text{ eV}), \ \text{Si–O–Si} \ (2.6\text{–}2.7 \text{ eV}), \ E_c \rightarrow \text{Si}^0 \ (2.1\text{–}2.2 \text{ eV}).$$

At the same time, no excitonic emission is detected from such nanohybrids that can be associated with relatively large nanoparticle size and/or due to nonradiative recombination with surface or bulk defects. The intensity ratio between different PL bands is varied depending on the used PLAL approach because of the change of corresponding radiative and nonradiative recombination mechanisms. Stronger PL for all nanohybrids prepared via the LcF can be due to smaller nanoparticles' size and larger involvement of oxygen molecules responsible for the emission. Similar shapes of the PL spectra for different nanohybrids can point out that the main emission mechanisms remain the same for all nanohybrids while the ratio between radiative and nonradiative recombination is changed.

It is worth noting that laser-induced structural modification can be employed not only for nanostructures obtained by PLAL but also for those prepared by chemical methods. In particular, the optical properties of fluorescent carbon quantum dots (C QDs) prepared by a chemical synthesis using a mixture of urea and anhydrous citric acid are significantly affected by laser irradiation in the presence of silver and gold solid targets [Ryabchikov and Zaderko, 2023]. This treatment results in the appearance of strong plasmonic properties of the chemically synthesized C QDs, whose efficiency and spectral position can be easily controlled by laser irradiation time and metallic target, respectively. Nevertheless, the presence of metallic elements in such nanohybrids leads to photoluminescence quenching that becomes stronger with increasing the metallic content [Ryabchikov and Zaderko, 2023]. However, a higher metallic content in these nanohybrids improves their laser-induced heating ability, making them promising for hyperthermia applications [Ryabchikov and Zaderko, 2023].

The presence of plasmonic metal with a low specific heat capacity also affects their thermal properties. So, the formation of a semiconductor-metallic structure leads to stronger fs laser-induced heating of colloidal solutions of semiconductor-based plasmonic nanohybrids [Ryabchikov, 2024]. This ability makes nanohybrids perspective candidates for photothermal therapy (PTT) applications against cancer cells [Huang, 2017]. By adapting the chemical composition of nanohybrids and, hence, their multi-modal performance, one can design nanoagents that might be able to apply several specific actions sensitized by external influence.

Employing nanohybrids for different healthcare applications requires a large number of corresponding nanostructures that can be obtained by multigram-scale production [Gurbatov, 2023]. It can be done via irradiation of Si NPs colloidal solution by a pulsed laser in the presence of $HAuCl_4$ leading to formation of Au NPs on the silicon nanoparticle surface through the thermal reduction of $[AuCl_4]^-$ species and further mixing of the silicon and gold phases upon remelting and recrystallization [Gurbatov, 2023]. This technique allows changing the chemical composition and corresponding optical properties of nanohybrids by varying the amount of $HAuCl_4$. Moreover, it also affects the structure of hybrid nanostructures, increasing gold contribution (XRD peaks at 2θ angle 38.1°, 44.4°, and 81.8°, see Figure 26.5a [Gurbatov, 2023]). Additionally, new gold crystalline planes appear due to the presence of reflexes at 64.7° and 77.6° angles accompanied by gold silicide formation (e.g., $A_{81}Si_{19}$): 2θ ≈ 32.2°, 39.7°, and 67.1° [Gurbatov, 2023]. Furthermore, the most intense XRD peak related to gold changes from Au (111) to Au(200), indicating the following most favorable orientation of Au(111)||Si(111) due to minimum crystal lattice mismatch between gold and silicon.

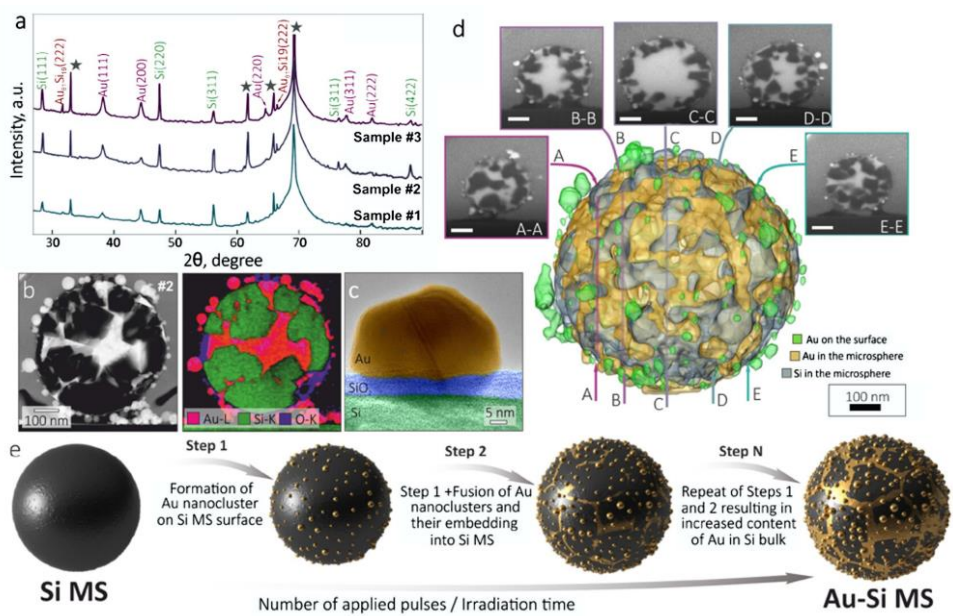

**Fig. 26.5:** (a) XRD patterns of Au-Si nanohybrids produced at different HAuCl$_4$ concentrations, (b) their TEM image and EDX mapping, (c) HR-TEM images of an isolated Au NP on the surface of Au-Si nanohybrid, (d) 3D model of an isolated Au-Si nanohybrid, (e) schematic illustration of Au-Si nanohybrids formation and morphology evolution during PLAL [Gurbatov, 2023]. Reprinted from [Gurbatov, 2023], Multigram-scale production of hybrid Au-Si nanomaterial by laser ablation in liquid (LAL) for temperature-feedback optical nanosensing, light-to-heat conversion, and anticounterfeit labeling, S.O. Gurbatov et al., ACS Appl. Mater. Interfaces **15**:3336–3347, Copyright 2023, with permission of American Chemical Society.

Apart from using monocrystalline wafers for producing hybrid Au-Si nanostructures, porous silicon targets prepared by electrochemical etching can also be used, allowing changing the size of Au-Si nanohybrids via modification of target porosity. In particular, a 3-fold size reduction of nanoparticles is obtained by changing a common monocrystalline Si wafer with the porous one affected by ns laser ablation [Gurbatov, 2024]. Further ns laser treatment in the presence of HAuCl$_4$ leads to the formation of Au-Si products, whose structure is affected by the concentration of the precursor. Indeed, variation of HAuCl$_4$ amount results in changing the structure of Au-Si nanostructures from "decorated" at lower precursor concentration to "hybrid" at greater one. Thus, decorating of silicon nanoparticle surface is obtained at lower HAuCl$_4$ concentration due to the thermal reduction of the metal precursor on the Si NPs surface, while the composite structure is formed at larger concentration with irregular Si nanocrystals wrapped by a gold matrix (Figure 26.6, [Gurbatov, 2024]). In the last case, the following processes can result in the formation of hybrid structures:

- thermal-induced reduction of the gold precursor molecules absorbed on the target surface upon its laser ablation;
- similar reduction process occurring on the surface of laser generated nanoparticles absorbing subsequent laser pulses;
- laser-assisted fusion of the NPs.

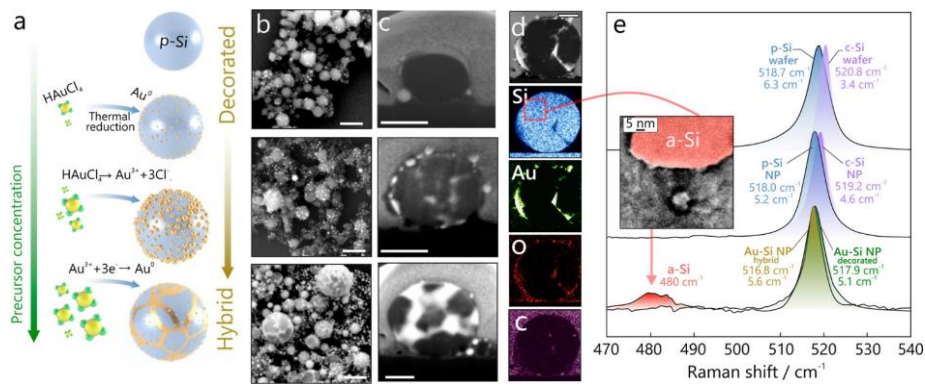

**Fig. 26.6:** (a) Schematic formation of Au-Si products and structural evolution upon increasing the HAuCl$_4$ precursor, (b) SEM images of Au-Si products obtained at different precursor amount, (c) cross-sectional SEM image of an isolated Au-Si nanostructures, (d) STEM image and EDX composition of an isolated Au-Si nanostructure, (e) Raman spectra of different silicon-based nanoparticles [Gurbatov, 2024]. Reprinted from [Gurbatov, 2024], Au−Si nanocomposites with high near-IR light-to-heat conversion efficiency via single-step reactive laser ablation of porous silicon for theranostic applications, S.O. Gurbatov et al., ACS Appl. Nano Mater. 7:10779−10786, Copyright 2024, with permission of American Chemical Society.

Here, organic solvent molecules containing –OH group can act as a reducing agent during the chemical reaction that occurs between thermally decomposed species (HAuCl$_4$ → Au$^{3+}$ + 3Cl$^-$) followed by the reduction of gold ions: Au$^{3+}$ + 3e$^-$ → Au$^0$. It is worth noting that such integration of gold nanoclusters onto a silicon core results in some increase of the Au-Si nanohybrid size. The aforementioned experimental variation provokes a considerable blue shift (~2.1 cm$^{-1}$) in the Raman spectra by changing the target from a monocrystalline to nanostructured one, accompanied by a strong signal broadening due to the phonon confinement effect or residual stresses. "Decorated" gold-silicon nanostructures exhibit a Raman signal similar to that from porous silicon nanoparticles, indicating the absence of the crystal lattice modification. However, a larger concentration of the precursor leads to the penetration of gold into the silicon core, forming irregular inclusions. It affects the Raman spectra of Au-Si nanohybrids, indicating the modification of the recrystallization process and additional contribution from the created amorphous phase detected at 480 cm$^{-1}$ and additionally confirmed by HR-TEM observation [Gurbatov, 2024].

A similar approach of the formation of gold-silicon nanohybrids is demonstrated in the following work [Kutrovskaya, 2017]. However, in this case, the addition of gold nanoparticles into Si NPs colloidal solutions prepared by CW laser ablation is realized instead of HAuCl$_4$. Using CW laser ablation leads to the formation of widely dispersed nanoparticles with large mean sizes (~100 nm) whose size increases with the laser irradiation power (Figure 26.7) [Kutrovskaya, 2017]. Further addition of small (10 nm) gold nanoparticles into prepared Si NPs colloidal solutions accompanied with laser irradiation from a nanosecond laser (1060 nm, 100 ns, 20 kHz, 1 mJ) results in the formation of hybrid nanostructures due to the electrostatic attraction of their constitutes [Kutrovskaya, 2017]. As a result, metallic nanoparticles are deposited on the surface of semiconductor ones, similar to the aforementioned case. The main mechanism of the formation of the hybrid structures is associated with the different charges of these two types of nanostructures. Here, Au NPs are negatively charged, while Si NPs are initially electrically neutral. Further laser-induced generation of free charge carriers can break chemical bonds at the nanoparticle surface, acquiring positive charges. Their further interaction can lead to the formation of stable hybrid Au-Si nanostructures, as illustrated in Figure 26.8 [Kutrovskaya, 2019].

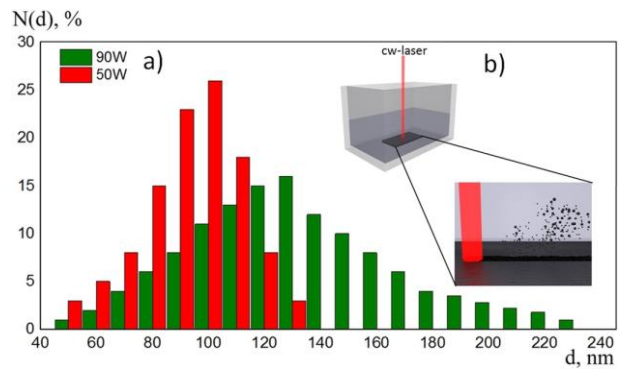

**Fig. 26.7:** Size distribution of nanoparticles formed by CW laser ablation at different laser power [Kutrovskaya, 2017]. Inset shows the scheme of CW laser ablation. Reprinted from [Kutrovskaya, 2017] The synthesis of hybrid gold-silicon nano particles in a liquid. Sci Rep 7, 10284, S. Kutrovskaya et al., Copyright 2017, with permission from Nature.

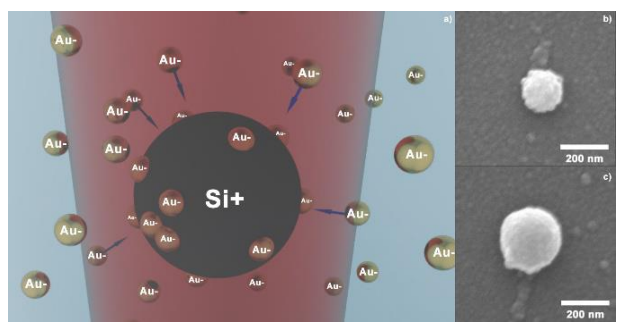

**Fig. 26.8:** Size distribution of nanoparticles formed by CW laser ablation at different laser power. The scheme of CW laser ablation [Kutrovskaya, 2019]. Reprinted from [Kutrovskaya, 2019], Hybrid gold-silicon systems with tuning optical properties, S. Kutrovskaya et al., J. Phys.: Conf. Ser. **1164**:012012, Copyright 2019, under Creative Commons Attribution 3.0. Retrieved from https://doi.org/10.1039/D4NR03120B.

To improve the performance of nanohybrids in healthcare applications, it can be required to adjust their chemical content or/and mean sizes, which can be done by changing PLAL experimental conditions. However, varying laser fluence affects neither the chemical content nor the size of hybrid nanoparticles, contrary to their single-component counterparts [Ryabchikov, 2019]. Nevertheless, a variation of the concentration of nanoparticles dispersed in a liquid medium considerably changes both the size distribution and chemical content of the nanohybrids [Ryabchikov, 2019]. Besides, it also affects the concentration of paramagnetic defects (~1 order of magnitude) as well as the intensity of their plasmonic properties that change the quality of the bacteria detection. Moreover, the efficiency of the plasmonic properties can be very easily adjusted by the duration of the laser ablation of metallic targets immersed in colloidal solutions of semiconductor nanostructures. Furthermore, controlling the size and coverage of gold nanoparticles on a silicon nanoparticle surface can also be done via the following procedure divided into a few steps (Figure 26.9) [Chaâbani, 2021]:
- heating of Si NPs to hydrolyze the surface;
- their functionalization with amine groups;
- binding of Au seeds with Si NPs surface;
- growth of gold shells.

By varying the amount of PVP molecules with respect to its concentration, three different types of heterostructures, such as Si/Au core–islands, Si/Au core–shell, and Si/Au core–shell oligomers, can be formed.

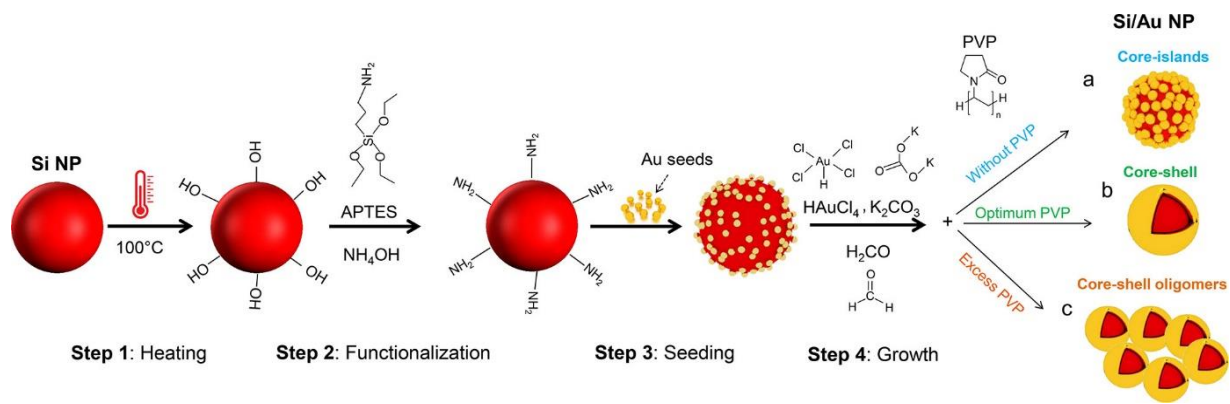

**Fig. 26.9:** Multistep synthesis of Au-Si nanohybrids with controlled size and structure [Chaâbani, 2021]. Reprinted from [Chaâbani, 2021], Si@Au Core–Shell Nanostructures: Toward a new platform for controlling optical properties at the nanoscale, W. Chaâbani et al., J. Phys. Chem. C **125**:20606–20616, Copyright 2021, with permission of American Chemical Society.

Apart from IV group semiconductors, metal oxide semiconductors such as $TiO_2$ can also be decorated with plasmonic metals by the PLAL technique, forming a hybrid nanostructure. Similar to Au-Si nanohybrids, Au-$TiO_2$ ones can also be formed due to the reduction of Au NPs onto the $TiO_2$ surface because of laser-induced melting and fusion (Figure 26.10) [Gurbatov, 2021]. The observed increase in size and spherical shape of hybrid Au-$TiO_2$ nanostructures in comparison with pristine titania NPs suggests that melting and fusion are key processes of the nanohybrid formation [Gurbatov, 2021]. Prolonged (2 hours) laser irradiation of $TiO_2$ nanoparticles having an irregular shape in the presence of $HAuCl_4$ precursor transforms them into spherically shaped Au-$TiO_2$. The increase of both $HAuCl_4$ concentration and irradiation duration provokes a larger number of gold nanoclusters adsorbed on the nanoparticle surface. Such nanohybrids can be formed due to the fusion of several pristine $TiO_2$ nanoparticles containing Au NPs on their surface, allowing gold to penetrate inside $TiO_2$ nanoparticles and doping Au-$TiO_2$ nanohybrids.

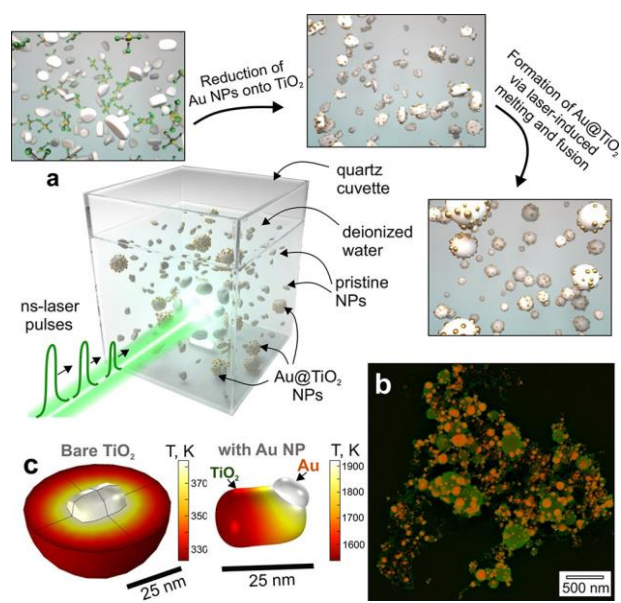

**Fig. 26.10:** Fabrication of Au-$TiO_2$ nanohybrids via PLAL. (a) Scheme of the fabrication process, (b) SEM image of nanoparticles produced upon irradiation of a suspension containing $TiO_2$ NPs, (c) Calculated temperature profile in the vicinity of a bare and Au-decorated $TiO_2$ NP upon irradiation with a single nanosecond laser [Gurbatov, 2021]. Reprinted from [Gurbatov, 2021], Black Au-decorated $TiO_2$ produced via laser ablation

These achievements highlight wide possibilities of the PLAL technique in the formation of different semiconductor-metallic nanohybrids using CW, ns, ps, or fs laser ablation, resulting in acquiring novel properties that can open up new horizons for semiconductor biomedical agents in the field of plasmonic biosensing, antibacterial applications, or temperature sensing.

## 2.2 Nanoparticles of High-Entropy Alloys

High-entropy alloys (HEAs) have emerged as a promising class of materials, distinguished by their multi-elemental composition and unique physicochemical properties [George, 2019; Tsai and Yeh, 2014]. They represent a rapidly evolving field in both fundamental and applied material sciences. On the one hand, from the fundamental point of view, such multi-principal element alloys offer new opportunities for understanding and controlling the material structure, stability, and functional properties through the control of configurational entropy [Miracle and Senkov, 2017]. On the other hand, HEAs are very attractive for various applications due to their unique mechanical, electrical, magnetic, and catalytic properties [Han, 2024a]. In particular, these alloys hold significant advantages for structural and material engineering as they provide a superior combination of strength and ductility [Yang, 2018] and have exceptional strength-to-weight ratio and corrosion resistance [Chang, 2020]. HEA studies have been primarily focused on bulk materials. However, multi-metallic nanoparticles (NPs) are of interest in a wide range of applications, especially since their properties can be different from those of bulk materials [Sun and Sun, 2024]. The reduced amount of HEA NPs compared to bulk HEAs is related to the additional challenges that the synthesis of HEA NPs poses. The multicomponent character of these NPs represents a challenge for chemical synthesis routes since each reagent has a different reactivity, including different reaction rates, threshold temperature, and possible interactions between them. The reaction path of each individual component of the HEA toward forming the desired alloy nanoparticle with homogeneous elemental distribution and controlled stoichiometry represents a challenge [Dey, 2023]. Specific synthesis routes have been developed, and alloy NPs with eight elements (Pt, Pd, Co, Ni, Fe, Cu, Au, Sn) were synthesized by carbothermal shock synthesis [Yao, 2018]. The remarkable method allows nearly uniform elemental distributions in single particles that are, furthermore, kinetically controllable in size, with a minimum diameter of 3 nm, by the shock duration. However, carbothermal shock synthesis only produces NPs immobilized on conductive, surface-oxidized carbon support materials, which limits the possibilities for large-scale applications. Another synthesis route that proved successful for the synthesis of quinary CrFeCoNiCu NPs of 9 nm diameter immobilized on graphene was mechanical ball milling of single metal powders mixed with graphene [Rekha, 2018]. This solid-state synthesis is more convenient than the carbothermal shock synthesis, but the authors did not achieve the targeted chemical distribution in the NPs. Niu et al. demonstrated a production route for CrCoNiCuAl NPs with 14 nm in average diameter based on the popular sol-gel auto combustion [Niu, 2017]. However, the size of the particles was larger than 100 nm, and the chemical composition was not evaluated in detail. Alternative methods such as microwave [Qiao, 2021], aerosol [Yang, 2020], high-temperature liquid shock [Cui, 2024], or liquid metal [Cao, 2023] -based synthesis have been proposed to expand the library of HEA NPs compositions.

Generally, it has been observed that HEA NPs synthesis is favored when the employed synthesis method achieves fast heating and cooling rates that can freeze the alloy composition directly after synthesis, minimizing elemental segregation that can lead to the variation of the stoichiometry and even the formation of elemental NPs instead of the desired alloy [Dey, 2023]. Laser-based HEA NPs synthesis methods provide fast heating. Laser scanning ablation in air has been shown to achieve the controlled synthesis of HEA NPs directly adsorbed on different substrates, highlighting that the formation process occurs within 5 ns, ensuring the combination of even dissimilar metallic elements with low solubility [Wang, 2022]. The preservation of the alloy composition after rapid laser synthesis has not been only proved through direct support on a substrate but alternative methods such as graphene encapsulation or the use of a shielding gas [Kim, 2024] have succeeded in synthesizing CrFeCoNiCu solid solution HEA NPs [Liu, 2024a] with productivity rates of up to 30 g/h. The laser synthesis approaches provide the required fast heating rates, however, encapsulation or direct deposition on substrates are essential to avoid elemental segregation during material cooling. HEA NPs laser synthesis in a solvent is an attractive approach to enhance the cooling rate and to freeze metastable phases and compositions produced during the fast heating generated by the laser interaction.

Over the past decades, the formation of colloidal solutions using pulsed laser ablation in liquids (PLAL), Figure 26.11, has attracted considerable attention due to its advantages over conventional NP synthesis methods, such as the reduction of chemical precursors. Besides the almost unlimited number of feasible target materials and NP compositions, PLAL-synthesized NPs in water exhibit surfaces free of any organic residuals. These advantages have given rise to the employment of PLAL as a well-established technique for the synthesis of alloy NPs [Waag, 2019]. For instance, the synthesis of magnetic ternary alloys such as Heusler-type alloys has attracted attention due to their potential magnetic shape memory effect and field-induced superior properties such as the magnetocaloric effect. Aksoy demonstrated the PLAL-synthesis of NiMnIn nanoalloys in the size range of 10-90 nm by ablating a NiMn-In target in water using a femtosecond laser [Aksoy, 2015]. However, theoretical results [Shih, 2018] indicate fundamental differences in the NPs genesis for ps and ns PLAL. Nanosecond pulses lead to NPs with a broad monomodal size distribution, while ps-PLAL typically leads to a bimodal size distribution explained by two different NP formation mechanisms. While smaller NPs are generated by nucleation and growth in an expanding metal-liquid mixing region, larger particles are formed due to hydrodynamic instabilities at the plume-liquid interface. Hence, the different NP formation kinetics can also affect the oxidation behavior of the formed NPs.

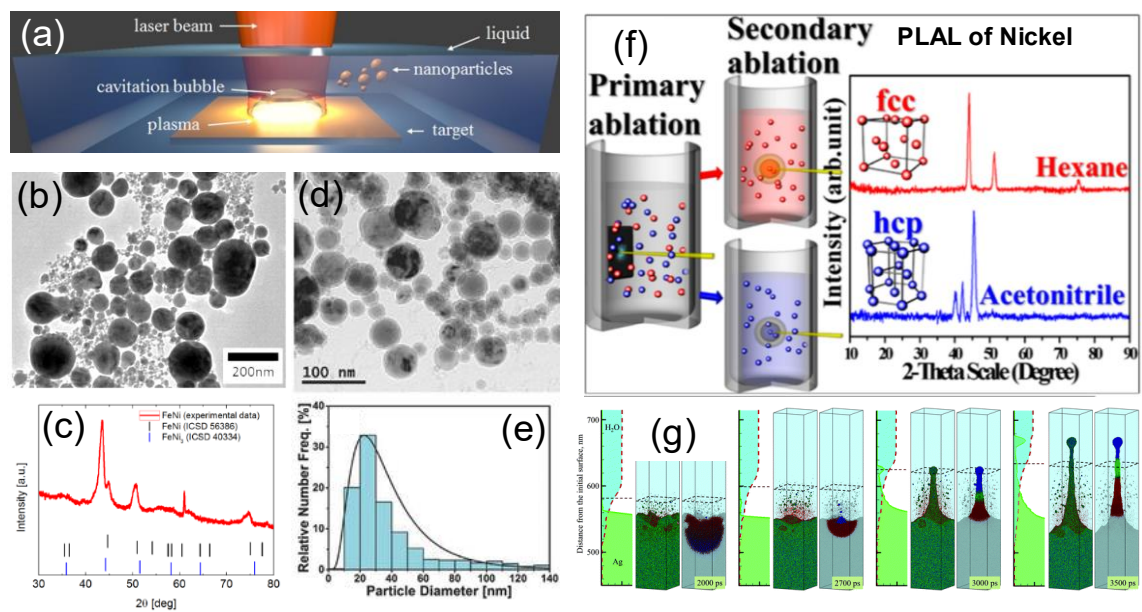

**Fig. 26.11:** Laser synthesis of alloy NPs (a) Illustration of PLAL [Gökce, 2015]; (b) TEM image of FeNi NPs obtained by PLAL [Khairani, 2023]; (c) XRD diffractogram of particles in (b) [Barcikowski, 2015]; (d) TEM image of laser synthesized FeNi NPs [Zhang, 2017]; (e) size distribution of FeNi NPs [Zhang, 2017]; (f) PLAL of Ni leading to different phases of Ni depending on the used solvent [Jung and Choi, 2014] (g) Atomistic simulation of PLAL of Ag showing the formation of two different populations of nanoparticles with different kinetics possibly leading to different oxidation behavior of each population [Shih, 2018]. (a) Reprinted from [Gökce, 2015], Ripening kinetics of laser-generated plasmonic nanoparticles in different solvents. B. Gökce et al., Chem. Phys. Lett. **626**:96–101 (2015). Copyright 2015, with permission from Elsevier. (b) Reprinted from [Khairani, 2023], Solvent influence on the magnetization and phase of Fe-Ni alloy nanoparticles generated by laser ablation in liquids, I.Y. Khairani et al., Nanomaterials **13**:227, Copyright 2023, under Creative Commons Attribution 4.0. Retrieved from https://doi.org/10.3390/nano13020227. (c) Reprinted from [Barcikowski, 2015], Solid solution magnetic FeNi nanostrand-polymer composites by connecting-coarsening assembly, S. Barcikowski et al., J. Mater. Chem. C Mater. **3**:10699–10704, Copyright 2015, with permission of Royal Society of Chemistry. (d) and (e) Reprinted from [Zhang, 2017], Laser synthesis and processing of colloids: Fundamentals and applications, D. Zhang et al., Chem. Rev. **117**:3990–4103, Copyright 2017, with permission of American Chemical Society. (f) Reprinted from [Jung and Choi, 2014], Specific solvent produces specific phase Ni nanoparticles: A pulsed laser ablation in solvents, H.J. Jung and M.Y. Choi, J. Phys. Chem. C **118**:14647–14654, Copyright 2014, with permission of American Chemical Society. (g) Reprinted from [Shih, 2018], Two mechanisms of nanoparticle generation in picosecond laser ablation in liquids: the origin of the bimodal size distribution, C.-Y. Shih et al., Nanoscale **10**:6900–6910, Copyright 2018, under Creative Commons Attribution 3.0. Retrieved from https://doi.org/10.1039/C7NR08614H.

PLAL can be employed to synthesize HEA NPs in the desired liquid media. Furthermore, the employment of picosecond lasers allows to increase in the production rate of the process compared to femtosecond and nanosecond pulses [Zhang, 2017b; Dittrich, 2019]. The suitability of PLAL to synthesize quinary transition metal alloy NPs by picosecond-pulsed laser ablation was confirmed for equimolar CrFeCoNiMn bulk sheets surrounded by ethanol to reduce the HEA NPs oxidation, Figure 26.12 [Waag, 2019]. The fast kinetics of the nanoparticle condensation in the laser-induced plasma and their surface passivation in the supercritical liquid confinement enable the synthesis of ultrasmall nanoparticles of less than 3 nm in diameter Figure 26.12b, which are homogeneous and very close to the chemical composition and structure of the HEA ablation targets. Indeed, the influence of the ablated target on the resulting HEA NPs composition and size has been evaluated, showing that single and polycrystalline bulk HEA targets do not provide a significant improvement of the process compared to targets prepared by elemental powder mixing and pressing. For every target, the resulting CrFeCoNiMn NPs exhibit oxidation below 20 at% with a similar size distribution and homogeneous elemental distribution [Tahir, 2024]. A key feature of PLAL synthesis is that the HEA NP colloids benefit from laser-induced electrostatic stabilization, which guarantees long-term colloidal stability without using any stabilizer. As proven by analytical ultracentrifugation Figure 26.12b, ultrasmall spatially isolated nanoparticles represent a major fraction of the synthesis product. The described features of HEA NPs PLAL synthesis with CrFeCoNiMn as model material have been confirmed for other HEAs such as HfNbTaTiZr [Jahangiri, 2023], AuAgCuPtNi [Lin, 2020], or AlCrCuFeNi [Rawat, 2024].

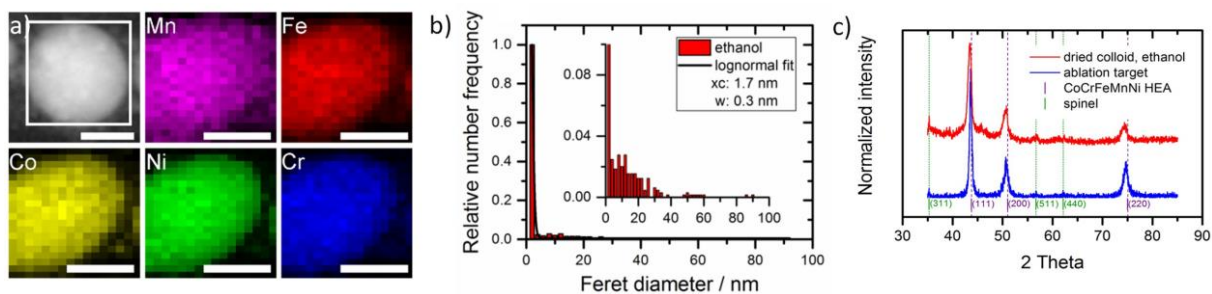

**Fig. 26.12:** Synthesis of CrFeCoNiMn NPs by PLAL [Waag, 2019]. a) Chemical composition of a single NP. TEM image and elemental maps show the homogeneous elemental distribution of Cr, Fe, Co, Ni and Mn. Scale bars are 25 nm. b) Size distributions of CrFeCoNiMn NP colloids determined by TEM. c) Lattice analysis by XRD. X-ray diffractograms of the ablation target (blue line) and a dried colloid produced in ethanol (red line). Reprinted and adapted from [Waag, 2019], Kinetically-controlled laser-synthesis of colloidal high-entropy alloy nanoparticles, F. Waag et al., RSC Adv. **9**:18547–18558, Copyright 2019, with permission of Royal Society of Chemistry.

While PLAL offers a versatile and simple method to generate HEA NPs, it generally exhibits low productivity. Nevertheless, upscaling strategies have been developed to increase PLAL productivity on an industrial scale. The approaches are based on spatially bypassing the vapor cavitation bubbles. In the first case, a laser with high laser repetition rates is combined with a high-speed polygon scanner, thus avoiding the bubble shielding effect by increasing the displacement speed and the inter-pulse distance on the target [Streubel, 2016]. In the second case, a diffractive optical element (DOE) is added to the PLAL system to split the beam into multiple beams. The reduction of the pulse energy per beam is compensated by decreasing the repetition rate of the laser pulse, hence achieving maximum productivity per each of the multiple beams while increasing the inter-pulse distance due to the repetition rate reduction that allows reducing pulse shielding by the cavitation bubbles [Khairani, 2024; Gatsa, 2024]. The employment of DOEs in PLAL has been tested for CrFeCoNiMn NPs production, confirming a significant productivity increase (Figure 26.13).

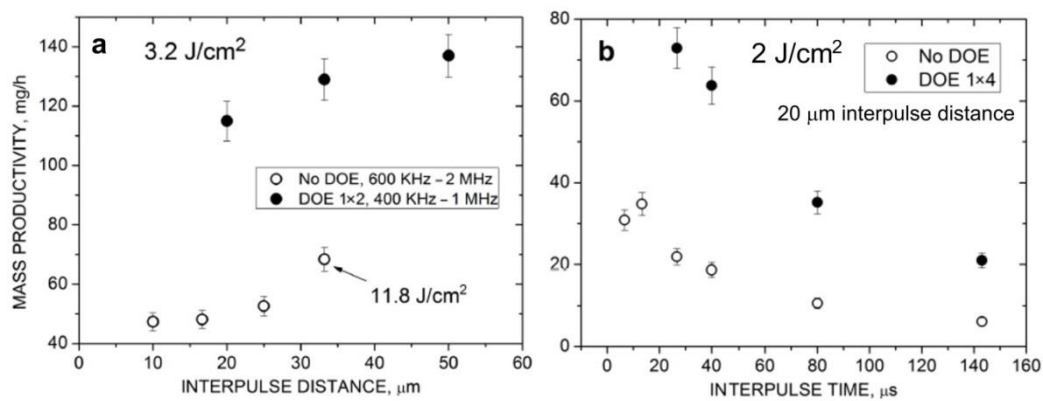

**Fig. 26.13:** Productivity of CrFeCoNiMn high entropy alloy NPs by ps PLAL in water with and without DOEs as a function of (a) inter-pulse distance and (b) inter-pulse time [Gatsa, 2024]. Reprinted from [Gatsa, 2024], Unveiling fundamentals of multi-beam pulsed laser ablation in liquids toward scaling up nanoparticle production, O. Gatsa et al., Nanomaterials **14**:365, Copyright 2024, under Creative Commons Attribution 4.0. Retrieved from https://doi.org/10.3390/NANO14040365.

## 2.3  Perspectives of Applications of Multicomponent Nanoparticles

According to their structure, multi-element nanoparticles can be classified [Koo, 2020] into five categories: metallic glass, alloy, intermetallic compounds, polyphase, and core-shell (Figure 26.14). In the case of HEA NPs, almost all the structural forms can be generated depending on the synthesis method and conditions [Waag, 2019; Löffler, 2021; Wang, 2022; Tahir, 2024]. The structural diversity of HEA NPs makes the field of studies and possible applications full of opportunities and challenges.

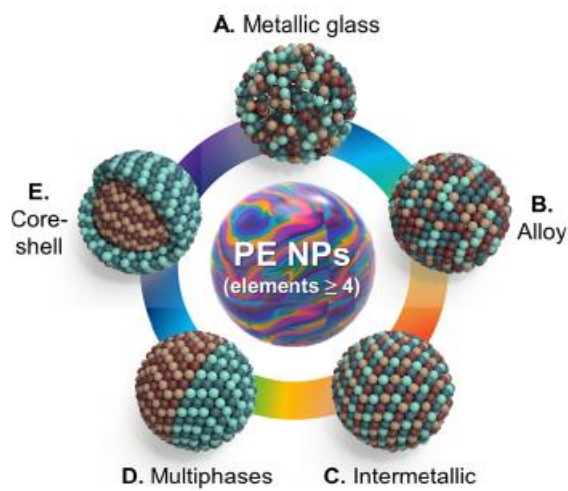

**Fig. 26.14:** Structure diagram of multi- (poli-) element nanoparticles (PE-NPs) [Koo, 2020]. Reprinted from [Koo, 2020], The design and science of polyelemental nanoparticles, W.T. Koo et al., ACS Nano **14**:6407–6413, Copyright 2020, with permission of American Chemical Society.

**Nanoparticles of High-Entropy Alloys**

Bulk high-entropy alloys exhibit improved mechanical, thermal, and chemical properties, with the possibility of using commonly available materials [Han, 2024b]. HEA NPs provide an extra advantage since the size reduction increases the surface-to-volume ratio of the particles, and hence surface surface-dependent effects, such as the catalytic response, can be increased. The discussed HEA NPs synthesis challenges have limited their incorporation in industrial processes, and most of the applications remain in the research field. However, the main applications field for the HEA NPs is currently electrocatalysis [Wang and Wang, 2022], specifically for the oxygen evolution reaction (OER) that is commonly employed for green hydrogen production, and Li-$CO_2$ batteries [Huang, 2022]. For the specific case of PLAL-produced HEAs, the focus has been primarily on the development of catalysts for OER.

CrMnFeCoNi NPs were synthesized by PLAL with controlled elemental content achieved by modifying the elemental powder ratio in the prepared target. A picosecond laser source was employed, and the Mn content was varied from 15 to 40 at% to evaluate its influence on the OER catalytic response of the produced NPs [Löffler, 2021]. It was observed that higher Mn content caused a reduction of the resulting high entropy alloy fraction, obtaining a more inhomogeneous elemental distribution and reducing the catalytic response of the produced NPs. The optimum Mn content was found to be 25%, exhibiting the highest current density, Figure 26.15.

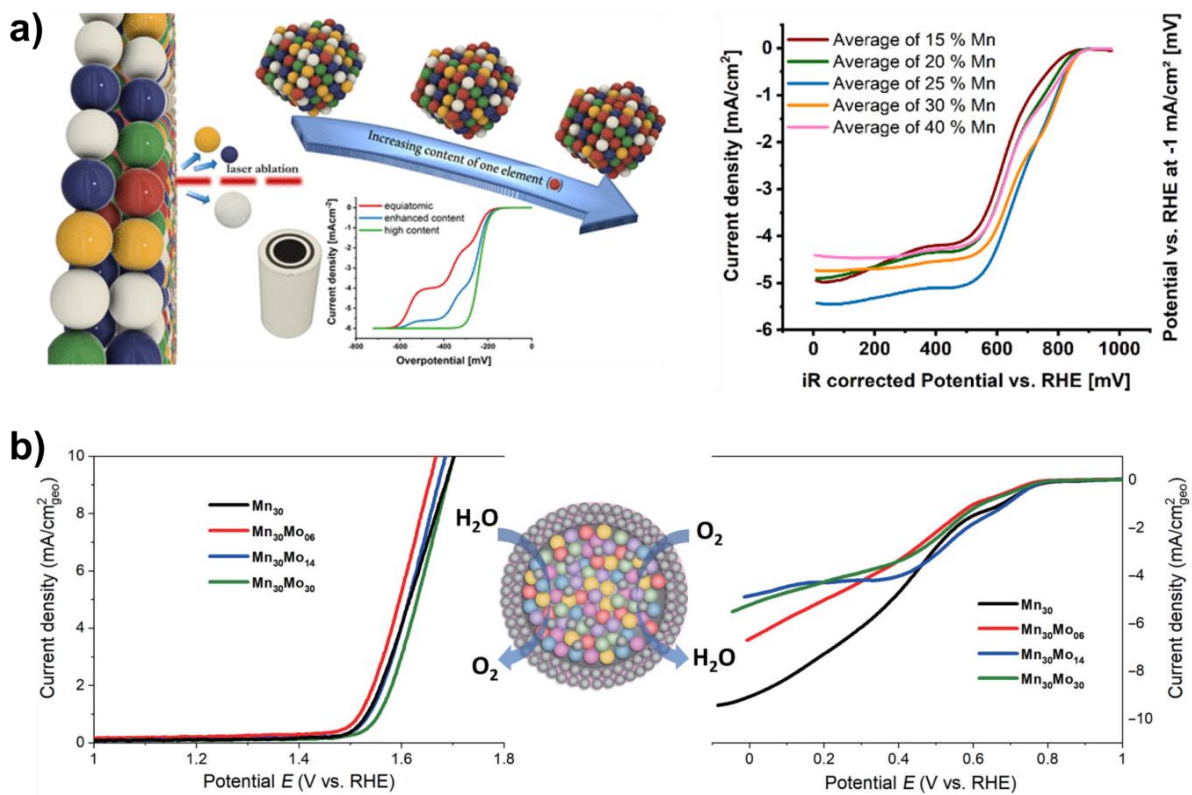

**Fig. 26.15:** High-entropy alloy NPs produced by ps PLAL with elemental control towards minimizing the overpotential required for OER. a) CrFeCoNiMn [Löffler, 2021] and b) CrFeCoNiMnMo [Johny, 2022]. (a) Reprinted from [Löffler, 2021], Comparing the activity of complex solid solution electrocatalysts using inflection points of voltammetric activity curves as activity descriptors, T. Löffler et al., ACS Catal. **11**:1014–1023, Copyright 2021, with permission of American Chemical Society. (b) Reprinted from [Johny, 2022], Laser-generated high entropy metallic glass nanoparticles as bifunctional electrocatalysts, J. Johny et al., Nano Res. **15**:4807–4819, Copyright 2022, with permission of Springer Nature.

In the case of PLAL-synthesized CrFeCoNiMnMo, the graphitic shell formed around the HEA NPs due to the irradiation in acetonitrile resulted in a metallic glass structure. The NPs were tested as catalysts for the complementary reactions of OER and oxygen reduction reaction (ORR) controlling the amount of Mo. The metallic glass NPs present enhanced OER activity compared to crystalline NPs, associating it with a denser defect distribution and higher concentration of unsaturated sites.

Due to the ever-increasing demand for energy globally, high-efficiency electrochemical energy storage has become an extremely important way for energy conservation. HEA NPs combine the advantages of large specific surface area with good electrochemical performance leading to a growing interest in exploiting these NPs in energy storage applications [Wang and Wang, 2024]. For instance, it has been demonstrated [González, 2016] that metal oxide layered HEA NPs exhibit excellent specific capacitance, cycling stability, and ultrafast charge transfer kinetics, paving the way for the improvement of supercapacitor performance.

Overall, HEA NPs are a rapidly evolving field with prospective applications in sectors such as high-temperature devices, medicine, supercapacitors, or catalysis. However, the current state of the art is still focused on the development of synthesis techniques that allow an industrial upscaling of production with elemental control and material versatility. PLAL offers material versatility, simplicity, and elemental control; however, the performance of the synthesized HEA

NPs by PLAL has been tested at lab scale with an enhancement of the catalytic activity, specifically as catalysts of the OER and ORR reactions.

**Hybrid Nanoparticles Combining Semiconductors and Metals**

Forming multicomponent nanoparticles via a combination of semiconductor and metallic nanoparticles can considerably improve their properties. The presence of plasmonic metal (Au or Ag) in semiconductor nanoparticles (Si NPs) results in their ability to identify not only organic dye molecules (rhodamine 6G) [Ryabchikov, 2019; Gurbatov, 2022] but also such bacteria as Listeria innocua ATCC 33090 and Escherichia coli W3110 with detection efficiency depended on chemical content of the nanohybrids [Kögler 2018]. Moreover, Ag-Si nanohybrids can detect plasmon-induced catalytic transformation of para-aminothiophenol (PATP) (lines at 1079 and 1575 cm$^{-1}$) to dimercaptoazobenzene (DMAB) due to appearance of additional 4 lines at 1145, 1192, 1394 and 1439 cm$^{-1}$ (Figure 26.16) [Gurbatov, 2022]. Increasing the concentration of AgNO$_3$ used for forming Ag-Si nanohybrids improves their SERS efficiency, nevertheless, leading to larger deviation of both SERS enhancement and heating efficiency limiting their application for reliable and precise measurements.

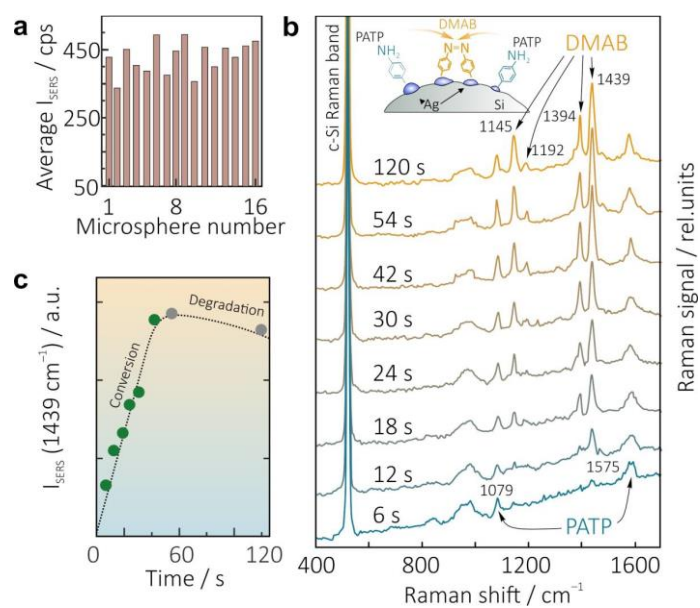

**Fig. 26.16:** (a) SERS yield of rhodamine 6G enhanced by Ag-Si nanohybrids, (b) time-resolved SERS spectra of RATP capping with Ag-Si nanohybrids (laser exposure time varies from 6 s to 120 s), (c) dependence of the Raman peak intensity (1439 cm-1) of RATP on laser exposure time [Gurbatov, 2022]. Reprinted from [Gurbatov, 2022], Ag-decorated Si microspheres produced by laser ablation in liquid: All-in-one temperature-feedback SERS-based platform for nanosensing, S. Gurbatov et al., Materials **15**:8091, Copyright 2022, under Creative Commons Attribution 4.0. Retrieved from https://doi.org/10.3390/ma15228091.

Hybrid nanoparticles can efficiently absorb incident near-IR laser radiation that can be further converted into local heat. Their laser-induced heating efficiency is increased compared to pristine semiconductor nanoparticles due to the presence of metals having low specific heat capacity values [Ryabchikov, 2024]. Moreover, various PLAL approaches also affect maximum temperature, which can be achieved by hybrid nanoparticles such as Au-Si, Au-SiC or Au-Ge, being higher for the laser co-fragmentation approach [Ryabchikov, 2024]. Comparing hybrid Au-Si and pristine Si NPs, one can observe stronger light-to-heat conversion performance of the hybrid ones. Variation of the chemical content of Au-Si nanohybrids can also change the light-to-heat transfer efficiency measured by Raman spectroscopy due to a more homogeneous distribution of plasmonic and semiconductor elements [Gurbatov, 2024]. In this case, laser-induced temperature increase provides a shift of the Raman position of the crystalline Si band (520.8 cm$^{-1}$) that can be further converted into the local temperature increase [Gurbatov, 2023].

Formation of hybrid metallic-semiconductor nanostructures can also affect cell viability due to changes of chemical content and laser irradiation power. So, viability tests of B16−F10 cells incubated with different amounts of NPs and assessed by AlamarBlue assay show slight toxicity of pristine Si and hybrid Au-Si nanoparticles. The presence of metallic species decreases the cell viability from 70 % to 60 % despite their significant difference in heating abilities (Figure 26.17). The performance of hybrid Au-Si nanostructures can be increased by increasing (i) their concentration or (ii) laser power density [Gerasimova, 2023].

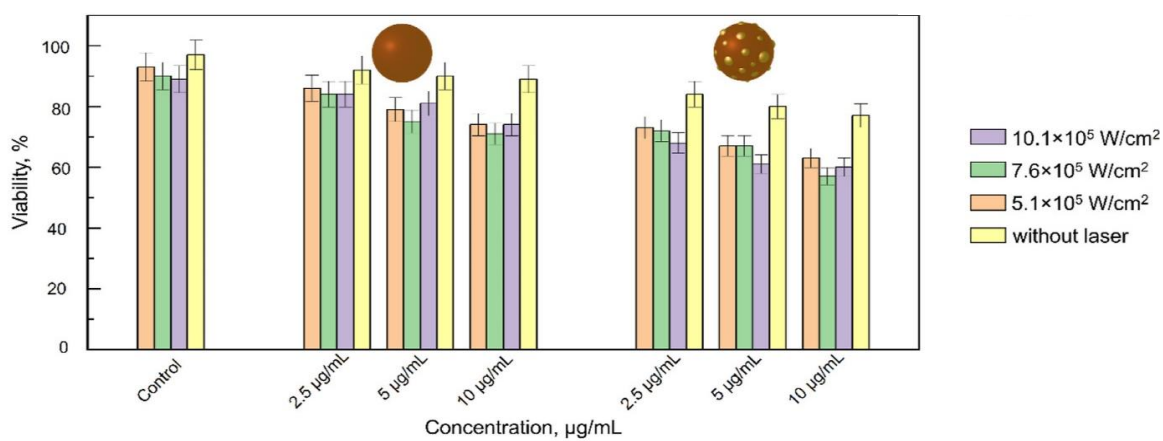

**Fig. 26.17:** Cell viability of pristine Si and hybrid Au-Si nanoparticles depending on their concentrations and used laser power [Gerasimova, 2023]. Reprinted from [Gerasimova, 2023], Single-step fabrication of resonant silicon-gold hybrid nanoparticles for efficient optical heating and nanothermometry in cells, E.N. Gerasimova et al., ACS Appl. Nano Mater. **6**:18848−18857, Copyright 2023, with permission of American Chemical Society.

Another interesting application of Au-Si nanohybrids is associated with their strong Raman and nonlinear photoluminescence signals. It makes them promising materials for state-of-the-art optical labels for anticounterfeiting applications based on a physically unclonable function (PUF) strategy [McGrath, 2019]. This approach is based on an uncontrolled process of the laser fabrication when created labels differ from each other, making them almost unclonable [Gurbatov, 2023]. The simple fabrication of such PUF labels, accompanied with reliable authentication procedure, make them promising candidates for both fundamental research and practical applications. A potential drawback of the PLAL synthesis related to randomness coupled with randomness of the drop-cast method provides unique conditions for designing unclonable tags with a low chance of their replications.

Unique fingerprints of such PUF tags can be based on different surface distributions of silicon Raman signal due to the random inclusion of semiconductor species in hybrid nanoparticles and related nonlinear photoluminescence distribution (Figure 26.18). Random distribution of both gold and silicon elements in the nanohybrids will vary characteristic fingerprints of the labels. By using just a small number of intensity levels ($N = 3$) for both characteristic signals, one can achieve a huge encoding capacity reaching ~101500 combinations. It permits reliable distinguish between true and duplicated labels with a low probability of false positives and negatives. One can also easily achieve larger values either by increasing pixel numbers analyzed in optical signals or by functionalizing Au-Si nanohybrids that can provide additional bands for nonlinear photoluminescence or Raman bands mapping.

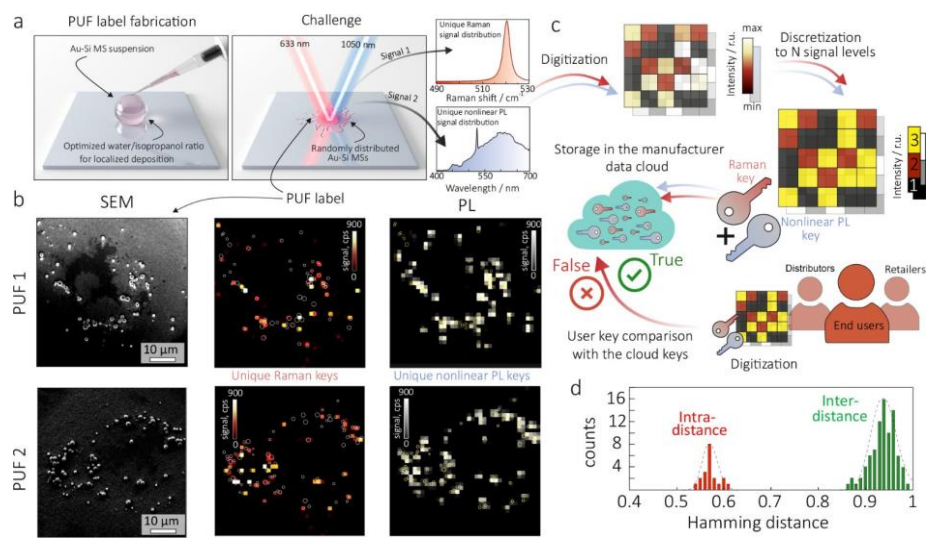

**Fig. 26.18:** Scheme of optical labeling using PUF tags. (a) fabrication process and optical readout by mapping Raman and nonlinear photoluminescence signal, (b) correlated SEM, Raman and nonlinear photoluminescence images of two labels composed of random arrangement of Au-Su nanohybrids, (c) label digitization and authentication, (d) Hamming intra- and interdistances estimated for 10 different PUF labels [Gurbatov, 2023]. Reprinted from [Gurbatov, 2023], Multigram-scale production of hybrid Au-Si nanomaterial by laser ablation in liquid (LAL) for temperature-feedback optical nanosensing, light-to-heat conversion, and anticounterfeit labeling. S.O. Gurbatov et al., ACS Appl. Mater. Interfaces **15**:3336–3347, Copyright 2023, with permission of American Chemical Society.

# 3. Blister-Based Laser-Induced Forward Transfer for High-Precision Printing of Nanomaterials

In this section, we discuss the problem of positioning nanomaterials, mostly focusing on 2D materials (2DMs), to desired locations on substrates for assembling nanomaterial-based devices. Various types of nanomaterials with unique optical, thermal, electronic, plasmonic, and magnetic properties are becoming cornerstones and building blocks of modern micro/nanoelectronics, sensing devices, and detectors, with potential applications in metamaterial designing. However, precise contamination-free positioning of nanosized structures and atom-thick layers remains highly challenging and is one of the focuses of this section. In subsection 3.1, a brief overview of methods of 2DM fabrication is given, showing the wealth of the 2DM family, which is extending with a demonstration of new functional properties. Subsection 3.2 summarizes existing methods of nanomaterial transfer for assembling devices or device parts. A BB-LIFT device enabling a precise printing of nanomaterials is described in subsection 3.3, with a demonstration of its capabilities in subsection 3.4.

## 3.1 General Ways of 2D Materials Fabrication and Necessity to Transfer to Substrates of Choice

Since the first exfoliation of graphene flakes from bulk graphite using Scotch tape [Novoselov, 2004], the field of 2D materials is swiftly developing with emerging new 2D structures having often extraordinary functional properties which make them promising for both exploring fundamental effects at the atomic scale and cutting-edge applications in electronics, biotechnology, energy harvesting, catalysis, and others. Figure 26.19 [Kumbhakar, 2021] illustrates some examples of 2DMs, graphene and beyond: transition metal dichalcogenides (TMD), hexagonal boron nitride (hBN), black phosphorus (phosphorene), and MXenes, and a number of their applications. The latter includes hybrid graphene–quantum dot phototransistor [Konstantatos, 2012], a single-layer MoS$_2$ FET (field-effect transistor) device [Yin, 2012], a MXene–titania thin film suitable for photoresistors with memory effect and sensitivity to the environment [Chertopalov and Mochalin, 2018], and a high-performance WSe$_2$/hBN photodetector [Jo, 2016]. These examples are only a few of the abundant 2DM families developed so far. Figure 26.19 also mentions 2D alloys, atomically thin metal sheets, and natural oxides. We can also list silicene, germanene, 2D silicon carbide, and many others. Furthermore, recently, via large-scale computations, 119 possible candidates for chemical exfoliation were determined and, as a consequence, Ru$_2$Si$_x$O$_y$ nanosheets were synthesized, which are different from the MXenes family [Björk 2024]. This underlines the importance of theory in new material design. Many reviews on the 2DM topic have been published. We refer readers to some recent review literature [Shanmugam, 2022; Wei, 2023; Liu, 2024b].

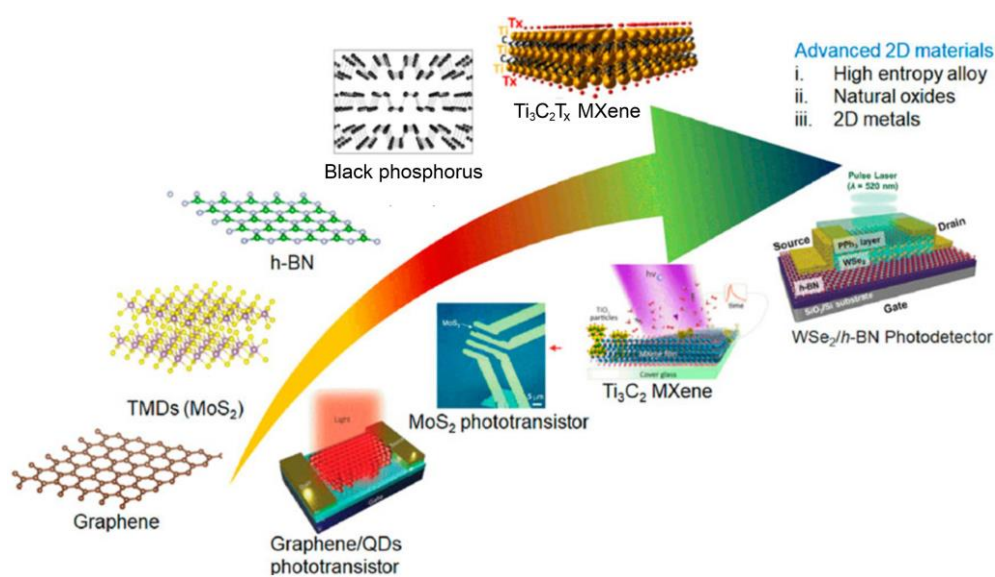

**Fig. 26.19:** Overview of different types of 2DMs and their optoelectronic applications [Kumbhakar, 2021]. Reprinted from [Kumbhakar, 2021], Advance optical properties and emerging applications of 2D materials. P. Kumbhakar et al., Front. Mater. **8**:721514, Copyright 2021, under Creative Commons Attribution 4.0. Retrieved from https://doi.org/10.3389/fmats.2021.721514.

There are multiple methods of synthesizing 2DMs, which can be optionally divided into top-down and bottom-up ones. The top-down methods relate to different types of exfoliations such as micromechanical exfoliation [Novoselov, 2004] and ultrasonic-driven liquid phase exfoliation [Han, 2014; Silva, 2022], which can be assisted by the liquid flow thus strongly increasing graphene productivity [Kwak, 2023] and using lithium intercalation process [Zeng, 2011]. As a top-down approach, a very efficient topochemical synthesis method of 2DMs can be considered, such as MXenes, hydrogenated germanene, and some other 2D layered structures [Xiao, 2018]. This method enables transformation into 2D structural forms of bulk materials that do not have van der Waals bonded polymorphs. The bottom-up techniques assuming assembling materials from atomic/molecular phases include chemical vapor deposition (CVD) [Shi, 2015], molecular beam epitaxy (MBE) [Dong, 2020], and wet chemical synthesis [Xu, 2021]. By varying the process parameters, these methods enable the generation of controlled layer numbers and large-scale growth on different substrates and limit the growth to desired locations.

Another family of single-atom-thick material in the border between 2D and 1D dimensions is graphene nanoribbons (GNRs), which are stripes of graphene with a typical width of several atoms (Figure 26.20 [Ruffieux, 2016]). Often, GNRs are referred to as quasi-one-dimensional materials having semiconducting properties whose bandgap is inversely proportional to their width, thus making this material highly attractive for nanoelectronic applications [Li, 2008; Tien, 2023]. An important feature of this material is the high chemical reactivity of its edges, which requires special edge passivation approaches to preserve electronic properties, especially for use and manipulation under ambient conditions [Ruffieux, 2016].

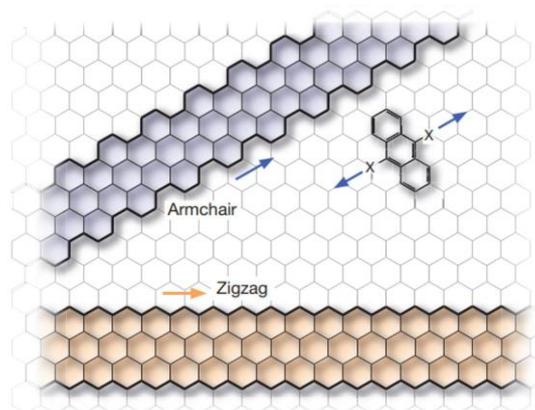

**Fig. 26.20:** Structure of armchair and zigzag graphene nanoribbons, and an exemplary anthracene-based molecular precursor for the bottom-up fabrication of armchair GNRs [Ruffieux, 2016]. Reprinted from [Ruffieux, 2016], On-surface synthesis of graphene nanoribbons with zigzag edge topology. P. Ruffieux et al., Nature **531**:489-493, Copyright 2016, with permission of Nature Research.

Due to the large surface-to-volume ratio, 2DMs have a remarkable potential to be further functionalized by changing their physicochemical properties via covalent (direct bond linkage) or noncovalent interactions [Brill, 2021; El-Said and Ghany, 2024]. Ways of covalent functionalization are summarized in Figure 26.21 [Jeong, 2022]. Noncovalent functionalization can be achieved via the adsorption of organic molecules or the decoration of 2DM sheets with inorganic nanoparticles [Hayes, 2020; Brill, 2021]. A striking example is when noncovalent functionalization induces covalent one [Drogowska-Horna, 2020]: modulation of charge-carrier density in graphene sheets due to their positioning on a periodically patterned a-SiO$_2$/Si substrate induces periodic chemical reactivity. Both types of 2DM functionalization provide further degrees of freedom for widening the application niches of these extraordinary materials [Duran, 2023].

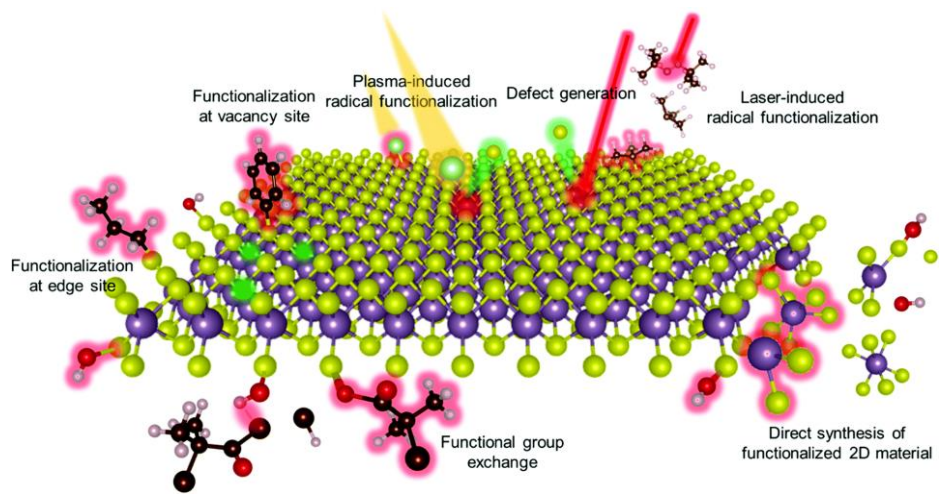

**Fig. 26.21:** Schematic illustration of functionalization sites and various techniques for covalent functionalization of 2DMs [Jeong, 2022]. Reprinted from [Jeong, 2022], Recent trends in covalent functionalization of 2D

As a whole, the field of 2D materials involves a sophisticated roadmap from their synthesis, often involving theory to real applications, as schematically shown in Figure 26.22. Abundant families of 2DMs whose physicochemical properties can be manipulated via functionalization require efficient, clean, and fast methods of their positioning on desired places on substrates for the fabrication of various types of innovative devices. In turn, material positioning/printing needs specific care to preserve their extraordinary electronic properties. The methods of printing developed in recent years will be discussed in the next subsection.

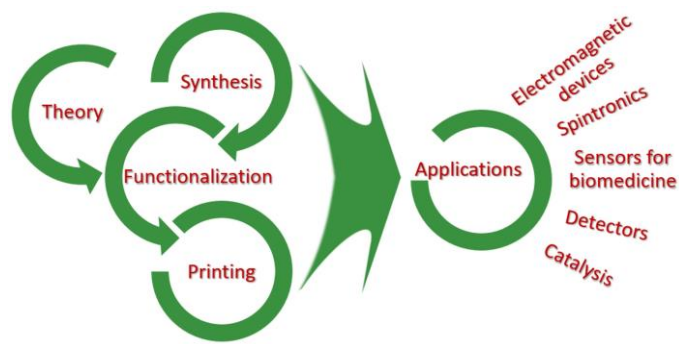

**Fig. 26.22:** Schematics for 2D materials roadmap from synthesis to real applications.

### 3.2 Methods of Transfer

2DM transfer for assembling electronic, photonic, or sensing devices can be considered as an additive manufacturing process in which nanomaterial is deposited to desired locations on a substrate layer-by-layer or into well-designed structural units. In a recent review, Cheliotis and Zergioti [Cheliotis and Zergioti, 2024] described in detail the main methods of 2DM transfer (Figure 26.23). The transfer methods can conventionally be divided into wet and dry ones and can be summarized as follows:

- *Wet transfer assisted by etching*. This is a multistep process that involves coating a polymer (supporting layer) on the CVD-grown 2DM surface (e.g., PMMA spin-coating [Zhao, 2022]), applying a solution of a chemical etchant, detaching the polymer/2D material from the etched growth substrate, locating 2DM on a target substrate, and finally removing the polymer layer by peeling it off or dissolving.
- *Etching-free wet transfer*. A more cost- and time-efficient alternative to the above technique is based on electrochemical bubbling for peeling off the CVD-grown 2DMs from the metal substrate of growth. For the first time, this method was applied by Wang et al. [Wang, 2011]. Successive stages of this method can shortly be described as follows [Cheliotis and Zergioti, 2024]. A 2DM on a growth substrate is coated by a layer of a polymeric material. The whole sample is placed into an electrolyte where an electrode is located. A DC voltage is applied between the electrode and metal substrate, causing an electrochemical reaction. Decomposition of water results in the generation of hydrogen bubbles, which aggregate at the 2DM/metal substrate interface, leading to 2DM detachment from metal. 2DM attached to the polymer and floating in the electrolyte is transferred onto the target wafer and left to dry, after which the polymer is removed, e.g., by immersing the sample in acetone.
- *Dry chemical-etchant-assisted transfer*. The main difference of this method from the wet transfer process with etching is that, after detaching from the growth substrate, the

2DM attached to a supporting substrate is allowed to dry before positioning on the target sample [Yang, 2014]. A supporting substrate usually consists of a polymeric and a thin compliant layer. After 2DM on the supporting substrate is detached from the growth substrate in an etchant solution and washed in deionized water for removal of any residues, it is allowed to dry. The next steps are placing 2DM on the target substrate with the help of the supporting substrate and peeling off the latter. In the peeling-off process, a controlled adhesion of the supporting substrate is of primary importance to ensure damage-free 2DM transfer [Yoon, 2022].

- *Mechanical exfoliation*. The advantages of this method are the possibility to avoid the use of chemicals and to reduce the number of steps for transfer. It is based on the difference of the 2DM adhesion to different surfaces [Yoon, 2012]. As the first step, an adhesive layer is deposited on the surface of the 2D material located on the growth substrate, on top of which a supporting layer is placed to prevent folding of the 2DM during the transfer. Using this geometry, the 2DM is carefully peeled off from the growth substrate and positioned on the target substrate. The final step is the removal of the adhesive layer and the supporting substrate. Depending on their properties, it can be done using acetone or deionized water [Moon, 2019].

- *Metal assisted transfer*. This method is usually used to transfer a 2DM from a dielectric growth substrate (e.g., $SiO_2$, $Al_2O_3$). For this aim, a thin metallic layer is deposited on the 2DM located on the growth substrate on the top of which a thermal release tape is placed (TRT) [Lin, 2015]. Using the TRT, the 2DM stack with the metal layer is mechanically peeled off from the growth substrate. The stack is pressed onto the target substrate by 2DM down. The TRT is removed using heating while the metal layer is selectively etched.

- *Laser direct transfer*. This method is used to assemble the nanomaterials into workable devices, assuming that the 2DM for transfer is already removed from the growth substrate and positioned on the top of the so-called donor substrate. This can be done using one of the above methods, e.g., the polymer-assisted wet transfer technique. For laser-induced forward transfer (LIFT), the donor substrate usually represents a transparent micrometer-sized film on the top of which a dynamic release layer (DRL) is deposited [Serra and Piqué, 2019; Cheliotis and Zergioti, 2024]. The DRL can be a metallic layer nanometers-thick or a triazine polymer layer as this layer serves for absorbing pulsed laser radiation coupling the DRL through the transparent substrate. Due to vaporization or bulging of the DRL, the 2DM is pushed along the laser beam and is transferred to the receiver substrate positioned parallel to the donor at a fixed distance. In the laser-induced backward transfer (LIBT) technique [Praeger, 2020], the laser beam irradiates the donor substrate through a transparent receiver substrate.

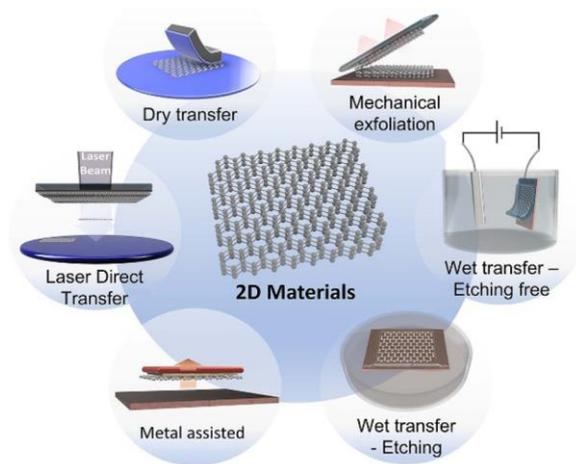

**Fig. 26.23:** The main transfer methods for 2D materials: wet transfer technique assisted by etching, wet etching-free transfer technique, dry transfer technique, mechanical exfoliation, metal-assisted transfer, and laser direct transfer [Cheliotis and Zergioti, 2024]. Reprinted from [Cheliotis and Zergioti, 2024], A review on transfer methods of two-dimensional materials, I. Cheliotis and I. Zergioti, 2D Mater. **11**:022004, Copyright 2024, under Creative Commons Attribution 4.0. Retrieved from https://doi.org/10.1088/2053-1583/ad2f43.

We refer the readers for more details to the review by Cheliotis and Zergioti [Cheliotis and Zergioti, 2024] and the extensive literature cited therein. But here, we will focus on a blister-based variation of the laser-induced forward transfer (BB-LIFT) technique, which can be considered as a green additive-manufacturing method that can enable contamination-free large-scale printing of 2D material and beyond. The uniqueness of BB-LIFT lies in the possibility of avoiding the direct laser heating of the transferred nanomaterial, thus minimizing any thermal damage [Kononenko, 2011; Goodfriend, 2018]. It has shown a big potential for printing such gentle materials as graphene nanoribbons [Komlenok, 2022]. A variation of BB-LIFT, blister-actuated LIFT is used for printing liquid droplets with a high lateral resolution [Brown, 2010]. Below, we present a patented optical device for precisely printing 2D materials for assembling 2DM-based components for workable devices [Goodfriend and Bulgakov, 2023, 2024].

### 3.3 BB-LIFT Device for Precise Transfer of Materials

Although the LIFT method has been demonstrated to be very efficient to transfer different 2D materials in a single-step process [Goodfriend, 2018; Praeger, 2020; Papazoglou 2021, Komlenok, 2020; Komlenok, 2022], most of the developed techniques do not provide precise positioning and placing these materials and other nano-objects required for fabrication of nano-devices. To overcome this drawback, a BB-LIFT printing device equipped with advanced visualization and scanning systems has been recently developed [Goodfriend and Bulgakov, 2023, 2024]. Figure 26.24 shows a scheme of the device. The nanomaterials to be transferred are located on a donor plate consisting of a transparent substrate (glass) and a dynamic release layer (metal film). The donor is laser-irradiated through the substrate to produce a transient deformation (blister) in the DRL that serves to gently desorb the nanomaterial without exposing it to laser radiation [Kononenko, 2009; Goodfriend, 2016; Goodfriend, 2018]. The desorbed nanomaterial is landed (printed) on a receiver substrate. In this device, both the donor and receiver are placed on XYZ stages, allowing their precise computer-controlled positioning. The optical system involving broadband light sources, focusing objectives, and CCD cameras located on both donor and receiver sides enables monitoring and control of the printing process. It is thus possible to select a particular 2D material flake on the donor to be transferred and a place on the receiver where it should be positioned. Simultaneous sc7777anning of the donor and receiver allows printing of ordered arrays or stacks of 2D nanomaterials.

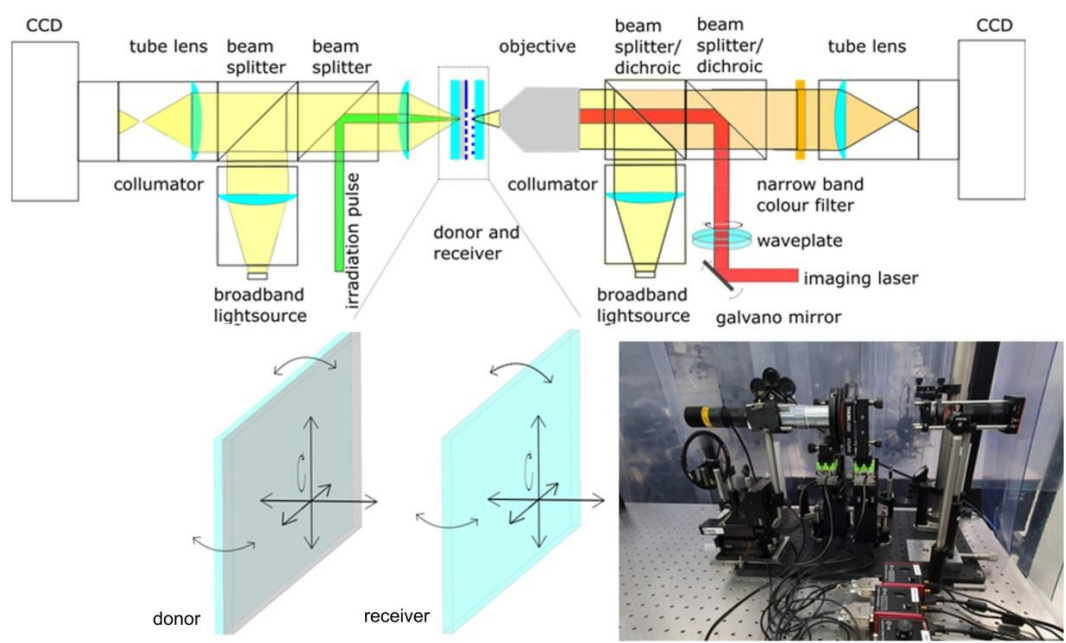

**Fig. 26.24:** Scheme of the BB-LIFT printing device with an optical monitoring system. The inset in the bottom right shows a general view of the device.

In addition, the printing device allows for determining the crystalline orientation of the transferred 2D lattices using the method developed by Maragkakis et al. [Maragkakis, 2019]. The technique is based on polarization-resolved imaging of the second harmonic generation (SHG) signal whose polarization is sensitive to the crystal orientation (Figure 26.25). Furthermore, the intensity of the SHG signal is dependent on the number of monolayers [You, 2019] that can be used for the evaluation of the flake thickness. The receiver can be irradiated by an additional imaging laser with controlled beam polarization (Figure 26.24). Adjusting the beam polarization helps to find the thickness and orientation of the deposited nanocrystals of 2D materials. If necessary, the crystals can be re-oriented to a specific angle requested by rotating the donor and receiver holders.

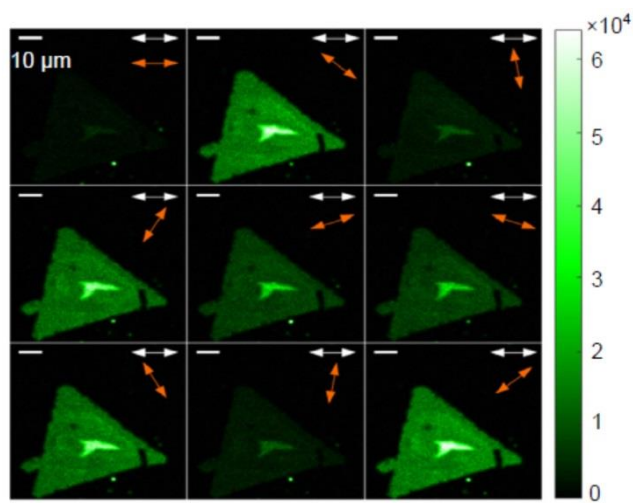

**Fig. 26.25:** Snapshots of SHG images of a CVD-grown $WS_2$ flake [Maragkakis, 2019]. The white double arrow shows the constant angle of the linear polarizer, while the orange double arrow shows the angle of the excitation

polarization. The rotation of the latter allows switching on and off of the SHG signal. Reprinted from [Maragkakis, 2019], Imaging the crystal orientation of 2D transition metal dichalcogenides using polarization-resolved second-harmonic generation, G.M. Maragkakis et al., Opto-Electron. Adv. **2**:190026, Copyright 2019 Authors, with permission from Maragkakis et al., 2019.

## 3.4 Demonstration of Capabilities of BB-LIFT for 2D Material Printing

Several transfers of 2D materials have been performed using LIFT printing devices including that described in the previous section. Here, we present some of the results obtained to assess the printing capabilities of such low dimension-materials printers. Additionally, we underline the points that are, in our view, important to reach micrometric precision and below, paving the way to faster meaningful assembly of 2DM.

### 3.4.1 Printing Capabilities

First attempts to transfer 2D materials by the BB-LIFT techniques were performed with CVD-grown single-layer molybdenum dichalcogenides (both nanosecond and femtosecond laser pulses were used) [Goodfriend, 2018] and graphene (nanosecond pulses) [Komlenok, 2020]. In both cases, the initial 2D material structure was nearly preserved after the LIFT, however, the transferred graphene fragments were crumpled with generated defects, while the $MoX_2$ crystals were partially fragmented. Praeger et al. performed the transfer of graphene using a backward version of the LIFT technique, but defects were also introduced during the printing process that downgraded the quality of the transferred material [Praeger, 2020]. The capacity of the BB-LIFT technique to provide defect-free transfer was demonstrated by Zergioti's group with graphene [Papazoglou, 2021] and molybdenum disulfide [Logotheti, 2023] using nanosecond laser pulses. Under optimal LIFT conditions in a single-shot laser mode, the reproducible seamless ejection of graphene monolayers from a donor nickel surface was achieved, and Raman spectra of printed graphene pixels indicated that no noticeable defects were introduced during the laser transfer process (Figure 26.26).

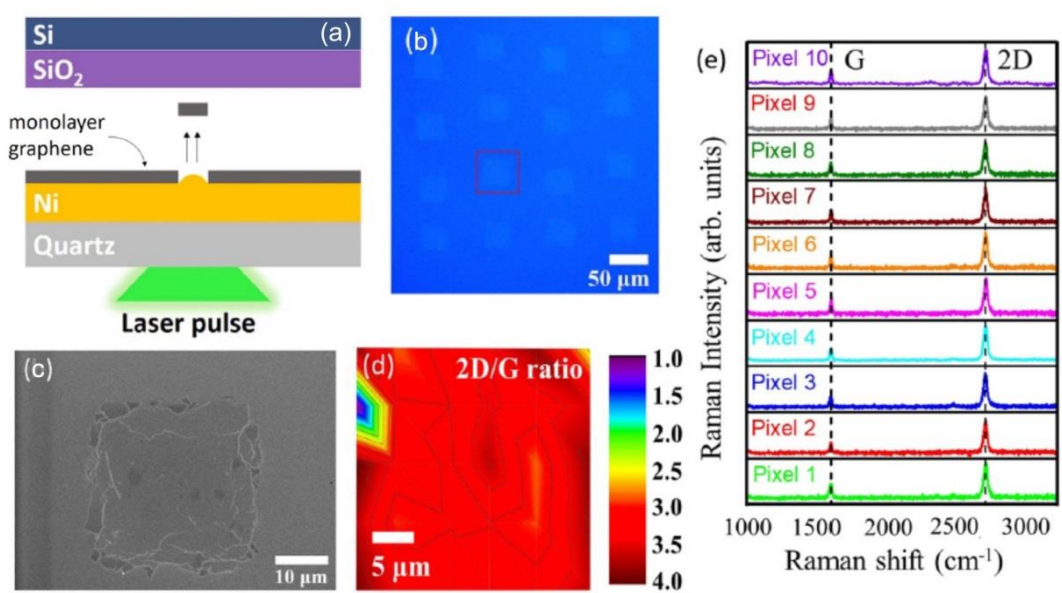

**Fig. 26.26:** LIFT printing of monolayer graphene on a $SiO_2$/Si substrate [Papazoglou, 2021]. (a) Schematic representation of transfer, (b) optical image of a printer array, (c) SEM image of graphene pixel highlighted in (b), (d) Raman color map of 2D/G peak intensity ratio within a printed graphene pixel, (e) Raman spectra of ten printed pixels. Reprinted from [Papazoglou, 2021], A direct transfer solution for digital laser printing of CVD graphene, S. Papazoglou et al., 2D Materials 8, 045017, Copyright 2021, with permission of IOP Publishing, Ltd.

One of the most challenging systems for printing among 2D materials is graphene nanoribbons (GNRs). GNRs are narrow, atomically precise stripes of graphene with pre-determined width and edge structure [Cai, 2010; Houtsma, 2021]. Quantum confinement and edge effects render GNRs semiconducting in contrast to the parent graphene material, which exhibits semimetallic behavior. This makes GNRs very promising for applications in nanoelectronic devices. However, their synthesis is achieved almost exclusively on noble-metal single-crystal surfaces, while, for most applications, they have to be transferred onto dielectric surfaces [Cai, 2010; Mishra, 2021]. With the precise and gentle structure of GNRs, this is a difficult task. Komlenok et al. demonstrated the potential of BB-LIFT to successfully print GNRs by the example of a 7-atom wide armchair GNR (7-AGNR), which is fairly stable and could be transferred from the synthesis place to the BB-LIFT donor surface [Komlenok, 2022]. To minimize the GNR heating and damage during the LIFT process, relatively thick titanium films with a thickness of up to 1 µm were used. Irradiation conditions were found when a gentle transfer of 7-AGNRs is realized in the air without destroying their structure, as was confirmed by Raman spectra (Figure 26.27). It should be noted, however, that in contrast to 7-AGNR, most of the GNRs interesting for applications, in particular, spinful systems [Mishra, 2021], are very reactive due to their unpaired electrons and can be easily oxidized in the presence of oxygen during the LIFT process. Therefore, their transfer has to be performed in an inert atmosphere or a vacuum.

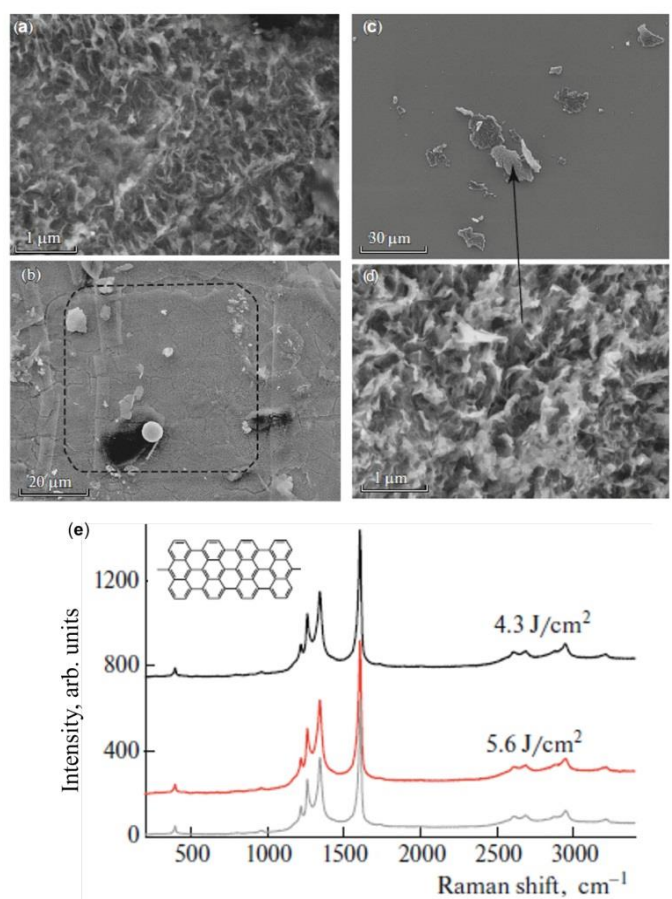

**Fig. 26.27:** LIFT printing of 7-AGNRs onto a SiO$_2$/Si substrate with 20-ns 248-nm laser pulses and a 1-µm DRL [Komlenok, 2022]. (a-d) SEM images of the donor surface before (a) and after (b) LIFT, and the receiver after LIFT (c,d), (e) Raman spectra of GNRs printed at different laser fluences and that of the original 7-AGNR. The inset shows the atomic structure of 7-AGNR. Reprinted from [Komlenok, 2022], Laser-induced forward transfer of graphene nanoribbons, M.S. Komlenok et al., Dokl. Phys. **67**, 228, Copyright 2022, with permission of Springer Nature.

The printing capacities of the BB-LIFT device described above (Sec. 3.3) were recently investigated by the example of hexagonal boron nitride [Goodfriend, 2025]. The transfer efficiency was examined for different transfer regimes comparing (a) femtosecond and nanosecond laser pulses, (b) "short" and "long" donor-to-receiver separation distances (less and greater than the estimated blister height), and (c) isolated laser pulses and series of overlapping pulses. It was demonstrated that the short distances enable efficient direct "stamping" of 2DM flakes to the receiver surface while, at longer distances, the transfer proceeds via mechanical ejection of nanomaterials. For both the stamping and ejection regimes, defect-free transfer of multilayer and monolayer hBN flakes was achieved (Figure 26.28). It was found, however, that the transfer with low-fluence ns pulses and separation distances smaller than the blister height provide the most favorable and reproducible conditions for BB-LIFT of 2DMs. Furthermore, the optical monitoring system of the printing device enabled the printing of ordered arrays of $WS_2$ 2DM flakes randomly distributed on the donor surface (Figure 26.29).

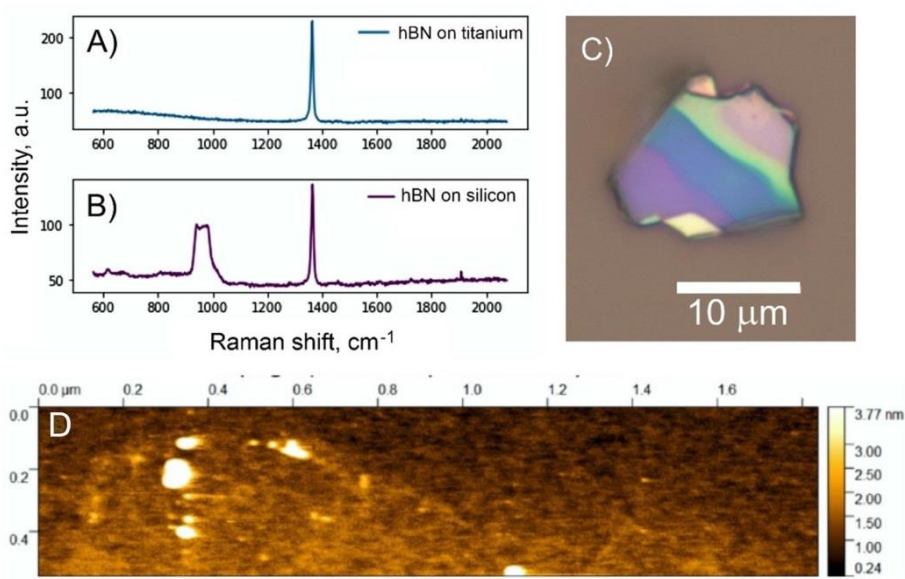

**Fig. 26.28:** Nanosecond BB-LIFT of hBN flakes on a $SiO_2$/Si substrate at a large donor-to-receiver distance (200 μm) [Goodfriend, 2025]. Raman spectra of hBN on the donor (A) and receiver (B) are identical (the additional peak in (D) is due to the substrate). (C) Optical image of a multilayer transferred flake (the coloration is due to a variation in the number of layers across the flake). (D) AFM image of a monolayer flake on the receiver. Reprinted and adapted from [Goodfriend, 2025], I. Mirza, A.V. Bulgakov, E.E.B. Campbell, N.M. Bulgakova, Laser-induced gas-phase transfer and direct stamping of nanomaterials: Comparison of nanosecond and femtosecond pulses. N.T. Goodfriend et al., Mater. Res. Express **12**:065004, Copyright 2025, under Creative Commons Attribution 4.0. Retrieved from https://doi.org/10.1088/2053-1591/ade497.

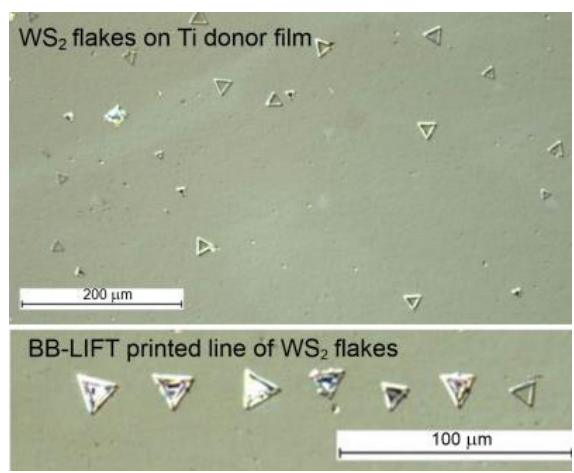

**Fig. 26.29:** WS$_2$ flakes transferred using a wet technique on the Ti DRL for printing (top) and a BB-LIFT-printed line of the flakes (bottom).

### 3.4.2 Aspects Related to the Donor in BB-LIFT

To achieve good quality of the BB-LIFT-transferred material, several conditions for the blister formation on the donor film have to be met. The blistering process should be reproducible from pulse to pulse. Second, it is crucial to avoid fragmentation/cracking of the blister with ejections of DRL fragments that can pollute the receiver. Finally, any degradation of the nanomaterial upon landing on the donor and/or due to heating of the laser-irradiated DLR is to be prevented.

The mechanisms responsible for blistering and material ejection are complicated and can involve the bulging off (or buckling) the DRL due to different thermal expansion of the metal thin film and substrate [Bulgakov, 2014] and/or by the ablation of the DRL surface directly exposed to the laser (at the substrate/film interface) which generates a confined vapor/plasma with a pressure sufficient to induce deformation of the remaining metal film [Kononenko, 2009; Brown, 2010]. The first scenario is predominant in the case of relatively long nanosecond laser pulses, while the second one is mostly realized with ultrashort pulses [Goodfriend, 2016]. Both mechanisms are strongly dependent on the thermophysical properties of the donor film, the main of which are considered below.

*Materials for DRL*

Brittle materials such as PMMA, some steels, and chromium are less prone to deformation and are thus likely to shatter when blistering. On the contrary, materials exhibiting high ductility like platinum, copper, or gold would deform more smoothly and be *a priori* suitable materials for BB-LIFT. On the other hand, metal films with high thermal conductivity can threaten the integrity of the materials to be transferred due to the temperature rise at the rear side of the film. This especially applies in the regime of long (nanosecond) pulses when the heat-affected zone is fairly large. As a consequence, for the transfer of very sensitive entities like biomolecules [Kattamis, 2009] or graphene nanoribbons [Komlenok, 2022], Au, Cu, Ag, and Al films should be generally avoided.

A high thermal expansion coefficient of the film is desirable to help the blister formation while limiting the film temperature at the interface with the 2DM. The higher the coefficient, the more bulging is expected at a given temperature. Complementarily, utilizing a metal with a high tensile strength can mitigate the risks of rupture of the film during blistering. Finally, good adhesion of the film to the substrate is an important factor. Gold, for example, is well known for having poor adhesion when deposited on glass without a buffer layer [Haq, 1969]. In such a case, the buckling of the film can exhibit chaotic delaminations on the areas around the laser spot, thus seriously impairing the reproducibility and the cleanliness of the BB-LIFT process. In all these senses, titanium seems to represent a good option. It has a good adhesion to inexpensive glass substrates, reasonable heat conduction and ductility, and a high thermal expansion coefficient, and titanium thin films have demonstrated excellent blistering performance in several LIFT experiments [Kononenko, 2009; Goodfriend, 2016; Goodfriend, 2018; Komlenok, 2022].

*Thickness of the DRL*

The laser fluence range for the DRL blistering and the repeatability of the LIFT process strongly depends on the thickness of the donor metal film [Kononenko, 2009; Brown, 2010; Domke, 2014]. In general, the thinner the film, the smaller the threshold fluence for the onset of blistering and transfer, and the smaller the fluence window where the transfer occurs without breaking up the blister (Fig. 26.30) [Kononenko, 2009]. To obtain reproducible blistering and thus reliable printing BB-LIFT, it is therefore important to employ donor films highly homogeneous in thickness. In addition, the thickness of the donor DRL strongly affects the velocity of the LIFT-ejected species, which was found in time-of-flight measurements to be inversely proportional to the thickness [Goodfriend, 2018]. Thus, reducing the thickness of a Ti film from 370 to 210 nm resulted in an increase in the mean velocity of the ejected carbon nanomaterials by almost an order of magnitude. The landing of fragile 2DM structures on the receiver at a high velocity can result in damage or fragmentation. Finally, the thickness directly impacts the evolution of the temperature rise on the rear side of the film where nanomaterials to be printed are located. Upon irradiation with nanosecond pulses, heat can reach the rear side of metallic films of hundreds of nm in thickness within a few ns before significant blistering has started. To prevent significant thermal effects on delicate transferred materials, it is therefore recommended to use fairly large DRL thicknesses, as was demonstrated with GNRs [Komlenok, 2022], and/or utilize ultrashort laser pulses in single-shot regimes [Goodfriend, 2025].

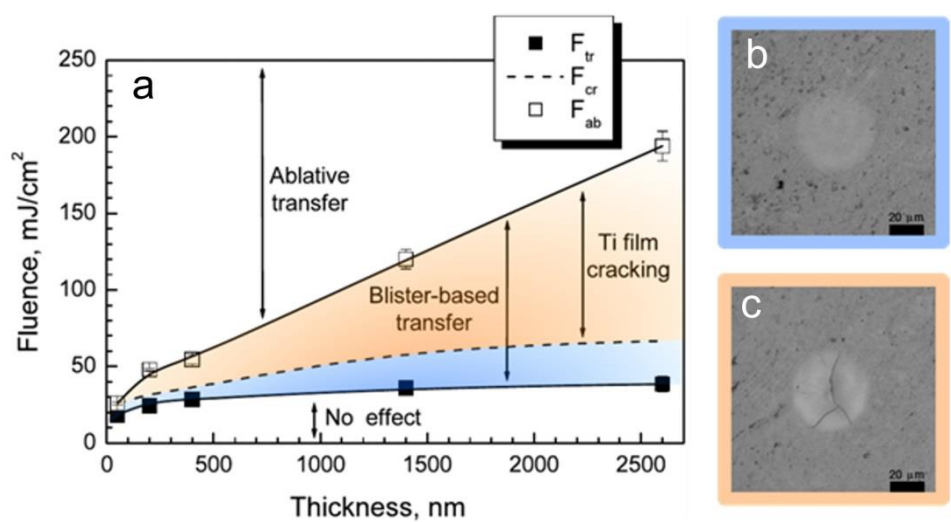

**Fig. 26.30:** (a) Fluence thresholds for blistering as a function of the thickness of a Ti film (50-ps, 532-nm laser pulses, quartz substrate, ~~40 μm spot diameter). Three thresholds are introduced for blistering, cracking, and ablation. The favorable BB-LIFT regime of blistering without cracking occurs in a narrow fluence range, which increases with the film thickness [Kononenko 2009]. (b) and (c) show optical images of blisters in a 200-nm Ti film without cracking (b, 25 mJ/cm$^2$) and with cracking (c, 40 mJ/cm$^2$). Adapted from [Kononenko 2009], Laser transfer of diamond nanopowder induced by metal film blistering, T.V. Kononenko et al., Appl. Phys. A **94**:531, Copyright 2009, with permission of Springer-Verlag.

### 3.4.3 Aspects Related to the Receiver

*Receiver preparation*

Substrates have a clear influence on the properties of 2DMs [Katsnelson and Geim, 2008; Wang, 2012]. The choice of the substrate for the LIFT-printed materials is thus principally dictated by the applications developed via the printing technique, with Si/SiO$_2$ wafers being routinely employed in the realization and study of graphene or 2DM-based transistors. Regarding the surface preparations, thorough cleaning in soap, isopropanol, and then demineralized water while in an ultrasonic bath is a routine procedure before attempting any transfers. Ozonation of the cleaned receivers by air plasma cleaners brings additional help in this respect in the case of dielectrics. Interestingly, we note that the elasticity of the receiving materials, while disregarded in the literature, appears to be a possible factor contributing to the integrity of the transferred delicate materials [Feinaeugle, 2013].

*Tailoring the receiver*

As mentioned in Section 3.1, functionalization of 2DMs can be achieved in a localized way, by tailored modification of the underlying substrate. For example, submicrometric laser-induced periodic surface structures (so-called LIPSS) were inscribed on a Si substrate covered by a smooth $SiO_2$ film [Drogowska-Horna, 2020]. When monolayer graphene was placed on such a substrate, the underlying LIPSS transferred their periodicity to the chemical reactivity of graphene via modulations of its doping level. Another approach involves transferring flakes of 2DMs directly on surfaces that have been patterned with LIPSS. The periodical topography modulations of the LIPSS are expected to periodically affect the local carrier properties due to alternating strain change induced by the underlying substrate. Figure 26.31 shows examples of such a 2DM-corrugated substrate assembly obtained with the BB-LIFT printer described in Section 3.3. Micrometric size $WS_2$ flakes were transferred onto $Al_2O_3$ substrates, which have been patterned using regular LIPSS scribing techniques (see Chapter 10 of this book [Römer, 2026] and Section 5 of this Chapter). The analysis of the Raman signal of the flakes shows a correlation with the LIPSS topographical modulation that demonstrates the possibility of using laser for both modification of the receiver and the blister-based transfer in view of functionalizing 2DMs.

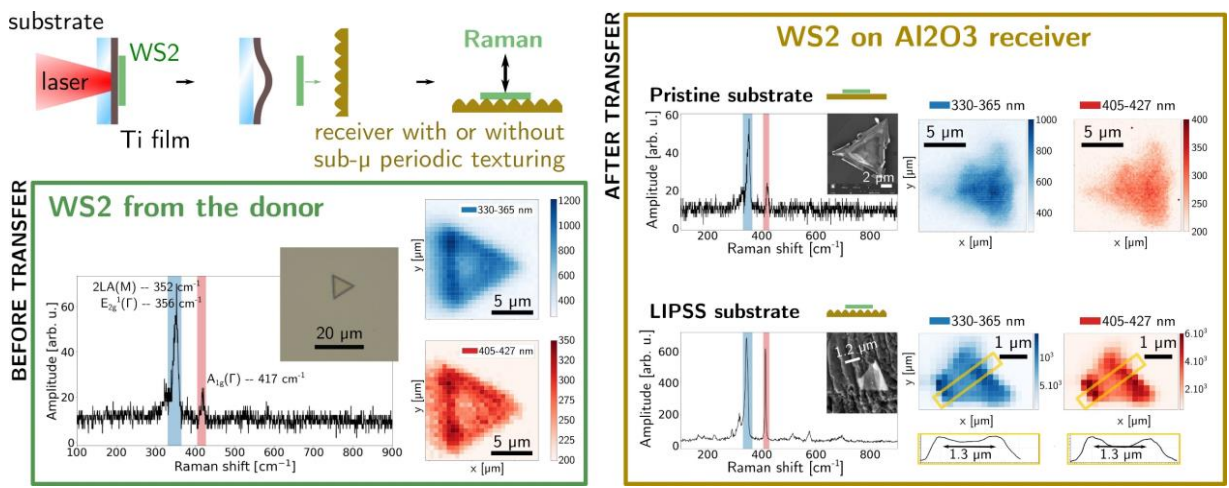

**Fig. 26.31:** Raman mapping of WS2 flakes on the donor (bottom left frame) and after transfer on $Al_2O_3$ (right frame). SEM images show a flake on pristine and on LIPSS-covered substrate. The corresponding micro-Raman mappings reveal modulations in the intensity of the vibration modes $E^1_{2g}$ (blue color) and $A_{1g}$ (red color) [Berkdemir, 2013] that correlate with the LIPSS topography. The distortion of the shape of the flakes in the Raman mapping on $Al_2O_3$ is due to thermal drift during the scans.

Finally, we underline that several other works reported how substrate engineering can enable particular functionalization of 2DM [Zhou, 2016; Kasischke, 2018; Tu and Yan, 2024]. In combination with a printing device, such as that based on BB-LIFT presented in Sec. 3.1, laser engineering of substrates opens the doors to a wider range of functionalization of nanomaterials.

# 4. Selective Annealing of Semiconductor Nanomaterials: Crystallization without Melting

This section focuses on one of the old problems of laser processing, the laser annealing of semiconductors. When semiconductor films are deposited on some surfaces from the vapor phase, they are usually in an amorphous state that does not have the long-range order typical of crystalline materials. Although amorphous semiconductors find applications in industries due to their adjustable electronic properties and flexibility, for large-area electronic devices such as thin-film transistors and solar cells, highly crystalline films are in big demand. Among different techniques of annealing for crystallizing amorphous materials (furnace annealing, rapid thermal annealing, flash lamp annealing), laser crystallization enables annealing only a thin controllable surface layer and the localized lateral dimensions, thus offering selective phase transformation with the advantage of not damaging the substrates or the devices where the film is built in. In Subsection 4.1, different regimes of annealing will be discussed, addressing the necessity of using high-power or high-intensity lasers. Subsection 4.2 is devoted to the fundamental problem of amorphous-to-crystalline phase transition, which at certain conditions, can proceed without the signs of melting. Finally, in subsection 4.3, some successful examples of the ultrashort laser annealing of nanostructured semiconductors will be presented.

## 4.1 Laser Annealing of Semiconductor Materials: Why High-Intensity Lasers are Needed

With the development of large-scale electronic devices based on micro- and nanoelectronics elements, the problem of improving the technology for the crystallization of semiconductor films on inexpensive non-refractory substrates continues to be relevant. There is a steady trend towards a constant increase in the sizes of microelectronics devices at high manufacturing precision that can be related to the so-called "inverse Moore's law" [Randall, 2018]. In these circumstances, ultrafast laser annealing is increasingly used for the crystallization of semiconductor films on non-refractory substrates [Zhou, 2024]. Selective laser annealing using UV lasers is utilized to improve the characteristics of modern memory elements (memristors) on non-refractory flexible substrates [Han and Shin, 2022]. Laser annealing is also used for the nanostructuring of solid solutions [Calogero, 2023], selective ablation of films on flexible substrates [Gallais, 2014], and the modification of 2D materials [Emelianov, 2024].

In the past years, mostly nanosecond lasers with typical pulse durations of 10-30 ns were used for laser-induced crystallization [Sameshima, 1990; Fogarassy, 1993]. At the same time, crystallization was carried out by either melting the film and its crystallization during the cooling stage (liquid phase crystallization (LPC) [Sameshima, 1990] or solid phase crystallization (SPC) [Efremov, 1996]. In the first approach, the main problem is the heterogeneity of the polycrystalline film relief during scanning processing, associated with its melting under a laser beam and subsequent crystallization. In the second approach, the growth of laser-induced crystalline nuclei and complete crystallization of the film requires additional furnace annealing at a temperature of 550 °C, which not all substrates can withstand.

With the development of laser technologies, femtosecond laser pulses started to be utilized for the crystallization of amorphous silicon films [Shieh, 2024; Lee, 2006]. The durations of such pulses are smaller than the characteristic times of energy exchange between the electrons and lattice in semiconducting materials. Hence, such regimes of high-intensity irradiation provide opportunities to study new aspects of phase transitions at highly non-equilibrium and nonlinear regimes of laser irradiation. The intensity of the incident light wave used for laser annealing can reach tens of petawatts per square meter. When using ultrashort laser pulses of such high intensities and, at the same time, with high photon energy (e.g., more than twice the bandgap of a semiconductor undergoing annealing), the effect of impact ionization, which is also called Auger generation can manifest itself [Kolodinski, 1995; Volodin, 2011]. The process means that more than one electron-hole pair is generated per absorbed photon, as shown in (Figure 26.32) [Volodin, 2011]. This effect can contribute to lattice destabilization of semiconductor materials, which is detected as non-thermal or cold melting [Stampfli and Bennemann, 1990; Sokolowski-Tinten, 1995; Sundaram and Mazur, 2002]. The cold melting effect in metastable materials such as amorphous semiconductors may trigger the phase transition to the stable crystalline phase. A striking example of femtosecond laser-induced crystallization of anodic $TiO_2$ nanotube layers into anatase [Volodin, 2023], where destabilization of the amorphous lattice may play an important role, will be considered in subsection 4.3 in more detail.

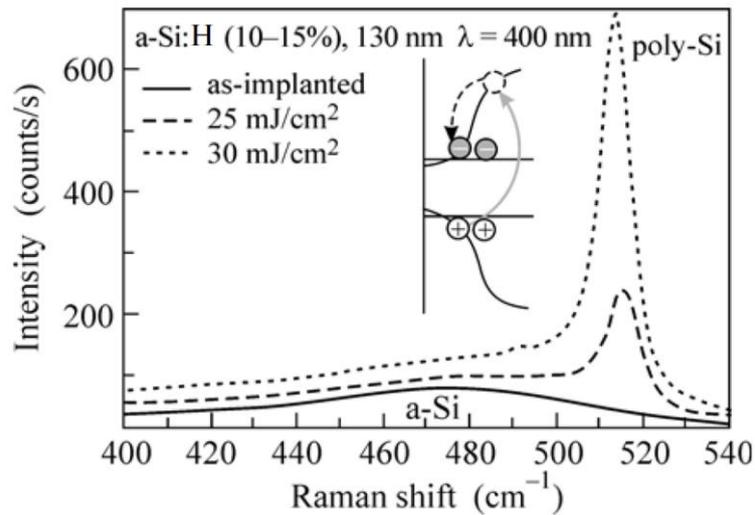

**Fig. 26.32:** Raman spectra of an as-deposited a-Si film with relatively low hydrogen concentration (solid line) and polycrystalline Si films after femtosecond laser irradiation (dashed and dotted lines) [Volodin, 2011]. The inset shows the schematic process of Auger generation. The irradiation was performed with 400-nm wavelength at 30-fs pulse duration. Adapted from [Volodin, 2011], Femtosecond pulse crystallization of thin amorphous hydrogenated films on glass substrates using near ultraviolet laser radiation, V.A. Volodin et al., JETP Letters **93**:603, Copyright 2011, with permission of Maik Nauka-Interperiodica.

## 4.2 Amorphous to Crystalline Phase Transition – Avoiding Melting

It is accepted that explosive crystallization of amorphous materials can proceed via four major mechanisms: solid-phase random (or nucleation) crystallization, solid-phase epitaxy crystallization, liquid-phase random (or nucleation) crystallization, and liquid-phase epitaxy crystallization [Geiler, 1985; Sinke, 1989]. These scenarios are shown schematically in Figure 26.33, left [Sinke, 1989]. Shortly, crystallization of amorphous samples can proceed either with or without a liquid intermediate layer via bulk nucleation and growth of crystalline phase or layer-by-layer (epitaxial) growth. In liquid-mediated crystallization, the thinner liquid layer is favorable for epitaxial crystal growth, while a thicker liquid layer leads to the nucleation mechanism, as schematically shown in Figure 26.33, right [Egan, 2019].

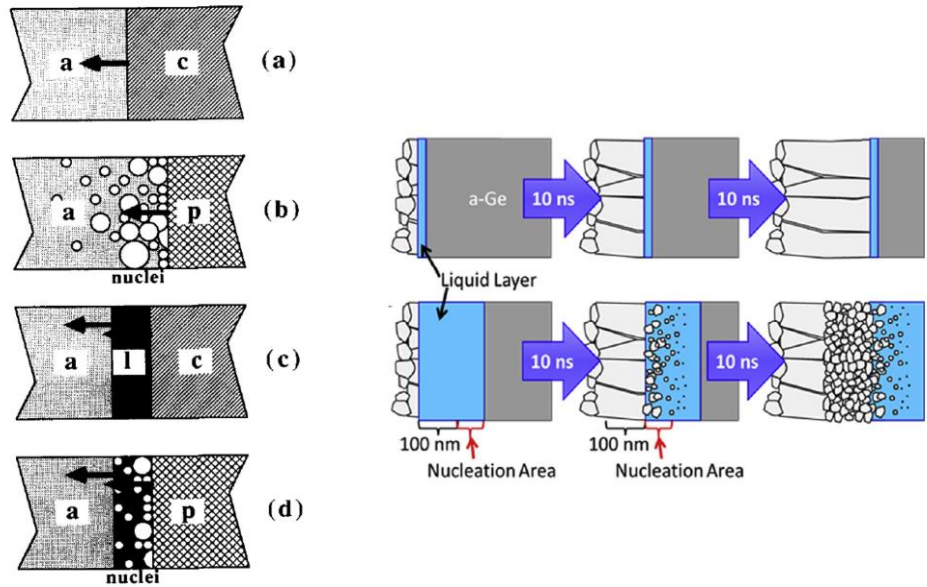

**Fig. 26.33**: Left: Schematic representation of the different modes of explosive crystallization: (a) solid-phase epitaxial crystallization (b) solid-phase random crystallization, (c) liquid-phase epitaxial crystallization, (d) liquid-phase random crystallization (a = amorphous, c = monocrystalline, p = polycrystalline, l = liquid) [Sinke, 1989]. Right: A schematic illustration of the competition between epitaxial growth and homogenous nucleation in the case of a thin (top) and thick (bottom) liquid layer. Assumes liquid-phase epitaxy occurs at ~10 m/s and liquid-phase nucleation occurs faster than 10 ns [Egan, 2019]. Figure on the left reprinted from [Sinke, 1989], Explosive crystallization of amorphous silicon: triggering and propagation, W.C. Sinke et al., Appl. Surf. Sci. **43**:128, Copyright 2009, with permission from Elsevier. Figure on the left reprinted from [Egan, 2019], A novel liquid-mediated nucleation mechanism for explosive crystallization in amorphous germanium, G.C. Egan et al., Acta Materialia **179**:190, Copyright 2019, with permission from Elsevier.

Different mechanisms of explosive crystallization have been extensively studied numerically based on one- and two-dimensional heat conduction approaches [Balandin, 1984; Smagin and Nepomnyashchy, 2009; Nikolova, 2014; Buchner and Schneider, 2015; Lombardo, 2018], including rate equations to describe the kinetics of the homogeneous amorphous-crystalline transition [Buchner and Schneider, 2015] or multi-well phase-field model [Lombardo, 2018], an analysis of thermal stress-induced modification of the activation energy [Elshin, 2015], and molecular dynamic simulations [Albenze, 2004; Albenze, 2006; Rogachev, 2019]. An example of solid-phase crystallization is demonstrated in Figure 26.34 for the case of amorphous CuTi thing film [Rogachev, 2019]. The degree of misfit (parameter W) characterizes a difference between the simulated structure and theoretically perfect crystal lattice. The simulations show that the velocity of the explosive crystallization from can range from the level of cm/s [Elshin, 2015] to several dozens of m/s [Rogachev, 2019] that looks to depend on many factors, including material properties and the local temperature.

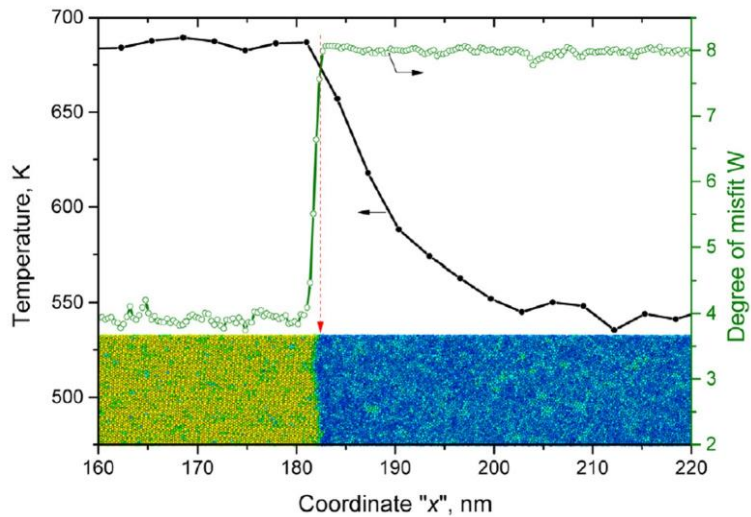

**Fig. 26.34**: Distribution of the temperature and the degree of misfit across the self-propagating front of crystallization in an amorphous CuTi thin film [Rogachev, (2019)]. The red arrow indicates the position of the front at the cross-section of the sample. Blue atoms are in amorphous structure, yellow atoms represent B11 crystal lattice, and intermediate tints of green correspond to intermediate degrees of ordering, according to the misfit parameter $W$. Reprinted from [Rogachev, (2019)], Explosive crystallization in amorphous CuTi thin films: a molecular dynamics study, S.A. Rogachev et al., J. Non-Cryst. Solids **505**:202, Copyright 2019, with permission from Elsevier.

For specific nanostructured amorphous materials, to keep them undistorted after laser annealing, the mechanism of solid phase crystallization is preferable as melting can lead to deformation of the initial nanostructure shape [Sopha, 2020; Mirza, 2023] or intermixing between adjacent nanolayers [Volodin, 2023; Mirza, 2023]. The conditions to achieve solid-phase crystallization are still a subject of investigations that will be discussed in the next subsection based on two successful examples, laser annealing of $TiO_2$ nanotube (TNT) layers and amorphous Ge/Si multilayer stacks.

## 4.3 Examples of Ultrashort Laser Annealing of Nanostructured Semiconductors

**Laser Annealing of $TiO_2$ Nanotube Layers**

The TNT layers are very promising nanostructures for catalytic applications [Macák, 2007; Lamberti, 2015; Sopha, 2018] due to their large surface area. They are usually fabricated via anodization process and can reach several hundreds of micrometers in thickness [Macák, 2005; Sopha, 2018]. Besides, the TNT layers are promising for sensing applications [Varghese, 2003]. The best catalytic and sensing activities of the TNT layers are demonstrated when their as-prepared amorphous form is converted into the anatase phase [Macak, 2007]. However, laser annealing is not a straightforward process, requiring a thorough choice of irradiation parameters. As an example, nanosecond laser annealing at 355 nm wavelength, although leading to the anatase phase formation without any signs of rutile or brookite, results in considerable distortion of the TNT layers due to the melting of their tops (Figure 26.35) [Haryński, 2020]. Nanosecond laser annealing with 266-nm wavelength also yields in TNT layers distortion, although a partial crystallization to anatase occurs [Wawrzyniak, 2020a]. It is interesting that, at certain irradiation conditions in an ns irradiation regime, an accurate closing of the TNT tops can be achieved (Figure 26.36) [Wawrzyniak, 2020b]. Using visible femtosecond laser irradiation (wavelength of 515 nm, pulse duration of 250 fs) also results in considerable distortion of the TNT layers with the appearance of both anatase and rutile phases (Figure 26.37) [Bernhardt, 2024].

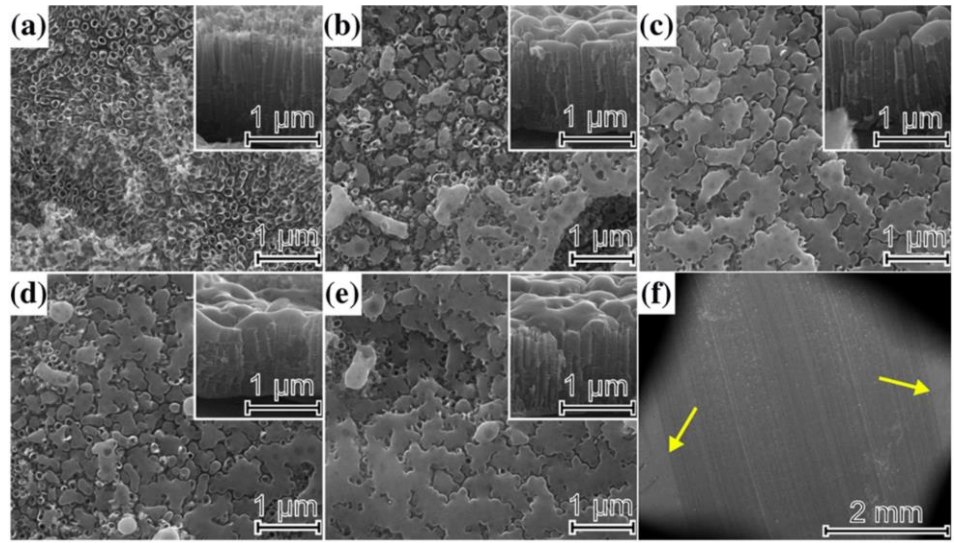

**Fig. 26.35:** SEM images with cross sections (insets) of laser-treated TiO$_2$ NTs with different laser fluences [Haryński, 2020]: (a) 10 mJ/cm$^2$, (b) 20 mJ/cm$^2$, (c) 30 mJ/cm$^2$, (d) 40 mJ/cm$^2$, and (e) 50 mJ/cm$^2$. (f) Laser trace obtained for a 40 mJ/cm$^2$ energy fluence whose boundary is marked by yellow arrows. Reprinted from [Haryński, 2020], Scalable route toward superior photoresponse of UV-laser-treated TiO$_2$ nanotubes, Ł. Haryński et al., ACS Appl. Mater. Interfaces **12**:3225, Copyright 2020, with permission of American Chemical Society.

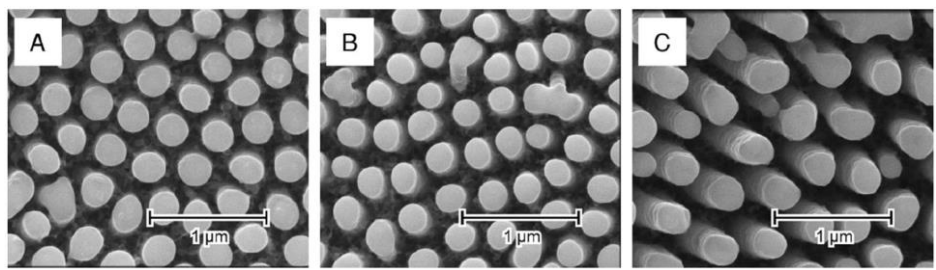

**Fig. 26.36:** SEM images of the TNTs anodized at 40 (A), 50 (B), and 60 V (C) after laser treatment ($\lambda$ = 355 nm, 30 mJ/cm$^2$) [Wawrzyniak, 2020b]. Reprinted from [Wawrzyniak, 2020b], Formation of the hollow nanopillar arrays through the laser-induced transformation of TiO2 nanotubes, J. Wawrzyniak et al., Sci. Rep. **10**:20235, Copyright 2020, with permission of Springer Nature.

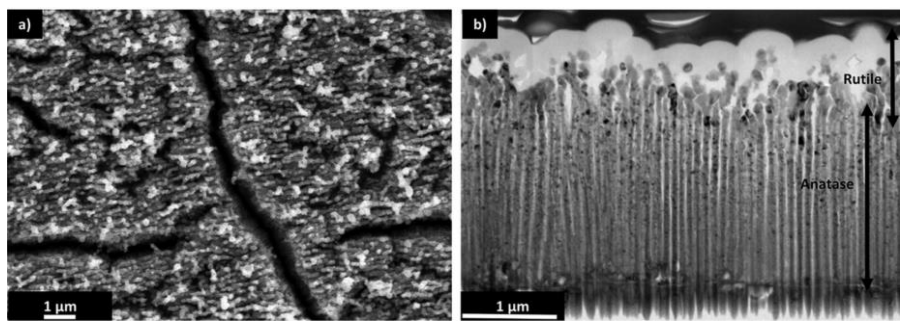

**Fig. 26.37:** a) SEM image from the top and b) TEM cross-section with rutile and anatase layer marked of NTs of sample laser treated at 376 mW and 5 mm s$^{-1}$ [Bernhardt, 2024]. Reprinted from [Bernhardt, 2024], Laser-crystallization of TiO$_2$ nanotubes for photocatalysis: Influence of laser power and laser scanning speed. A. Bernhardt et al., Laser Photonics Rev. **18**:2300778, Copyright 2024, with permission from Wiley.

However, successful selective transformation of the amorphous TNT layers without distortion and visible signs of melting can be achieved with ultrashort laser irradiation under specific irradiation conditions. Qiao et al. [Qiao, 2020] synthesized the TNT layers via a double-step anodization process so that the TNTs were organized into a honeycomb-like structure, as shown in Figure 26.38. These peculiar nanotube layers were irradiated by 50-fs laser pulses at 800-nm wavelength. In a certain range of laser intensities, crystallization to the anatase phase was achieved similar to that obtained using 400 °C thermal annealing.

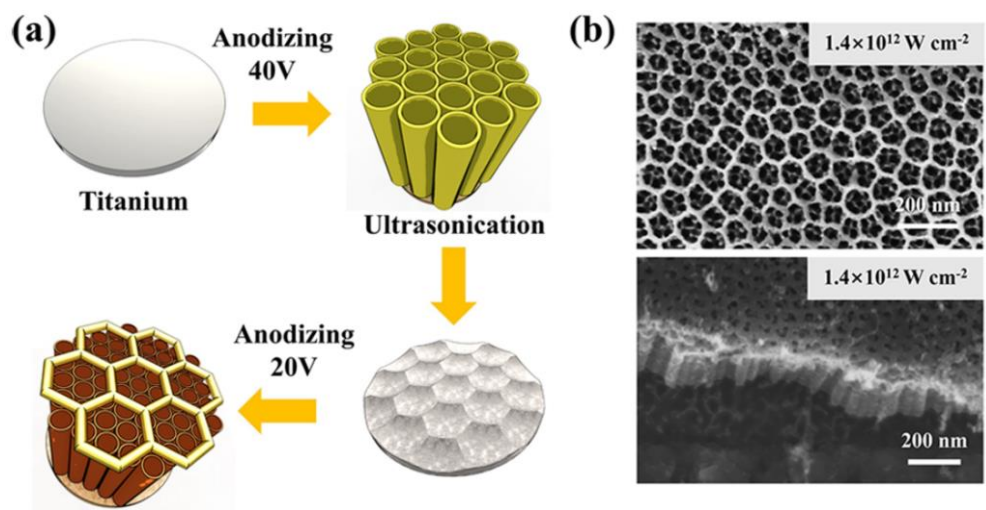

**Fig. 26.38:** Schematics of the preparation procedure of TNTs by the two-step anodizing process (a) and SEM images of TNTs irradiated by the femtosecond laser with the energy intensity of $1.4 \times 10^{12}$ W/cm$^2$ (b) [Qiao, 2020]. Reprinted from [Qiao, 2020], Femtosecond laser induced phase transformation of TiO$_2$ with exposed reactive facets for improved photoelectrochemistry performance, M. Qiao et al., ACS Appl. Mater. Interfaces **12**:41250, Copyright 2020, with permission of American Chemical Society.

Qiao et al. [Qiao, 2020] attribute the crystallization mechanism to nonequilibrium conditions realized in the TNTs upon ultrafast laser excitation of the electron plasma followed by electron-lattice thermalization at a few-ps timescale when the titania lattice does not have time to melt, thus leading to solid phase transformation. However, in the recent work by Bernhardt et al. [Bernhard, 2024], even a more favorable regime for electron plasma excitation (higher photon energy, longer pulse, giving more space for the development of avalanche ionization) yields a considerable distortion/melting of the TNT layers. It can be speculated that the driving force of amorphous-to-crystalline phase transformation without visible melting signs is the solid-phase explosive crystallization as considered by Mirza et al. [Mirza, 2023].

Indeed, evaluations of stresses and temperatures realized upon laser annealing of the TNT layers with the forth harmonics of picosecond diode-pumped thin-disk laser source PERLA-C developed at the HiLASE centre (fundamental laser wavelength of 1030 nm, 2 ps pulse duration) [Novák, 2016] indicate that stress-induced explosive crystallization is the most probable mechanism of the observed phase transformation [Mirza, 2023]. The schematic of the process is shown in Figure 26.39a. The stress formed in only 2-ps time on the very top of the nanotubes in the laser-light absorption zone generates a stress (or shock) wave propagating one-dimensionally along each nanotube. If such a wave ignites crystallization in the most stressed nanotube top, the heat of ~22.6 kJ/mol released upon crystallization can result in a self-propagating crystallization wave [Rogachev, 2017] along the nanotube. The activation energy $\Delta E_{act}$ for amorphous-to-anatase transformation is 69 kJ/mol, which is lower than that for anatase-to-rutile conversion (129 kJ/mol), which is in favor of anatase formation at relatively mild stresses (see scheme in Figure 26.39b [Mirza, 2023]). This is also supported by Ostwald's rule [Ostwald, 1897], according to which the phase transformation proceeds first not to the most stable phase (rutile in the case of $TiO_2$) but to a less stable polymorph which is closest in energy to the original state (anatase for $TiO_2$ although it is a metastable phase). On the shock wave path, the sequence of phase transitions depends on whether the pressure is reduced slowly or rapidly or, in other words, if the new crystalline phase is frozen or can further transform into a more stable phase. This scenario results in the successful transformation of amorphous TNT layers to the anatase phase while preserving the initial nanomaterial morphologies without any melting signs [Mirza, 2023].

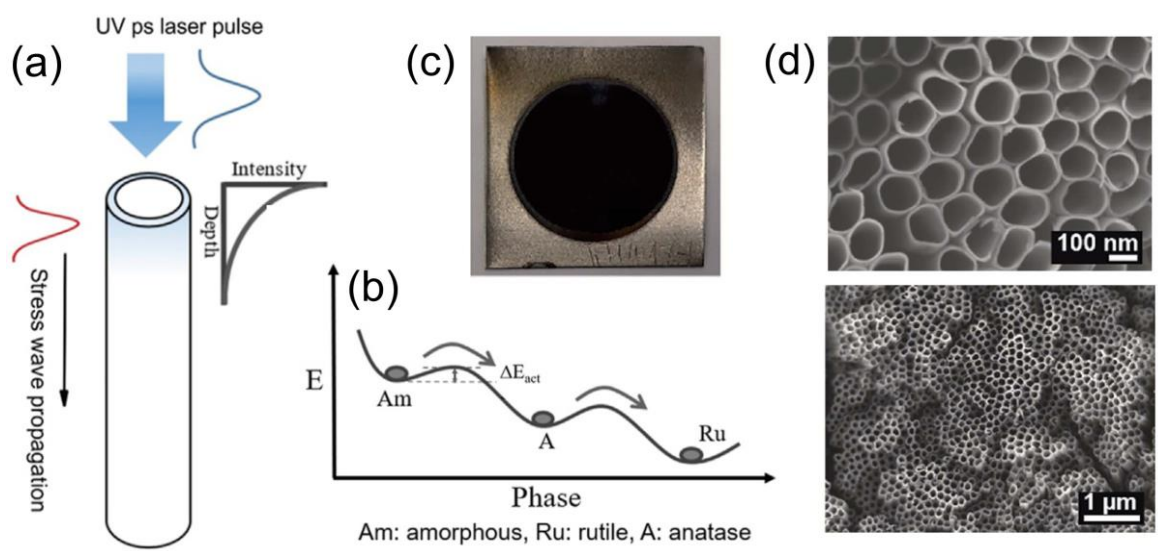

**Fig. 26.39:** (a) Schematics of a stress/shock wave initiation upon ultrashort laser annealing of the TiO2 nanotubes. (b) A sketch for phase transformation from the amorphous state to anatase and rutile. (c) An optical image of the laser-annealed TNT layer at the optimal conditions (peak fluence of ~1 mJ/cm$^2$, ~10$^6$ pulses per sight). (d) Magnified views (SEM images) of the sample shown in (c). No signs of melting are visible [Mirza, 2023]. Reprinted and adapted from [Mirza, 2023], Non-thermal regimes of laser annealing of semiconductor nanostructures: crystallization without melting, Mirza et al., Front. Nanotechnol. **5**:1271832, Copyright 2023, under Creative Commons Attribution 4.0. Retrieved from https://doi.org/10.3389/fnano.2023.1271832.

Comparing the laser irradiation regimes applied in works [Mirza, 2023] and [Qiao, 2020], single-photon ionization using fourth harmonics of the *powerful* PERLA-C laser [Mirza, 2023] is much more favorable for large-scale annealing. The large elliptical irradiation spot on the sample surface with the size of 2.9×2.5 mm$^2$ enabled to anneal the surface area of 1 cm$^2$ in ~14 min with the beam scanning velocity of 0.25 mm/s along the scanning line and the shift of 0.5 mm between the lines [Mirza, 2023]. The tight focusing regime of irradiation using a commercial 800-nm laser at fundamental harmonics involving multiphoton mechanism of $TiO_2$ ionization (irradiation spot diameter of 1.69 μm, the laser scanning velocity of 1 mm/s, the distance between scanning lines of 1 μm) [Qiao, 2020] would require ~28 hours to anneal 1-cm$^2$ sample area.

**Selective Laser Annealing of Amorphous Ge/Si Multilayer Stacks**

Another spectacular example of explosive amorphous-to-crystalline solid-to-solid transition is high-intensity (femtosecond) selective laser annealing of amorphous Ge/Si multilayer stacks [Volodin, 2023]. Selective actions are usually considered a planar selectivity when the laser beam is processing only selected parts of a planar sample [Kwon, 2014] or the whole sample, which has regions with different reflectivity properties (reflective or antireflective) so that the laser anneals only needed parts [Tsukamoto, 1993]. In our case, selectiveness refers to annealing of multilayer stacks, which consist of materials with different absorptivity of laser radiation and different thermodynamic and mechanical properties that could result in phase transformation of some layers while leaving other layers intact.

First attempts of selective crystallization of amorphous germanium (a-Ge) layers in multilayer Ge/Si structures were made in 2019 [Volodin, 2019]. In this work, the action of nanosecond ruby laser at 693 nm wavelength resulted in the intermixing of Ge and Si layers with their crystallization. Unsuccessful crystallization can be explained by the fact that laser radiation at the used wavelength is absorbed in both Si and Ge layers, leading to melting the layers during the relatively long pulse (70 ns).

In 2020, attempts were made to apply femtosecond laser pulses [Kolchin, 2020]. A laser at a wavelength of 1250 nm and a pulse duration of 150 фс was used to anneal multilayer structures consisting of ten pairs of silicon (10 nm thick) and germanium (5 nm thick) amorphous layers on a glass substrate. The effect of selective laser crystallization of Ge layers was not achieved. Instead, solid crystalline SiGe alloy was formed.

In a series of works [Bulgakov, 2023; Mirza, 2023; Volodin, 2023], ultrashort laser annealing of multilayer structures consisting of 4 silicon and 3 germanium (40-nm and 15-nm thick, respectively) alternating amorphous layers on silicon and glass substrates was investigated. The samples were fabricated by plasma-enhanced chemical vapor deposition (PECVD). A cross-section image of the studied structure is shown in Figure 26.40, left.

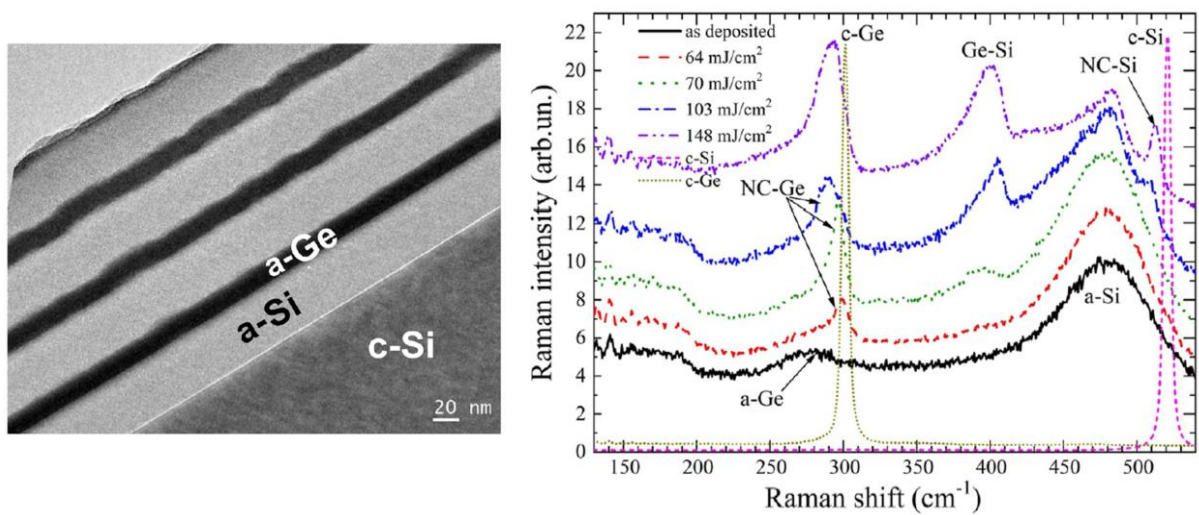

**Fig. 26.40:** Left: Transmission electron microscope image of as-deposited a-Si/a-Ge multilayer stack. Ge layers are dark, while Si layers are light grey. Right: Raman spectra of the pulse-laser annealed Si/Ge multilayer stack shown on the left [Mirza, 2023]. For reference, the peaks of crystalline silicon and germanium (c-Si and c-Ge, respectively) are also added. The annealing was performed at a wavelength of 1.5 μm and a pulse duration of ~70 fs. At laser fluences slightly above the modification threshold, nano-crystallization of Ge is observed without visible effects of melting (red dashed line). With increasing laser fluence (spectra from bottom to top), intermixing of Ge and Si at film interfaces starts pointing to melting effects that are followed by the appearance of the crystalline silicon peak and strong intermixing of layers. Reprinted and adapted from [Mirza, 2023], Non-thermal regimes of laser annealing of semiconductor nanostructures: crystallization without melting, Mirza et al., Front. Nanotechnol. **5**:1271832, Copyright 2023, under Creative Commons Attribution 4.0. Retrieved from https://doi.org/10.3389/fnano.2023.1271832.

The most successful crystallization of germanium nanolayers without any signs of intermixing with germanium was achieved using femtosecond laser Astrella from Coherent in combination with optical parametric amplifier TOPAS from Light Conversion. The output laser beam was tuned to the wavelength of 1.5 μm for which silicon is transparent while germanium is well absorbing. Laser pulse duration was ~70 fs. Figure 26.40, right presents the Raman spectra of laser annealed samples. At certain laser fluences slightly above the modification threshold of the samples, a well-defined peak of crystalline germanium is observed without any signs of Si-Ge bonds that indicate the absence of intermixing between the adjacent Si and Ge layers. With increasing laser fluence, intermixing becomes noticeable (see line for 103 mJ/cm$^2$ in Figure 26.40, right), and crystalline silicon (c-Si) peak emerges at even higher fluences.

The two given examples demonstrate capabilities of high-power laser radiation for gentle selective modification of amorphous semiconductors to desired crystalline phases. In the first example, high-power ultrashort UV laser pulses are necessary to rapidly transform the amorphous TNT layers on large areas that was achieved in a single-photon ionization regime. In the second case, high-intensity mid-IR irradiation secures nonlinear selective crystallization of germanium, keeping silicon layers intact.

# 5. Upscaling Surface Nanostructuring

Micro- and nanostructured surfaces of materials of different kinds represent the building blocks of many technologies from tribology and anti-freezing applications to marking, energy harvesting, sensing, and biomedicine. The majority of applications require a high-throughput large-area structuring in a well-controlled manner. This section discusses the problems and solutions for upscaling surface structuring for industrial demands. Subsection 5.1 gives a short overview of existing and possible ways of upscaling laser nanostructuring. In subsection 5.2, a striking example of using diffractive optical elements for surface structuring is given that allows rapid large-area inscription of various micro-/nanoreliefs by demand. A novel application of laser structuring of skies for controlling their sliding on snow is described in subsection 5.3.

## 5.1 Ways to Upscale Laser Nanostructuring

One of the most popular regimes of surface nanostructuring is the formation of laser-induced periodic surface structures, which is already used in applications in technologies and biomedicine [Bonse, 2017; Römer, 2026]. The direct laser formation of subwavelength structures without a complicated lithography patterning process is attractive due to its cost-effectiveness and the fact that its throughputs are approaching the industrial demand level [Gnilitskyi, 2016; Gnilitskyi, 2017; Gnilitskyi, 2021; Römer, 2026]. A comparison of the developed method throughput with other nanoimprinting methodologies is shown in Figure 26.41a (adapted from [Imboden and Bishop, 2014]). The peculiarity of nanostructuring developed in these works is that, using the single pulse laser processing at a high repetition rate (600 kHz), highly regular LIPSS (HR-LIPSS) were formed on metallic and semiconductor surfaces in the regimes of strong ablation when only a few laser pulses overlap in the same irradiation area. A high laser-beam scanning velocity along the sample surface was achieved using a combination of a galvoscanner and a translation stage [Gnilitskyi, 2017]. It should be noted that, in the regimes of low overlapping of laser pulses, the mimicking of LIPSS from the structured area to the new one can be significantly amplified by the tree-wave interference effect, as was shown recently by Mirza et al. [Mirza, 2025] (see schematics in Figure 26.41b). Thus, upscaling of LIPSS formation on large areas can be achieved using single ultrashort laser pulses in *strong ablation regimes* and at *high repetition rates* using *accelerated laser-beam scanning*. The application of a smart two-dimensional *polygon mirror scanner* can provide even higher acceleration of laser-beam scanning for large-scale surface structuring at high throughputs [Roessler and Streek, 2021].

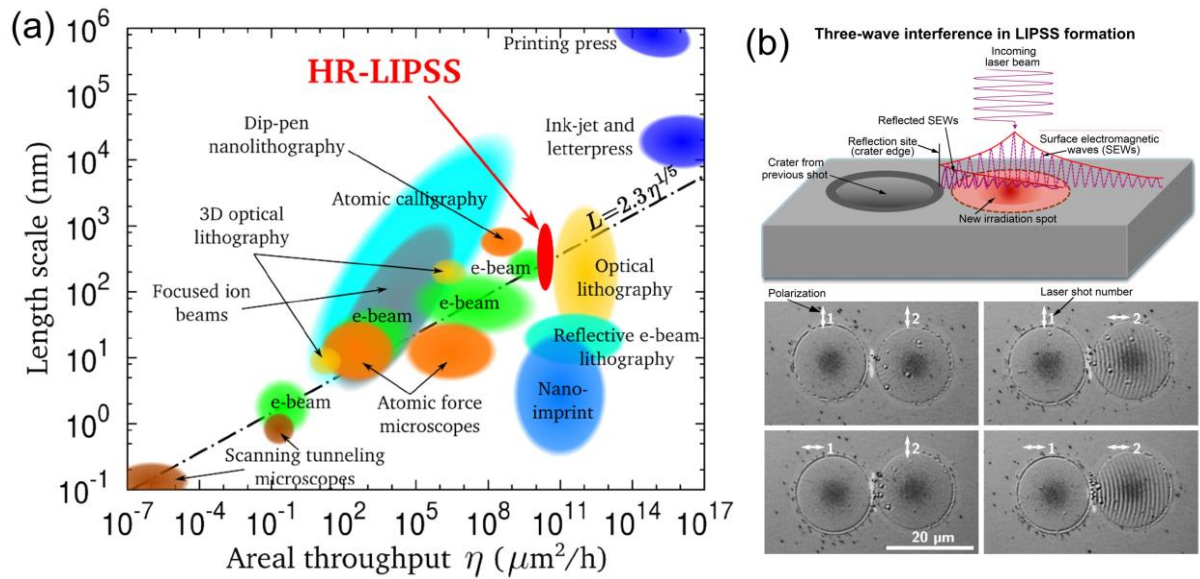

**Fig. 26.41:** (a) Summary of the nano-manufacturing techniques [Imboden and Bishop, 2014]. The black dot-dashed line stands for Tennant's law [Tennant, 1999]. The red ellipse corresponds to the processing method developed in [Glilitskyi, 2017]. (b) Top: Illustration of the three-wave interference mechanism upon fabrication of the LIPSS. Bottom: Differential interference contrast (DIC) optical microscope images of two-shot irradiation patterns on Si (100) surface with pulses of different mutual polarization [Mirza, 2025]. The spot number corresponds to the order of irradiation. (a) Reprinted and adapted from [Imboden and Bishop, 2014], Top-down nanomanufacturing, M. Imboden, D. Bishop, Phys. Today **67**:45, Copyright 2014, with permission of American Institute of Physics. The figure at the bottom of (b) is reproduced and adapted from [Mirza, 2025], Coherence effects in LIPSS formation on silicon wafers upon picosecond laser pulse irradiations, Mirza et al., J. Phys. D: Appl. Phys. **58**:085307, Copyright 2025, under Creative Commons Attribution 4.0. Retrieved from https://doi.org/10.1088/1361-6463/ad9d51.

It can be anticipated that, using high-power lasers, it would be possible to accelerate surface structuring. However, with single laser beams of high irradiation spots, the nanostructures may lose their quality. Indeed, as was shown in [Öktem, 2013], the spatial coherence of the excited surface waves responsible for nanostructure formation can be preserved more easily on a small irradiation area. A smart solution developed recently is the use of *diffractive optical elements,* which split a high-power beam into many smaller beams suitable for precise surface structuring as reviewed in the next Subsection.

## 5.2 Diffractive Optical Elements Enable Industrial Throughput in Surface Nanostructuring

The scalability of laser-based surface miro/nanostructuring to industrial-scale throughput demands innovative beam delivery solutions, particularly when aiming for sub-micrometer resolutions on large areas. In response, diffractive optical elements (DOEs) have emerged as key enablers for upscaling throughput [Yang, 2024; Kiefer, 2025]. A scheme of surface structuring using a DOE is shown in Figure 26.42 [Indrišiūnas, 2022]. DOEs promise to close the speed gap by parallelizing ultrashort-pulse laser processing when enough pulse energy is available. Instead of one scan spot, DOEs can split the beam into arrays from dozens to tens of thousands of sub-beams to multiply throughput [Gillner, 2019; Hauschwitz, 2021a] or shape the beam into complex shapes to allow "laser stamping" of complex patterns [Katz, 2018].

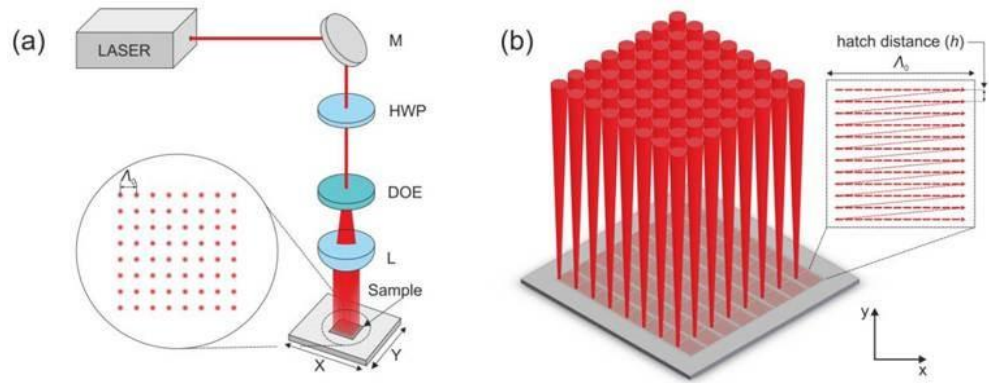

**Fig. 26.42:** (a) Schematic of an 8 × 8 beam-splitting DOE setup for parallel laser microstructuring [Indrišiūnas 2022]. The ultrashort-pulse beam is reflected by mirror (M), passes through a half-wave plate (HWP), then traverses the diffractive optical element (DOE), and is focused by lens (L) onto the sample, forming a 64-spot array (inset). (b) Illustration of the resulting multi-spot interaction on the substrate with user-defined hatching, enabling high-throughput periodic patterning over larger areas. Reprinted from [Indrišiūnas 2022], Large-area fabrication of LIPSS for wetting control using multi-parallel femtosecond laser processing, S. Indrišiūnas et al., Materials, **15**:5534, Copyright 2022, under Creative Commons Attribution 4.0. Retrieved from https://doi.org/10.3390/ma15165534.

Industry has increasingly adopted DOE-based solutions to meet new demands from sub-micrometer periodic structures on consumer electronics to functional tribological patterns on metallic mechanical components [Bruening 2021; Schille and Loeschner 2021]. However, implementing large multi-spot arrays at powers exceeding tens or hundreds of watts brings new complexities: managing thermal load on the DOE, ensuring uniformity among thousands of sub-beams, mitigating nonlinear distortions, and preventing overexposure or stitching defects on the substrate. This section focuses on how DOE elements may enable scalable nanostructuring throughputs for bio-medical and industrial applications and overcome these challenges.

### 5.2.1 DOE Fundamentals and Beam Parallelization

A diffractive optical element exploits the wave nature of light to *diffract* an incoming beam into one or more predetermined diffraction orders. By precisely designing micro- or nanostructured phase masks (often via iterative computational approaches such as the Iterative Fourier Transform Algorithm, IFTA), engineers can prescribe how the beam is split or shaped in the far field [Khonina, 2024]. Unlike refractive lenses or mirrors, DOEs do not rely on varying refractive indices over large optical path lengths but rather on carefully tailored phase delays in thin, micro-patterned surfaces.

Diffractive optical elements (DOEs) enable the formation of complex intensity distributions, including customized patterns, arrays, and non-diffracting beams such as Bessel and Airy beams [Niu, 2018], by precisely modulating the phase of the wavefront [Doskolovich, 2021]. These characteristics make DOEs essential for various applications, including laser material processing, beam shaping, focusing, optical trapping and manipulation, microscopy, and lithography [Komlenok, 2021]. Furthermore, DOEs are integral to wavefront control and adaptive optics (AO) systems [Salter and Booth, 2019]. By accurately shaping the phase profile of incident light, they can correct optical aberrations, leading to improved imaging resolution, enhanced beam quality, and better focal spot control [Dong, 2024].

Commonly employed DOEs in high-throughput material processing include Beam splitters (multi-spot elements) for the generation of a discrete array of beams from a few spots up to tens of thousands (e.g., 201×201 matrix) [Hauschwitz, 2021b; Kato 2005]. Top-hat and vortex DOEs for the production of uniform "flat-top" intensity profiles or ring-shaped beams can minimize thermal effects and improve structure uniformity to optimize ablation, LIPSS formation, or photo-polymerization [Maibohm, 2020]. Focusing DOEs (diffractive lenses) can provide controlled focusing properties (e.g., elongated Bessel-type beams or multi-foci for volumetric processing) [Cheng, 2021].

Integrating DOEs with industrial lasers has two major advantages: (i) *parallelization*, achieved by splitting the energy into multiple beamlets for simultaneous processing of large surface areas, and (ii) *complex beam tailoring*, allowing for advanced texturing strategies such as interference-based patterning, local intensity modulation, and depth-of-focus engineering [Brodsky and Kaplan, 2020; Weber and Graf, 2021].

To implement these elements for parallelization in *surface nanostructuring,* it is necessary to consider several aspects. The first is the available power and pulse energy with respect to the required number of sub-beams, each receiving a fraction of the total laser energy and diffraction efficiency. DOE diffraction efficiency determines how much of the incident energy is allocated to the usable spots versus higher-order losses. Multilevel or continuous-phase DOEs typically achieve >90% efficiency, while simpler binary DOEs may be limited to ~70–80% [Brodsky and Kaplan, 2020; Khonina, 2024]. In the case of material selection, fused silica or crystalline substrates are often favoured due to high laser-induced damage thresholds (LIDT) [Hilton 2016; Kawamura 2020]. Hence, compared to conventional refractive or reflective beam shaping, DOEs allow finer wavefront engineering at relatively thin device thicknesses—a major advantage for integrating with high-power laser systems [Komlenok, 2021].

**Comparison to Alternative Nanofabrication Methods**

Electron-Beam Lithography (EBL) remains the gold standard for sub-50 nm resolution but is extremely slow (often ~cm$^2$ per day) and not cost-effective for large surfaces [Sharma 2022]. Nanoimprint Lithography (NIL) offers faster replication once a stamp is made, but it lacks the direct-write flexibility of laser-based approaches and may introduce issues with stamp wear and contamination [Torii, 2022]. Mechanical Texturing handles large surfaces quickly but typically falls short of sub-µm resolution [Soni and Neigapula, 2024].

DOE-based laser processing often provides a compromise between EBL-level resolution and mechanical texturing speeds. For functional surfaces with a feature size in a range of ~300 nm to 5 µm, DOEs stand out as a potentially *viable, cost-effective* solution. Notably, throughputs reaching only a few minutes per square meter are becoming feasible for structured features in the 0.3–5 µm range [Faas, 2019; Hauschwitz, 2021a].

**5.2.2 High-Throughput Applications and Performance Benchmarks**

**Industrial and Photonic Surfaces**

In recent years, multiple authors have reported DOE-based beam splitting methods for surface micro- and nanostructuring, often achieving impressive throughput for industrial applications. For example, Gillner et al. [Gillner, 2019] demonstrated efficient micromachining with high-power ultrashort pulses using 144 sub-beams, enabling high-speed parallel texturing on metals at energies suitable for industrial production lines. Faas et al. [Faas, 2019] harnessed a 525 W picosecond laser with long focal-length scanning to generate LIPSS on stainless steel, reaching up to 12.5 cm$^2$/min. Meanwhile, Hauschwitz et al. [Hauschwitz, 2020] applied DOE-based 784 sub-beam splitting to fabricate patterns on invar, targeting OLED display manufacturing.

From a tribological standpoint, rotors, bearings, and piston rings also benefit from DOE-driven texturing, where multi-spot femtosecond ablation produces micro-dimples or grooves that enhance oil retention and reduce friction and wear [Danilov, 2023; Schille, 2020]. Beyond tribology, photonics and energy-harvesting devices can also exploit DOE-based processes to produce uniform sub-wavelength features at a large scale. For instance, Darwish et al. demonstrated a multi-beam approach for embedding rare-earth-doped phosphor nanoparticles into polymer films for luminescent solar concentrators [Darwish, 2018]. Kirubaraj et al. [Kirubaraj, 2018] used multi-beam interference lithography to form 1D, 2D, and 3D periodic patterns as small as 0.07 μm, beneficial for light trapping in solar photovoltaics. Further, multi-beam laser interference has enabled direct silicon etching for antireflective and self-cleaning surfaces, lowering solar-weighted reflection while improving hydrophobicity [Zhao, 2015]. Such strategies prove vital for OLED displays, solar panels, and wafer-scale photonic modules, offering large-area coverage with fine feature control—an essential step toward robust high-volume manufacturing.

**Biomedical Device Surfaces**

To improve antibacterial properties, femtosecond-induced nanopatterns (e.g., ripples or spikes) are often used [Lutey, 2018]. Yet throughput in such applications often remains limited to only a few cm$^2$/min. Consequently, DOE-based nanostructuring is increasingly used to accelerate large-area fabrication for biomedical surfaces. For instance, in a roll-to-roll embossing context, a 16-spot multi-beam approach harnessing a 500 W picosecond laser enabled the production of 1−5 μm antibacterial structures on copper surfaces at high ablation rates [Bruening, 2020]. After replication onto plastic substrates, the structured topography demonstrated freedom from chemical additives and potential for mass-manufactured bioactive surfaces.

Holographic lithography augmented by DOEs also holds promise for large-scale bio-implants. Stankevičius et al. fabricated PEG-DA scaffolds suitable for muscle-derived myogenic stem cells by splitting a 515 nm femtosecond beam into four identical sub-beams [Stankevicius, 2012]. Their approach drastically cut fabrication time while maintaining periodic microstructures over multi-millimeter areas, confirming robust cell attachment and demonstrating how DOE-based interference can speed scaffold production.

In a different domain, Bai et al. integrated a diffractive top-hat shaper into an IR-MALDESI mass spectrometry system for biological tissue analyses, enabling uniform ablation profiles and clearer molecular imaging [Bai, 2022]. By distributing the laser energy more evenly, they avoided oversampling and enhanced both 2D and 3D tissue analysis throughput.

Extending from 2D surfaces to 3D microstructures, Maibohm et al. [Maibohm, 2020] employed a 9-beam multi-spot two-photon polymerization (2PP) process to create large-area scaffold arrays with sub-micrometer features (Figure 26.43). The resulting periodic configurations supported living HeLa cells, showing potential for cell migration, separation, or tissue engineering applications. Notably, the DOE-enabled parallelization cut typical 2PP build times substantially, underscoring the key role of multi-beam design in bridging the gap between specialized laboratory techniques and broader-scale biomedical engineering. Most recently, Hauschwitz et al. demonstrated how combining DOE-based multi-beam splitting with dynamic beam shaping for nanogrooves and micropillars on stainless steel boosted production speed by over 500%, significantly improving cell proliferation and osteointegration [Hauschwitz, 2023]. Altogether, these efforts underscore the versatility of DOE-assisted processing in achieving high-throughput, functionally enhanced surfaces for improved biocompatibility and precision in medical implants and devices.

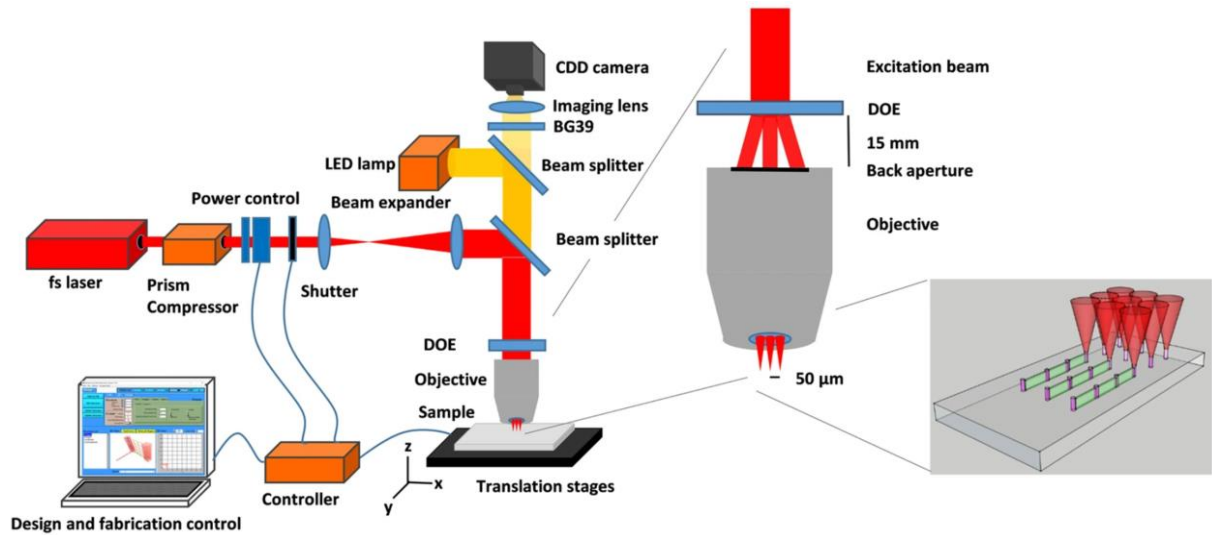

**Fig. 26.43:** Schematic of a two-photon polymerization (TPP) setup based on stage scanning with a femtosecond laser [Maibohm, 2020]. The beam is expanded and then passed through a diffractive optical element (DOE) near the microscope objective, creating multiple beamlets for parallel direct-write fabrication. The inset (right) illustrates the writing process, where nine focused sub-beams trace 3D structures in a photosensitive polymer. A CCD camera and beam splitter enable real-time monitoring, while the motion of translation stages is controlled to precisely pattern the sample. Reprinted from [Maibohm, 2020], Multi-beam two-photon polymerization for fast large area 3D periodic structure fabrication for bioapplications. C. Maibohm et al., Sci. Rep. 10, 8740, Copyright 2020, under Creative Commons Attribution 4.0. Retrieved from https://doi.org/10.1038/s41598-020-64955-9.

**Combining DOE with Other High-Throughput Technologies for Industrial Lines**

Despite their standalone utility, DOEs can be further leveraged by pairing them with additional high-throughput laser-processing technologies. Two prominent approaches include Direct Laser Interference Patterning (DLIP) [Lasagni, 2017] (see also Chapter 9 [Lasagni, 2026]) and polygon scanning [Schille, 2016], which already enable large-area texturing at high speeds, can each benefit from the unique beam-shaping and parallelization capabilities of DOEs. Hence, to harness the full potential of high power sources in the kW-range, *multi-beam* or *beam-shaping* methods, DOEs must be combined with additional strategies like polygon scanning. This synergy may yield multi-fold productivity gains; for instance, Schille et al. [Schille, 2020] reported 1300 $cm^2$/min for LIPSS production on a stainless steel sample by combining polygon scanner and DOE beamsplitter, demonstrating how DOE-based parallelization combined with fast scanning strategies can dramatically improve overall processing efficiency (Figure 26.44).

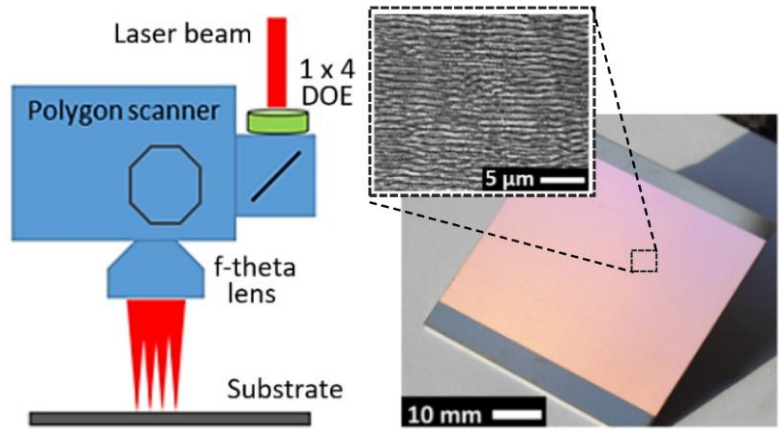

**Fig. 26.44:** Schematic illustrating how a polygon scanner, in combination with a 1×4 DOE and an f-theta lens, enables high-speed multi-beam scanning for LIPSS generation [Schille, 2020]. (Right) Photograph of the resulting stainless steel sample, where the inset shows uniform ripples (LIPSS) at the microscale, achieved at throughput rates exceeding 1300 cm²/min. This synergy between polygon scanning and DOE-based beam splitting highlights the potential for rapid, large-area texturing. Reprinted from [Schille, 2020], High-rate laser surface texturing for advanced tribological functionality, J. Schille et al., Lubricants **8**:33, Copyright 2020, under Creative Commons Attribution 4.0. Retrieved from https://doi.org/10.3390/lubricants8030033.

**Direct Nanostructuring Through a Microscopic Objective and DOE-Assisted Aberration Compensation**

While multi-beam splitting and long-focal-length scanning solutions dominate large-area industrial texturing, certain high-resolution direct-write applications call for focusing the laser beam through high-numerical-aperture (NA) microscopic objectives [Bernardeschi, 2021]. These systems, often deployed for specialized micro- or nano-scale patterning, can benefit substantially from diffractive optical elements not just for *parallelization* but also for beam correction [Kiefer, 2025]. One notable challenge in short-pulse laser machining through complex objectives (particularly multi-element microscope lenses) is chromatic aberration, where different wavelengths or spectral components of the ultrashort pulse do not converge to the same focal plane. This mismatch can degrade the spot size, reduce peak fluence at the focal spot, and ultimately compromise feature resolution. By designing a custom DOE to pre-compensate for the objective's chromatic dispersion and other wavefront distortions, one can preserve near-diffraction-limited focusing across the entire laser bandwidth [Dong, 2024; Kiefer, 2025]. In practice, such DOEs incorporate phase patterns calculated to offset the lens-induced spectral delay, thus aligning all significant pulse components into a smaller, sharper focal spot. The net result is enhanced ablation efficiency, smoother feature edges, and potentially higher aspect ratios for nanopatterns.

**5.2.3 Challenges and Mitigation Strategies**

Despite breakthroughs in multi-beam and high-speed scanning, multiple challenges must be addressed. Even with thousands of sub-spots, large-scale texturing generally requires scanning by moving the beam or the substrate. Overlap mismatches, or "stitch lines," can degrade both the functionality and appearance of the processed area [Moskal, 2023]. In addition, as sub-beam counts rise above several tens or hundreds, even slight variations in diffraction efficiency create large fluence fluctuations and intensify the risk of sub-spot nonuniformity [Arnoux, 2021]. Mitigation often involves designing multilevel phase elements, applying robust fabrication controls, and integrating sensor-based feedback to detect and compensate for intensity variations. Some systems actively measure the incident beam pattern and adjust an upstream deformable mirror to maintain uniform sub-spot fluence [Cumming, 2013]. High-numerical-aperture or wide-field f-theta lenses must also be carefully matched to the DOE, particularly when large splitting angles risk compromising diffraction-limited performance across a wide field [Brodsky and Kaplan, 2020; Khonina, 2024].

Compounding these issues, ultrashort lasers at average powers above 0.5 kW pose additional risks for DOE damage if energy is not judiciously distributed and cooled. Crystalline substrates, such as $CaF_2$ or sapphire, may endure intense repetitive pulsing more reliably [Hilton, 2016; Kawamura, 2020], but water- or air-cooling is frequently necessary at elevated repetition rates to avert thermal buildup. Moreover, multi-spot interference or closely spaced beamlets can cause complex heat-affected zones and chemical cross-talk (proximity effects) [Arnoux, 2021], where local polymerization or ablation departs from the intended pattern. Strategies to diminish these include radical quenchers or specially tailored materials, multi-pass scanning that partially offsets each subsequent exposure, or simulation-based optimization to reduce undesired overlap. Finally, nonlinear effects such as self-focusing, Kerr lensing, and multiphoton absorption can degrade beam quality at especially high pulse intensities [Diels and Rudolph, 2006]. Slight defocusing of the incident beam onto the DOE and implementing specialized anti-reflective (AR) coatings help counteract excessive intensities in the substrate, thereby minimizing local hotspots and preserving the integrity of the final structures. Table 26.1 provides a concise summary of these challenges.

Table 26.1. Main challenges in DOE-based high-throughput laser nanostructuring and potential mitigations

| Challenge | Causes / Manifestations | Potential Solutions |
|---|---|---|
| (1) Stitching Errors | Overlapping passes or partial coverage due to scanning errors or misalignments | - Precision motion systems (sub-μm stage accuracy)<br>- Hybrid polygon + galvo scanning<br>- Real-time feedback (OCT/camera) |
| (2) Non-Uniform Beam Distribution | Inconsistent diffraction efficiency or local hot spots with large sub-beam arrays | - Iterative DOE design, e.g., multi-level surfaces<br>- In-situ beam diagnostics to balance sub-spot intensities<br>- Adaptive SLM add-ons |
| (3) Thermal Load and DOE Damage | High-average powers (>100 W) can heat or damage the DOE's microstructure | - Fused silica or crystalline DOE substrates<br>- Water-cooled DOE mounts<br>- Distributing sub-spot intensities to avoid local hotspots |
| (4) Proximity Effects | Multi-spot polymerization or ablation leading to overlapping heat zones or radical diffusion | - Larger inter-spot spacing or multi-pass approach<br>- Fast scanning to minimize local overexposure |

|  |  | - Photoresists with inhibitors (TPP) |
| --- | --- | --- |
| (5) Nonlinear Distortions | Kerr effects, self-focusing, or multiphoton absorption inside the DOE at high intensities | - Slight defocusing or beam expansion upstream<br>- High-LIDT coatings/dopants in DOE substrate<br>- Minimizing direct focusing in the DOE |

### 5.2.4 Future Outlook and Conclusion

The development of DOE-assisted laser nanostructuring is rapidly shifting from proof-of-concept laboratory setups to more robust, industry-ready solutions. Building upon the multi-spot beam shaping and parallel processing strategies detailed in earlier sections, future research will likely emphasize scaling to multi-kW ultrafast lasers and thousands of diffractive beamlets, pushing throughput far beyond current records. Addressing the corresponding challenges will require further innovation in damage-resistant DOE materials, adaptive wavefront control, and real-time in situ diagnostics.

Increasingly, these solutions will integrate with polygon scanners, roll-to-roll systems, and advanced scanning optics, aiming to achieve continuous high-speed coverage of very large surface areas. As more pilot lines demonstrate stable, 24/7 operation with high beam uniformity and minimal stitching artifacts, the technology should mature rapidly for both biomedical (e.g., implant functionalization, antibacterial surfaces) and industrial (e.g., tribology, electronics, optics) applications. Additionally, progress in two-photon polymerization (2PP) and other volumetric processes could extend parallelization from 2D to 3D, paving the way for mass production of intricate micro- or nano-architectures.

Looking ahead, machine learning and adaptive DOE concepts may further refine beam shaping on-the-fly, compensating for drift, thermally induced distortions, and variations in the substrate or environment. These cumulative advances promise to bridge the gap between specialized lab-scale functionalities and the high-throughput demands of real-world manufacturing, solidifying DOE-based laser nanostructuring as a promising technology in the next generation of precise, large-are surface nanostructuring.

## 5.3 Nanostructuring for Sliding on Snow

Among various material properties modified by the laser micro/nanostructuring of surfaces, such as optical response, coloration, and wettability, the tribological properties of materials occupy a specific place, as the laser action can significantly modify surface frictional behavior [Bonse, 2016; Pan, 2022]. Concerning alpine skies, their bases must meet certain criteria to enable easy maneuverability and high control over change of direction and velocity. Skis must also be light weight to glide fast on the snow and be able to absorb vibrations caused by various kinds of snow surface irregularities. The common structure and design of a ski is sharp, side-cut metal edges and plastic bases. For better performance and reduced friction, skies bases are also subjected to stone grinding and waxing. Figure 26.45 shows a micrograph of a stone-grinded ski base [Maggiore, 2022].

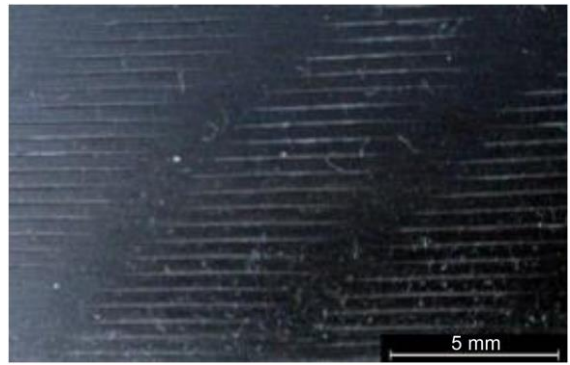

**Fig. 26.45:** A stone grinded UHMWPE ski base [Maggiore, 2022]. Reprinted from [Maggiore, 2022], Sliding on snow of AISI 301 stainless steel surfaces treated with ultra-short laser pulses, E. Maggiore et al., Appl. Surf. Sci. Adv. 7:100194, Copyright 2022, with permission from Elsevier.

Such skies allow skiers to keep a precise and controlled trajectory on the snow surface. The conventional materials for ski base and edges are dense, ultrahigh molecular weight polyethylene (UHMWPE) and carbon steel. Carbon steel (C60) provides good wear and abrasion resistance, acceptable toughness, and ductility. On the other hand, UHMWPE doped with graphite or silicon nanoparticles provides good thermal and electrical conductivity for fast sliding. Since 1970, after the introduction of these economic materials, there has not been a significant improvement in the exploration of other ski base materials due to their easy, well-established manufacturing process and reliable mechanical and thermophysical properties.

Recently, alternative materials such as steel mesh [Nordin and Styring, 2014], laser-textured stainless steel, and Al alloy [Ripamonti, 2018] as well as laser nanostructured stainless steel were studied for ski bases [Ripamonti, 2018; Magiore, 2022]. Laser nanostructured steel was tested for its tribological properties and hydrophobicity [Maggiore, 2023; Bonse, 2015]. Here, we will summarize the results of laser micro-nanostructured surfaces that have been tested for ski bases.

Figure 26.46 presents different kinds of laser-induced structures on steel that were tested as ski base material, such as laser-engraved lines [Maggiore, 2023], LIPSS [Maggiore, 2022], and dimples [Ripamonti, 2018]. The friction coefficient in all these samples shows slightly different behavior. The average friction coefficients of bare, laser-engraved, and LIPSS-covered steel surfaces are presented in Figure 26.47 for different snow temperatures [Maggiore, 2023]. For –3 C°, a comparison of the average coefficient of friction of all tested materials with UHMWPE is given. For all temperatures, the LIPSS-covered AISI 301 sample with sliding direction perpendicular to the LIPSS orientation shows the lowest friction coefficient μ value. In general, LIPSS-covered samples show a lower μ value compared to laser-engraved surfaces. The effect is explained by the reduction of the capillary drag forces, where laser-processed surfaces break the continuity of the quai-liquid layer in contact with ski bases and reduce the extent of capillary drag. The bare steel shows a noticeably higher μ value at higher snow temperatures (–3 C°). This was attributed to the thicker liquid water layer under the ski base that can considerably affect the gliding of a bare steel surface [Maggiore, 2023].

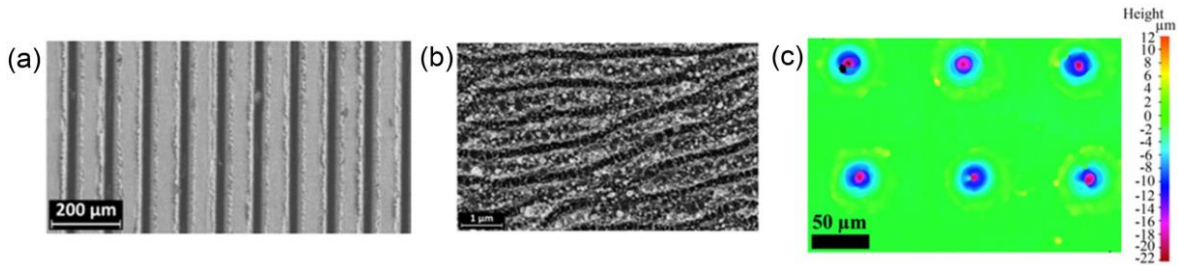

**Fig. 26.46:** Different kinds of laser structured steel surfaces investigated as ski base materials. (a) Laser engraved lines [Maggiore, 2023]; (b) fs laser-induced LIPSS structures [Maggiore, 2023]; and (c) laser-induced dimple structures [Ripamonti, 2018]. The figures (a) and (b) are reprinted from [Maggiore, 2023], Laser-treated steel surfaces gliding on snow at different temperatures, E. Maggiore et al., Materials **16**:3100, Copyright 2023, under Creative Commons Attribution 4.0. Retrieved from https://doi.org/10.3390/ma16083100. The figure (c) is reprinted with permission from [Ripamonti, 2018], Innovative metallic solutions for alpine ski bases, F. Ripamonti et al., J. Vac. Sci. Technol. B 36, 01A108, Copyright 2018, with permission from Ripamonti et al. (2018).

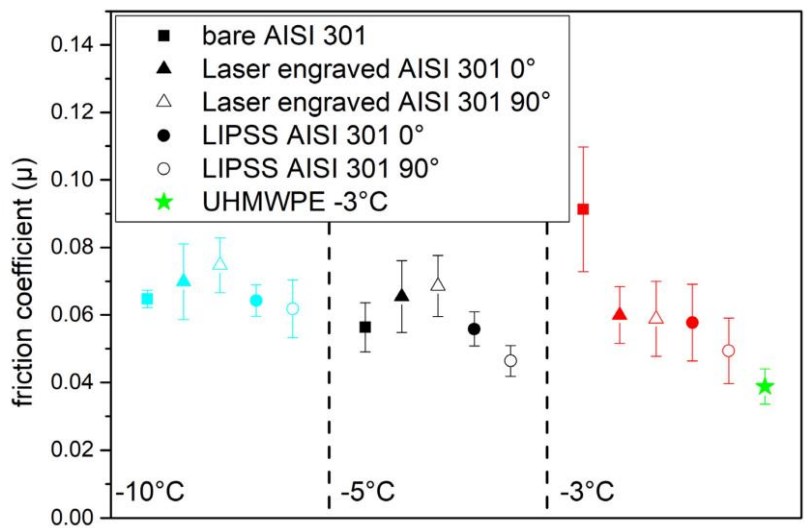

**Fig. 26.47:** Average μ values of differently treated AISI 301H samples at different temperatures and for parallel (full symbols) and perpendicular (open symbols) gliding on snow [Maggiore, 2023]. For reference, the green star is the average μ value for parallel gliding at −3 °C of stone grinded and waxed UHMWPE. Reprinted from [Maggiore, 2023], Laser-treated steel surfaces gliding on snow at different temperatures, E. Maggiore et al., Materials **16**:3100, Copyright 2023 Authors, under Creative Commons Attribution 4.0. Retrieved from https://doi.org/10.3390/ma16083100.

Figure 26.48a shows an image of a snow tribometer equipped with a sample holder and styrofoam container filled with snow [Maggiore, 2023]. An electrical motor with a controller regulates the angular speed of the annulus container. A cylindrical weight can also be inserted to press the sample holder onto the snow surface. Figure 26.48c presents SEM micrographs of the LIPSS prepared on a steel surface using femtosecond and picosecond lasers. Two steal samples (Hi-fs and Hi-ps) were irradiated with 247 fs and 7 ps pulses, respectively, using a commercial Yb:KGW laser system (PHAROS, Light Conversion) operating at a central wavelength of 1030 nm. LP-ps1 and LP-ps2 were processed by a 1064-nm laser with a 3 ps pulse duration. In all samples, some bifurcations can be seen (a phenomenon where an LIPSS ridge splits into two ridges [Bonse, 2015]). In the ps-processed samples, the LIPSS look to be more continuous. A high-resolution SEM image (Figure 26.48c, right) demonstrates that, along with the Low Spatial Frequency LIPSS (LSFL) whose periodicity is comparable with the laser wavelength, the High Spatial Frequency LIPSS (known as HSFL) and nanoparticles are present. Figure 26.48b shows the μ values for the measured samples [Maggiore, 2023]. Among all

LIPSS-covered samples, the friction coefficients for Hi-ps and LP-ps2 are nearly half of that of bare steel, although still they are larger by ~ 30 % as compared with UHMWPE. In terms of durability, unwaxed laser-treated samples show sufficiently low µ values (although still higher than that for state-of-the-art UHMWPE). In conclusion, experimental findings show that laser-treated AISI steel is a perspective alternative base for full-scale skis. We underline that the efficient low-cost laser treatment of ski bases can be achieved using high-power lasers with upscaling techniques, as described above.

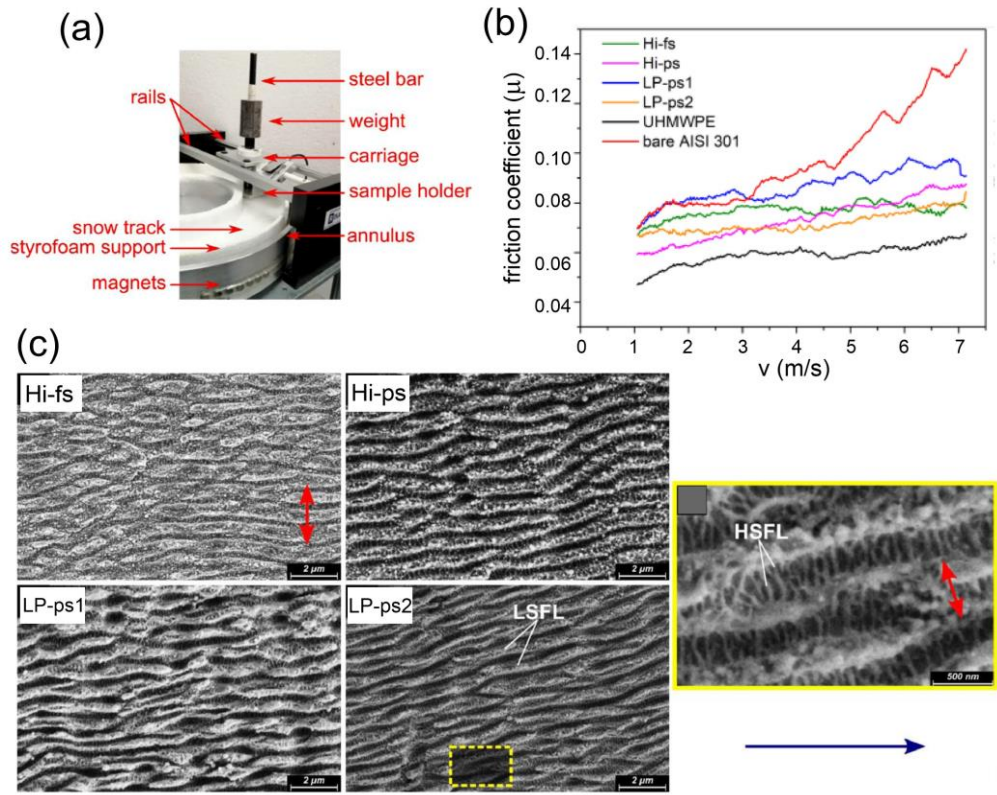

**Fig. 26.48**: (a) Snow tribometer with the Styrofoam container filled with snow inserted inside the aluminum annulus. The sample is held in place by the carriage, and it leans above the snow surface [Maggiore, 2022]. (b) Friction coefficient µ as a function of sliding velocity v for the studied samples. (c) SEM micrographs of LIPSS fabricated using fs and ps laser irradiation. The blue arrow indicates the sliding direction of samples for all cases. The red arrows in the inset indicate the direction of laser light polarization. Reprinted from [Maggiore, 2022], Sliding on snow of AISI 301 stainless steel surfaces treated with ultra-short laser pulses, E. Maggiore et al., Appl. Surf. Sci. Adv. 7:100194, Copyright 2022, with permission from Elsevier.

# 6. Open Access Program of the HiLASE Centre

Over the years, the large laser infrastructures shifted from very secretive and closed centers into open and living communities. Open access provided by these centers was so successful that purely open access facilities were constructed. Subsection 6.1 gives a short overview of the Laser Centres that provide Open Access options. The laser lines for Open Access experiments in the HiLASE Centre of the Czech Republic are specified in Subsection 6.2. The HiLASE center is not fully open access, but it pioneered open access for industrial applications. In Subsection 6.3, examples of Open Access experiments carried out in the HiLASE Centre for both scientific and industrial applications are provided.

## 6.1 Importance of Open Access Programs of Large-Scale Laser Infrastructures

After the demonstration of laser in 1960 [Maiman, 1960], rapid development of laser technology was sparked. In the early years, progress was made in small laboratories that focused on the development of different lasers or pilot applications of lasers. After many simple discoveries were made, large laser facilities emerged (LLNL, AWRE, Arzamas-16) [Parker, 2002; Danson, 2021; Karlov, 2010], facilitating research that required more powerful lasers that could be afforded by only a few major laboratories (laser systems like Argus [Argus], HELEN [Cooke, 1982], or Iskra [Iskra] were developed delivering kJ-level energy). Such laboratories were mostly connected with the military, and hence they were highly confidential. Later it became clear that the scientific community would greatly benefit from access to such facilities. As a result, non-military versions of such lasers were constructed (Omega [Omega, 2010], Vulcan [Danson, 2021], Asterix III [Bederlov, 1976], or Delfin [Karlov, 2010]). They were open to experiments from the scientific community, but access was rather difficult and usually restricted. Sources of funding were limited to some groups that were geographically close or had signed a contract of cooperation. The European Union decided to increase cooperation with the idea of joining efforts of large-scale laser facilities that resulted in establishing the Lasernet program [Lasernet], which reduced fragmentation of laser-based research infrastructure. However, joint development was still difficult since the budget was limited. The Lasernet evolved later to the Laserlab, which was a joint consortium of many research infrastructures with increased budget and competences. Laserlab-Europe grew from 17 partners in 2003 to 35 partners in 2024 [Lasernet]. Laserlab offers transnational access even to non-EU members. Laserlab was so successful that other consortia of research infrastructures were founded (LEAPS [LEAPS] for accelerator-based photon sources, Radiate [Radiate] for ion beams, and many others), and topical programs to fund the consortia are regularly introduced (Re-Made, RIANA, Lasers4EU) [Re-Made; RIANA; Lasers4EU]. The consolidation of research in EU also inspired research in the USA, where in 2018, LaserNetUS [Lasernet] was founded, joining leading laser laboratories in the USA and Canada. Many new research infrastructures offer their own user program under Laserlab-Europe. In the Czech Republic, the joint laboratory of the Institute of Physics and Institute of Plasma Physics of the Czech Academy of Sciences called PALS [PALS] paved path towards centers HiLASE [HiLASE] and ELI Beamlines [ELI], the two facilities that offer user access to next generation of high average power pulsed lasers. Under the umbrella of ELI ERIC, several purely open-access facilities were constructed, namely ELI Beamlines, ELI Alps, and ELI NP were opened in the second decade of the new millennium.

Laserlab-Europe does not only support user experiments, but joint developments are also funded with the aim to improve capabilities of the consortium, including laser pulse contrast enhancement, increase of harmonic conversion efficiency, and many others.

## 6.2 Overview of Laser Lines for Open Access Experiments at the HiLASE Centre

The HiLASE Center was founded in 2011 with the aim to develop new types of lasers for research and industry. The Centre successfully delivered a picosecond low-energy high repetition rate thin disk PERLA lasers developed in-house [Smrž, 2017] and high-energy nanosecond laser Bivoj developed by STFC as DiPOLE100 with the HiLASE participation [Arnoult, 2018]. HiLASE is not a typical user facility since it offers only 10-20% of beamtime to external users through Open Access, the rest of beamtime is used for internal research.

The Perla lasers were built on the concept of thin disks developed at the Institut für Strahlwerkzeuge (IFSW) of the University of Stuttgart, Germany [Giesen, 1994]. The working principle of a thin-disk laser is presented in Figure 26.49 [Schuhmann, 2016]. Although the energy or average power of our lasers is below the state of the art represented mostly by TRUMPF disk lasers, the beam quality is much better and allows for massive parallel processing by beam splitting by diffractive optical element [Hauschwitz, 2021b] (see Subsection 5.2). The output parameters range from a few mJ 1 ps laser pulses at a repetition rate of 100 kHz to more than 20 mJ 1 ps laser pulses at a 1 kHz repetition rate. Fundamental output at 1030 nm wavelength can be converted to higher harmonics (515, 343, 257, 206 nm) [Smrz, 2021] or to lower frequencies in mid-IR (1.5-12 μm) [Duda, 2022].

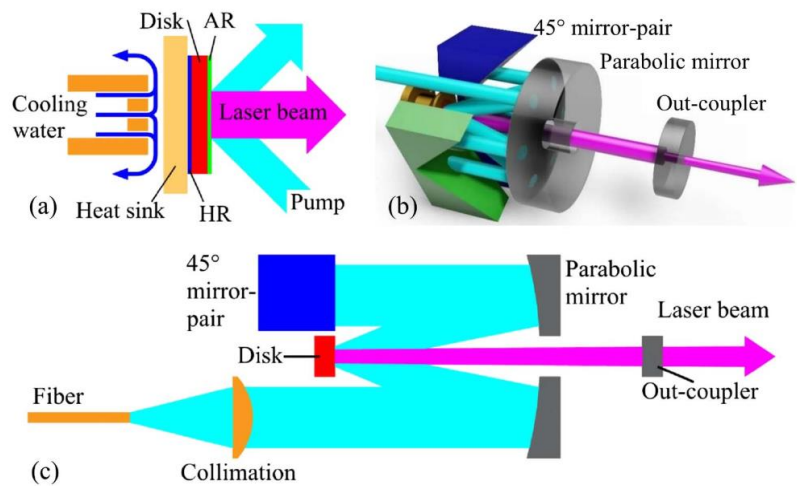

**Fig. 26.49**: Working principle of a thin-disk laser and thin-disk laser pump arrangement [Schuhmann, 2016]. (a) Scheme of the thin-disk active medium mounted on a water-cooled heat sink. Lasing and cooling occur along the disk axis. (b) 3D schematic of the pump optics: heat sink (gold), parabolic mirror (gray), and prisms acting as mirror pairs (green and blue). (c) Pump light multipass arrangement. The light from a homogenizer is imaged onto the disk via a parabolic mirror. The multipass is realized via the disk, the parabolic mirror, and the 45° mirror pairs. The pump beam propagation in the multipass and the laser beam are given in cyan and magenta, respectively. HR and AR stand for high reflection and antireflection layers, respectively. Reprinted from [Schuhmann, 2016], Thin-disk laser pump schemes for large number of passes and moderate pump source quality, K. Schuhmann et al., Appl. Opt. **54**:9400, Copyright 2016, with permission from Optical Society of America.

The Bivoj laser is based on DiPOLE/Mercury multi slab technology [Bayramian, 2008; Banerjee, 2015; Mason, 2015]. Multislab amplifier designs provide a long gain length while maintaining a high surface-to-volume ratio from which to remove residual heat. Heat extraction through the faces of each slab is shown in Figure 26.50 where a schematic of a multislab gas-cooled laser amplifier is presented [Mason, 2015]. The Bivoj delivers 150 J in a temporarily shaped nanosecond pulse with a 10 Hz repetition rate at a wavelength of 1030 nm [Divoky, 2021]. The pulse shape can vary between 2-14 ns with 150 ps resolution. The output beam can be sent to experimental areas through a high-power, large-aperture Faraday isolator [Slezak, 2023]. Alternatively, it can be converted to higher harmonic frequencies (515 nm, 343 nm) [Divoky, 2022; Pilar, 2023] and then sent to experimental areas. These polarization-sensitive processes can be utilized due to successful mitigation of thermally induced polarization changes [Slezak, 2022]. The Bivoj is a versatile tool that can be used to prototype novel methods for science and industry before more affordable and less flexible lasers are developed for specific applications.

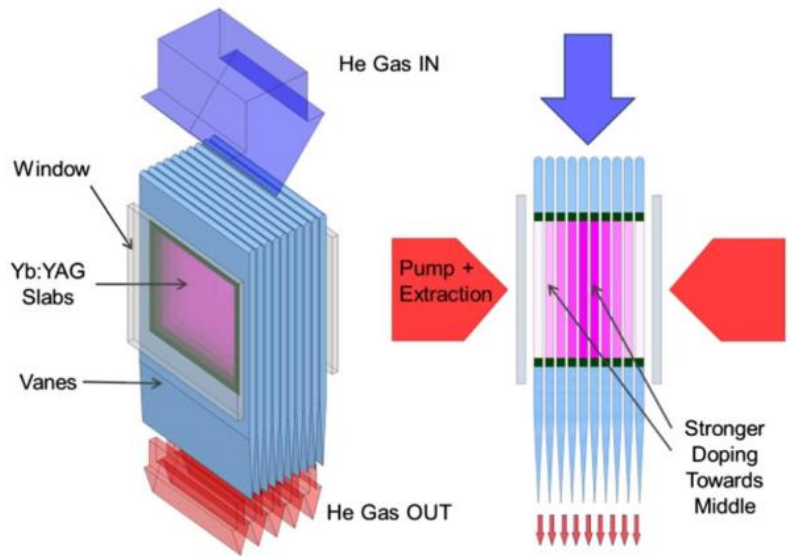

**Fig. 26.50**: Concept of a multislab gas-cooled laser amplifier [Mason, 2015]. Reprinted from [Mason, 2015], Scalable design for a high energy cryogenic gas cooled diode pumped laser amplifier, P.D. Mason et al., Appl. Opt. **54**:4227, Copyright 2015, with permission of Optical Society of America.

The Perla and Bivoj lasers were built with an integrated computer control system that includes archiver logging most of the laser output parameters in real time. These parameters can then be linked to achieved results and used for a better understanding of the processes involved.

**Photon pressure energy meter**

Measuring some laser parameters like energy becomes tricky at high powers and with large-aperture beams. The beam is usually sampled via the leak from a high-reflecting mirror. This decreases fluence and average power incident on the energy meter. However, thermally induced polarization changes the transmission through the mirror during the laser operation and affects the calibration factor of the energy meter. This can be remedied by using normal incidence on the sampling mirror. But for large beams, it is difficult to place such mirrors as they require a long path in front of them to separate the incident and reflected beams. Even if it is achieved, heating the mirror under the high average power load changes the transmission even at normal incidence (for example, if 99.99% mirror decreases its reflectivity by 0.01% upon heating, the transmission doubles, while reflectivity is unaffected). Another solution would be to place an uncoated wedge into the beam. However, this will introduce an energy loss of approximately 4% and so it is not preferred. Therefore, an experiment to measure the radiation pressure of the pulsed beam was conducted [Williams, 2022], where pulse energy was identified by using a precise scale with a nanogram resolution, which measured the pressure of photons incident on a high-reflectivity mirror (see Figure 26.51). The precision of energy measurements was improved at least 5 times in comparison to standard energy measurements using a pyroelectric energy meter.

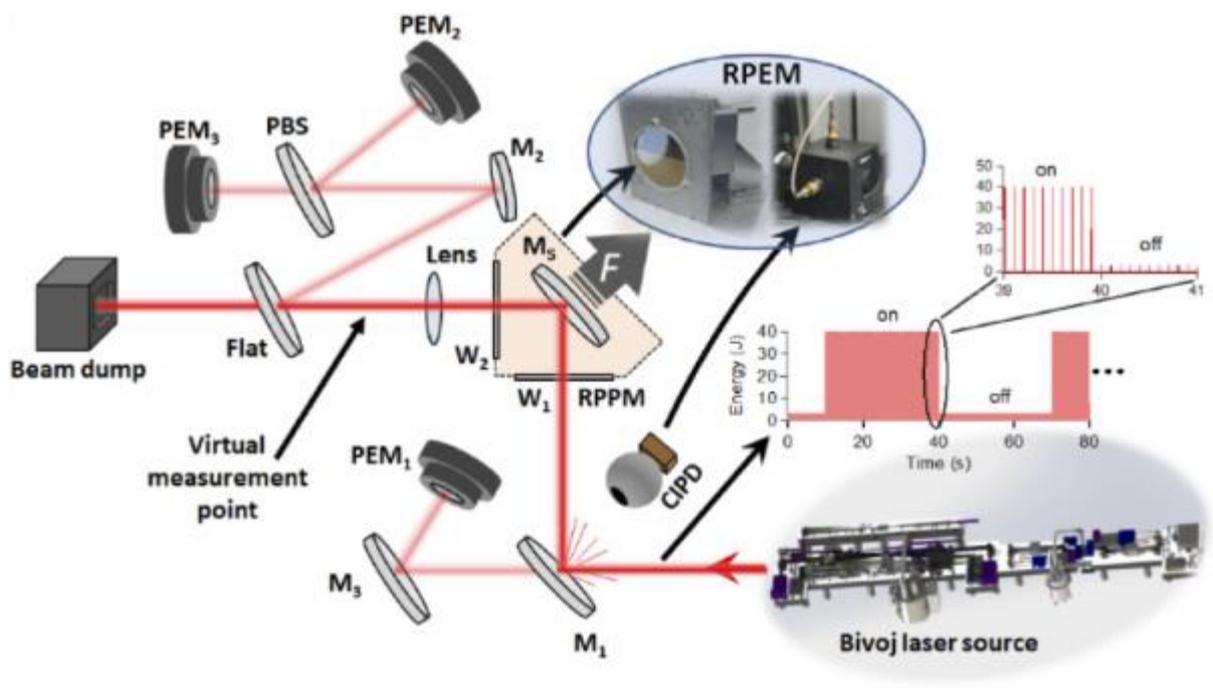

**Fig. 26.51**: A scheme of measurements of pulse energy from the "Bivoj" laser source [Williams, 2022]. The laser produces pulses at 10 Hz with a duration of 10 ns. In the "on" state, the laser final amplification stage is energized, yielding up to 100 J/pulse, and in the unenergized "off" state, only ~ 3 J leakage pulses are emitted (inset). M1, M2, and M3 are dielectric-coated turning mirrors; W1 and W2 are anti-reflection-coated windows. The optical flat samples the pulse energy for measurement by pyroelectric energy meters (PEM2 and PEM3) using the polarizing beam splitter (PBS) for polarization insensitivity. PEM1 is an uncalibrated redundant monitor. RPPM is the radiation pressure power meter measuring the light's average force F on the sensing mirror MS. CIPD is the charge integrator photodiode sampling the light scattered from M1. RPEM is the radiation pressure energy meter comprising RPPM and CIPD. The virtual measurement point is the location at which all meters estimate laser pulse energy. Reprinted from [Williams, 2022], Extreme laser pulse-energy measurements by means of photon momentum, P.A. Williams et al., Opt. Express **30**:7383, Copyright 2022, with permission of Optical Society of America.

**Laser shock peening**

The 10 J power amplifier has been used for user experiments since 2017. The first application of this laser was Laser Shock Peening (LSP), which improves metal material wear resistance. Although the process is known for a long time [Breuer, 2007; Azhari, 2016], different groups used different lasers with different laser parameters, and their results were not comparable and sometimes contradicting. In general, the LSP process offers a much deeper compression residual stress layer in comparison with other techniques like shot peening [Gujba, 2014] or liquid jet peening [Chillman, 2006]. The wear resistance improvement is also slightly (50%) [Azhari, 2016] or substantially better up to several times (~300%) [Kaufman, 2025] (see a scheme of the laser peening process in Figure 26.52). Using a unique pulse of the Bivoj laser and beam-shaping capability, we were able to duplicate the results of other groups and optimize the laser parameters for various materials.

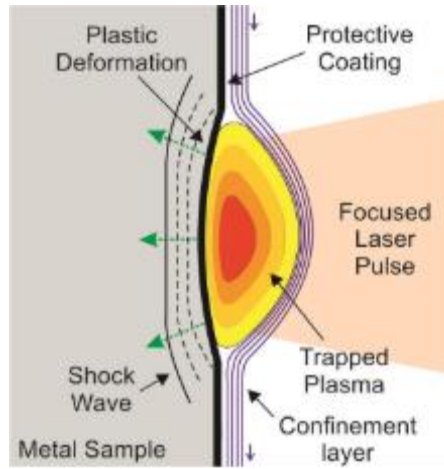

**Fig. 26.52**: Principle of laser shock peening [Kaufman, 2021]. Reprinted from [Kaufman, 2021], Effect of laser shock peening parameters on residual stresses and corrosion fatigue of AA5083, J. Kaufman et al., Metals 11, 1635, Copyright 2021, under Creative Commons Attribution 4.0. Retrieved from https://doi.org/10.3390/met11101635.

## 6.3 Examples of Open Access Campaigns on HiLASE Beamlines

The high average power of the lasers developed at the HiLASE Centre can also show advantages in the fabrication of compact particle sources based on laser irradiation of specific targets and in studies of space propulsion concepts based on lasers. These two applications have been investigated at HiLASE in the frame of the Open Access program, and sometimes experiments were also supported by Laserlab-Europe. On the Bivoj beamlines, the combined high pulse energy and significant repetition rate made it possible to explore the repetitive irradiation regimes that can be practical for efficient space propulsion [Phipps, 2014; Phipps, 2021]. On the PERLA beamlines, the kHz repetition rate with reasonable energy per pulse enabled exploration of irradiation regimes that can be suitable for the realization of relatively compact and controlled (switchable) source of alpha particles, as well as for laser-induced momentum transfer in the frames of space propulsion.

**Generation of alpha particles *via* laser-induced proton-boron fusion**

Lasers have proved very efficient for the generation and acceleration of particles. Multi-terawatt and petawatt laser irradiations of solid targets have shown how ions and particularly protons can be accelerated to energies beyond tens of MeV (see for example [Clark, 2000; Wilks, 2001] and reviews such as [Macchi, 2013]) over several tens of micrometers *via* target normal sheath acceleration (TNSA). In the regimes of lower peak power (gigawatt to terawatt), particle acceleration by laser-plasma interactions may also be achieved, at a lower level though, provided the focusing of the laser beam allows intensities on targets close to $10^{16}$ W/cm$^2$ [Torrisi, 2014].

The latter concept was employed at HiLASE in an Open Access experimental campaign supported by Laserlab-Europe [Istokskaia, 2023]. The goal was to use the PERLA-B beamline with only 10 GW peak power to generate alpha particles *via* proton-boron (p-B) fusion reactions. The nuclear reaction at play involves a proton and a $^{11}$B nucleus yielding mainly an alpha particle and an unstable $^8$Be nucleus that further decays into two alpha particles [Becker, 1987]. The net balance for each reaction is thus the production of three alpha particles, $^4$He, sharing 8.68 MeV of released energy:

$$p + {}^{11}B \rightarrow {}^{8}Be + {}^{4}He \rightarrow 3\ {}^{4}He + 8.68\ \text{MeV}$$

This reaction has great potential for clean energy production as well as for medical applications due to the high yield of alpha particles and the absence of emitted neutrons. For the energy production however, the early schemes considering a confined plasma at equilibrium were suffering from the needs of high temperature to reach a sufficiently high reaction parameter between the ions (protons and boron), detrimental radiation losses at these temperatures and, in the case of inertial confinement, very high density needed to achieve modest expected gain [Moreau, 1977; Tahir and Hoffmann, 1997; Scheffel, 1997]. The evolution of the paradigm towards out-of-equilibrium approaches gave more hope [Rostoker, 1997; Kulcinski and Santarius, 1998; Rostoker, 2003], and now concepts of generators have been reported [Kurilenkov, 2016; Kurilenkov, 2021]. In parallel, with the emergence of short and ultrashort laser pulses providing high peak power in the beginning of the 21$^{st}$ century, generation of highly energetic ions have become routinely possible and laser-driven p-B reaction schemes started to be explored in view of a future promising alternative for clean energy production as well as for the controlled production of radioisotopes and tumor treatments [Yoon, 2014; Qaim, 2016; Szkliniarz, 2016; Cirrone, 2018].

Mostly two types of target configurations (see Figure 26.53, left) have been employed in the laser-driven concept: the pitcher-catcher and the in-target configurations [Margarone, 2020; Margarone, 2022]. In the pitcher-catcher geometry, two targets are utilized: the first is the source of protons to be accelerated (mainly by TNSA) towards the second target, which contains boron. The in-target scheme involves a single target containing both nuclei. The two approaches have involved numerous strategies in the target manufacture or irradiation dynamics such as boron-polyethylene composite target [Belyaev, 2005], solid hydrogenated Si target dopped with boron [Picotto, 2014; Margarone, 2015], boron nitride targets [Giuffrida, 2020], double (combination of ns and ps) pulse approach [Labaune, 2013] or a single femtosecond laser pulse accelerating a proton beam [Baccou, 2015].

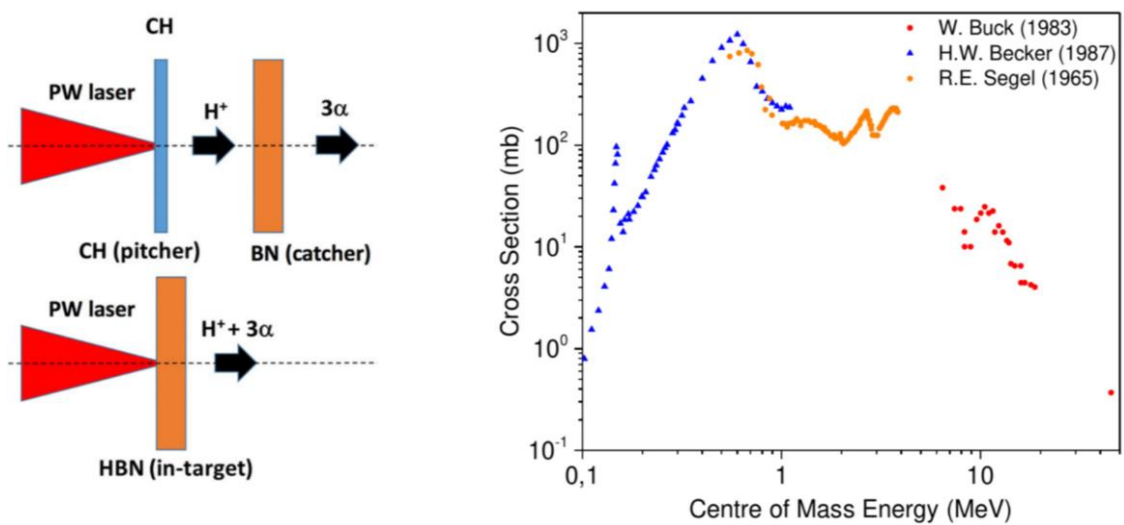

**Fig. 26.53:** Left panel: two schemes usually considered in laser-induced p-B reaction (in this case with BN), the pitcher-catcher (top) and the in-target (bottom) schemes [Margarone, 2020]. Right panel: cross-section of the p-B fusion reaction (note the logarithmic scales). The figure on the left is reprinted from [Margarone, 2020], Generation of α-particle beams with a multi-kJ, peta-watt class laser system, D. Margarone et al., Front. Phys. **8**:343, Copyright 2020, under Creative Commons Attribution 4.0. Retrieved from https://doi.org/10.3389/fphy.2020.00343; the figure on the right is courtesy of G.A.P. Cirrone and L. Giuffrida.

The p-B reaction requires overcoming the Coulomb barrier, and thus, a significant energy of the protons with respect to the boron for the cross section is important. The cross-section shown in Figure 26.53, right, peaks around 600 keV. Interestingly, the cross section exhibits a secondary resonance for energies around 148 keV [Cirrone, 2018].

This resonance at relatively modest energies was exploited during the Open Access campaign [Istokskaia, 2023] to achieve, via laser-induced p-B fusion reactions, several thousand neutron-less fusion reactions per second. The experiment consisted of an in-target approach by focusing, under high vacuum, the 1030 nm, 1.5 picosecond, and up to 20 mJ laser pulses to a size of a few microns onto a solid target. The boron-containing target (BN in this case) was covered with a 600 nm thick layer of polymer to create a plasma where the protons could be accelerated to sufficient energies, allowing for fusion reactions to take place. The thin polymer film was realized by plasma-assisted vapor phase deposition [Choukourov, 2016] and was composed of carbon and hydrogen in a ratio of hydrogen-to-carbon close to 1.9. It was the source of hydrogen and, hence, of protons to feed the p-B reactions.

Several passive CR39 track detectors were employed to monitor the generated alpha particles. They were positioned in front of the target to accumulate the imprints of alpha particles generated at various angles to reconstruct the angular distribution of particles. The laser beam was focused by a lens placed in a vacuum with 100 mm focal length and protected from target ablation debris by a thin transparent pellicle. Before a series of experiments, the focus position was adjusted by optimizing the ion signal from a dummy target. The ion signal was measured by a Faraday cup time-of-flight detector. Once the optimal position of the surface with respect to the lens was found, a boronated target was placed instead of the dummy targets with its surface at that same optimal position.

The intensity on the target reached a few $10^{16}$ W/cm$^2$. A pre-pulse, with about 300 times lower energy than in the main pulse, was impinging on the target about 14 ns prior to the main pulse and created a pre-plasma of H and B that was then heated up by the main pulse. The analysis of CR39 track detectors revealed the production of alpha particles up to 6700 per shot. The energy of alpha particles spread between 1 MeV and at least 4.5 MeV, with the maximum around 3.5 MeV. The angular distribution peaked around the normal to the target surface with 104 particles per steradian per shot and remained (103 particles per steradian per shot) up to the probed angles of 48 degrees from the normal.

It must be noted that the target was mounted on a specifically designed spiral rotation holder to allow irradiating a fresh target spot at each pulse while taking advantage of the kHz repetition rate of the PERLA B laser. The accumulation of 1000 shots per second could thus partly compensate for the modest energy per pulse (compared to, for example, petawatt-class laser installations) and yielded up to 106 alpha particles per second without full optimization (H content of the target, prepulse parameters, etc.). The good beam quality of PERLA beamlines also helped to achieve a decent intensity when focusing on the target. It is expected that the further development of compact laser sources combining kHz repetition rate and sub-Joule or Joule-level pulse energies could bring the alpha particle production rate to values close to $10^{10}$/s, comparable with pettawatt-class installations [Istokskaia, 2023].

**Experiments on laser-ablation for space propulsion**

Space missions regained a large interest since the last decade, and the laser-ablation propulsion is one of the hot topics in this area. A recent experiment was performed on the HiLASE Bivoj beamline in the frame of Open Access and supported by Laserlab Europe that was aiming at investigations of the repetitive regimes of laser irradiation with high energy per pulse for space propulsion. After a brief introduction on lasers for space applications, some results will be presented which were obtained during that campaign, taking advantage of the 10 Hz repetition rate and high energy per pulse of the Bivoj beamline.

Only a few years after the invention, lasers were already envisioned, on paper, for space propulsion as can also be seen in the reports for NASA in the 1970s and 1980s (see, e.g., [Myrabo, 1983]). Due to the laser ability of long-distance energy transfer, the source of energy powering the propulsion could be located on the ground and independent of the spacecraft itself. Consequently, the specific impulse with such an approach could be close to the ideal case for which a spacecraft is propelled for infinite times. Initial laser-based concepts of propulsion can be classified into three categories and combinations of them: laser-thermal [Nored, 1976], laser-electric [Landis, 1991; Bozek, 1993], and light-pressure based [Forward, 1962; Marx, 1966; Redding, 1967; Simmons and McInnes, 1993]. In the laser-thermal approach, the laser is meant to contribute to the heating up (or even ignition) of a solid, liquid, or gas propellant. In the laser-electric scheme, the laser energy is received by a transducer that converts the incident power to electricity, which is then delivered to a conventional electric thruster (like ion thrusters). The laser light-pressure concept simply considers the momentum transfer to a space vehicle due to the photon pressure, providing a force $F$ on the sample related to the incident power $P$ as $F = P/c$ (c is the speed of light) which can be doubled in the case of a reflecting surface. In parallel, a fourth category emerged, rather discretely in the beginning, employing a powerful laser for inducing ablation and thus gaining recoil pressure from the ablated products [Askar'yan and Moroz, 1963; Kantrowitz, 1972; Pirri and Weiss, 1972; Bunkin and Prokhorov, 1976]. A few decades later, Myrabo reported the launch of a flyer, up to ~70 m high, utilizing laser-ablation propulsion with microsecond $CO_2$ laser pulses [Myrabo, 2001]. It is emphasized that, despite the heating and vaporization enthalpy to overcome by the laser supply to reach ablation, the ablation-induced momentum transfer can be much more efficient than the simple photonic pressure. Indeed, typical momentum coupling coefficients, $C_m$, of appropriate materials are in the range from a few tens of N to 10 kN per megawatt of incident average power. This is well beyond the ~7 mN obtained for a megawatt laser by photonic pressure [Phipps, 2010].

Laser-ablation momentum transfer finally settled as a promising approach of laser-based propulsion and correction of trajectories of small space objects with an emphasis to address the issue of space debris accumulation around the Earth [Schall, 1991; Campbell, 1996; Phipps, 1996; Phipps, 2014]. As of February 2025, the European Spatial Agency reports 40500 space debris objects larger than 10 cm, 1.1 million between 1 cm and 10 cm, and 130 million debris between 1 mm and 1 cm size [ESA, 2025]. Unexpected break-ups, explosions of satellites (e.g., due to aging of a satellite with remaining fuels in the harsh space environment), collisions between satellites, or already existing debris generate an unpredictably growing number of them. This is of particular concern in the low Earth orbits as it can seriously threaten the current and future missions in space and reduce the lifetime of satellites already in operation. Significant efforts worldwide are being invested to remediate space debris and mitigate their impact [Esmiller, 2014; Scharring, 2018], in which laser-induced momentum transfer may be one of the keys. The idea is to irradiate a piece of debris with a ground-based or spaceborne laser to ablate it partially and thereby induce a change in its velocity, thus lowering its orbit and forcing it to re-enter the atmosphere for disintegration. Compared to capturing nets or space tugs, the laser solution benefits from its ability to deal with many debris with one instrument (thus limiting the cost of remediation missions) as well as from the possibility to target both the small, medium, and large debris [Phipps, 2014].

To induce a sufficient change in the velocity of a debris, the energy of the laser must be efficiently transferred into ablation and thus to the momentum of the target. The efficiency of the transferred momentum is measured by the coupling coefficient, $C_m$, expressed in *N/W*, ratio of the momentum acquired by the irradiated target *M* to the invested laser energy *W*. Since the ablation is material and fluence dependent for a given wavelength and pulse duration, the coupling coefficient of a target with a given laser depends on the incoming fluence. It is important to note that the $C_m$ does not behave linearly with the fluence or intensity. Its typical evolution with the fluence or intensity is shown in Figure 26.54, left. On the one hand, at low fluences, only heating and possibly melting of the target occurs. Thus, no ablation takes place, and the $C_m$ is zero. On the other hand, for fluences far above the material ablation threshold, most of the pulse energy contributes to the decomposition of the matter into plasma, and the efficiency drops. Therefore, for each material and irradiation regime (fs-, ps-, or ns- pulse durations and different wavelengths), it is crucial to characterize the coupling coefficient for different fluences to identify the optimum momentum coupling regimes.

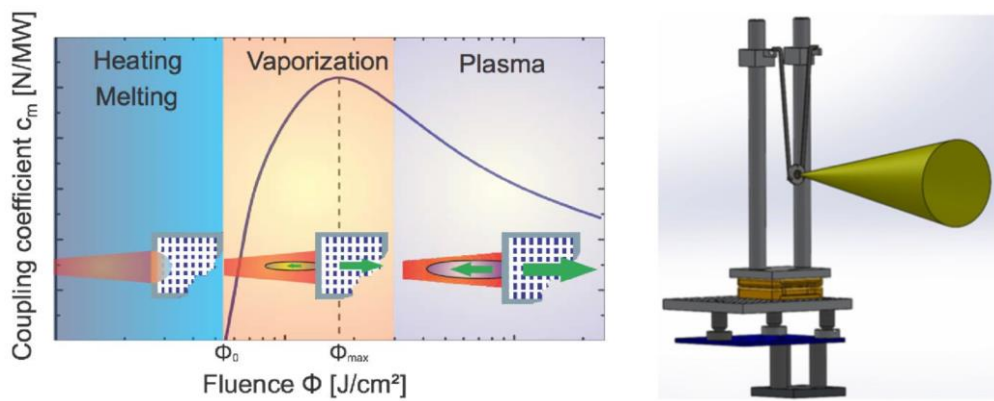

**Fig. 26.54:** Left: typical evolution of the coupling coefficient as a function of laser fluence incoming on the target [Esmiller, 2014]. Right: example of a pendulum designed for laser-induced momentum measurement. The cone indicates the laser beam [Phipps, 2017]. The figure on the left is reprinted from [Esmiller, 2014], C. Jacquelard, H.-A. Eckel, E. Wnuk, Space debris removal by ground-based lasers: main conclusions of the European project CLEANSPACE, B. Esmiller et al., Appl. Opt. **53**:45, Copyright 2014, with permission from Optical Society of America. The figure on the right is reprinted from [Phipps, 2017], Laser impulse coupling measurements at 400 fs and 80 ps using the LULI facility at 1057 nm wavelength, C.R. Phipps et al., J. Appl. Phys. **122**:193103, Copyright 2017, with permission of American Institute of Physics.

Additionally, pulsed lasers with significant repetition rates are needed for laser-ablation space debris removal [Phipps, 2014; Esmiller, 2014; Igarashi, 2022] to maximize the action during the limited time when a debris passes in the range of the laser site. Measuring the effects of the repetition of the laser pulse action is, therefore, also important.

To address some problems outlined above, the experiments were carried out recently using the HiLASE Bivoj beamline in the frame of Open Access and supported by Laserlab Europe [Boyer, 2024]. Several types of materials were irradiated at high fluences in the nanosecond, 10-Hz regime, the momentum transferred to targets upon shooting was measured, and the behavior and integrity of materials were investigated. To measure the momentum, the targets were mounted on a solid pendulum, as shown in Figure 26.54, right. The pendulum placed in a vacuum chamber was loaded with weights to ensure small amplitudes of its oscillations. The velocity of the pendulum can be tracked, for example, by a Doppler velocimeter or by a laser pointer deflected by a mirror mounted on the pendulum stem [Phipps, 2017]. From the fit of the evolution of the pendulum velocity, it is possible to retrieve the force and the impulse imparted to the pendulum during the short pulse irradiation. Other methods involve direct mounting of the target on a pressure gauge or piezoelectric sensors [Guo, 2015] or using a torsion pendulum [Gamero-Castaño, 2001; Rocca, 2006]. Additionally, the target is often monitored by a fast camera to follow its speed evolution for cross-checking or to observe and correlate possible ejecta with the post-irradiation analysis of the samples.

The experimental campaign on the Bivoj beamline employed the techniques mentioned above to obtain the single shot coupling coefficients at fluences of several tens of GW/cm$^2$. The analysis of the data and results on the $C_m$ have not yet been communicated at the time of the publication of this book. After irradiation, the morphology of the ablation craters on the samples was analyzed and spallation events assessed. Spallation is to be avoided since it can significantly diminish the lifetime of the target (number of pulses it can withstand) and may additionally generate millimetric size debris that can be harmful to the space environment. Preliminary analysis of the experimental campaign showed the limit of application of such high intensities in the 10-ns regime or irradiation for polymers like polyoxymethylene (POM). For such polymers in spite of good coupling coefficients, spallation occurred upon irradiation close to 70 GW/cm$^2$, thus seriously limiting the specific impulse of the target for propulsion while additionally generating millimetric size debris. Similar results were obtained for POM samples doped with aluminum that are even more attractive for laser-ablation momentum transfer due to their large $C_m$.

The Bivoj beamline, with its ns, 100 J level pulse energy and 10 Hz repetition rate, can offer valuable conditions for investigating space propulsion concepts based on laser-induced ablation, and more experimental campaigns are foreseen in this respect. Similarly, campaigns on the PERLA beamlines (not reported here) were performed to investigate the kHz repetition rate yielding coupling coefficients for polymers at 1 kHz. The results (not published yet) are of interest since the ps irradiation regime, depending on the material, is a priori a good candidate for significant coupling coefficients while drastically limiting the thermal coupling that induces melting instead of ablation [Phipps, 2017].

## 7. Conclusions

In this Chapter, present applications of high-power and high-intensity lasers have been overviewed, which have already occupied the specific niches in the fields of synthesis of new materials, printing of microelectronic devices, material surface micro/nanostructuring and selective annealing of semiconductor nanostructures, and generation of alpha particles. High-power lasers have tremendous potential for space propulsion, not only as a trust mean for spacecrafts but also for solving the problems associated with space debris accumulations around the Earth. They have wide horizons of expanding their use in different sectors of research and industrial applications, for instance attosecond laser pulses (see, e.g., [Wheeler, 2012] and Section 4 in Chapter 1 of this book [Nolte, 2026], high-intensity mid-IR laser pulses that can find applications in biomedicine or semiconductor industry (Section 4), few-circle laser pulses for ultra-clean material processing [Lenzner, 2000]. High-power ultrashort laser pulses can facilitate fundamental understanding of materials under extreme conditions (high energy coupling into material at ultrashort time in micro/nanovolumes) [Glenzer, 2007] that can bring discoveries of new effects, opening perspectives for novel technological breakthroughs. Regarding the technological application of high-power lasers, their use in material processing is still under development, with already successful results obtained, as discussed above. The ways to facilitate and accelerate laser material processing are being rapidly developing, see Chapters 17 [Förster and Neuenschwander, 2026] and 18 [Manek-Hönninger and Lopez] of this book where burst-mode processing is described, Chapter 22 [Schille and Löschner] where advanced optics for increasing productivity rates are overviewed, and Chapter 34 [Hennig, 2026] where different engraving strategies are discussed. The application of machine learning in laser texturing is envisioned to further widen the horizons for industrial applications of lasers, in particular, high-power lasers (Chapter 21 of this book [Steege, 2026]).